\documentclass[a4paper,11pt]{article}

\usepackage{mathtools}  
\usepackage{mathrsfs}
\usepackage{amsmath}
\usepackage{amssymb}
\usepackage[margin=0.75in]{geometry}
\usepackage{textcomp}
\usepackage{array}
\usepackage[usenames, dvipsnames]{color}
\usepackage[font=small]{caption}
\usepackage{floatrow}
\usepackage{subfig}
\usepackage{cite}
\usepackage[utf8]{inputenc}

\newcolumntype{C}{>{$}c<{$}}

\newcommand{\kahl}{K\"ahler}
\newcommand{\be}{\begin{equation}}
\newcommand{\ee}{\end{equation}}
\newcommand{\bea}{\begin{eqnarray}}
\newcommand{\eea}{\end{eqnarray}}
\newcommand{\ba}{\begin{align}}
\newcommand{\ea}{\end{align}}
\newcommand{\bi}{\begin{itemize}}
\newcommand{\ei}{\end{itemize}}
\newcommand{\w}{\ensuremath{\wedge}}
\newcommand{\ep}{\varepsilon}

\newcommand{\mc}{\mathcal}
\newcommand{\mbb}{\mathbb}

\newcommand{\h}{\hat}

\newcommand{\dd}{\mathrm{d}}
\newcommand{\dD}{\mathrm{D}}

\newcommand{\fs}{Y_4}
\newcommand{\hs}{X_3}
\newcommand{\hsz}{X_2}
\newcommand{\hb}{B_2}
\newcommand{\fb}{\mc{B}_3}
\newcommand{\bfs}{\hat{Y}_4}
\newcommand{\bfb}{\hat{\mc{B}}_3}

\newcommand{\fsg}{Y_{n+1}}
\newcommand{\hsg}{X_{n}}
\newcommand{\hbg}{B_{n-1}}
\newcommand{\fbg}{\mc{B}_{n}}
\newcommand{\bfsg}{\hat{Y}_{n+1}}

\newcommand{\hp}{\pi_H}
\newcommand{\fpr}{\pi_F}
\newcommand{\hsec}{\sigma}
\newcommand{\idb}{S}

\newcommand{\ec}{\chi}
\newcommand{\hbc}{\mc{C}}
\newcommand{\bim}{\mathfrak{g}}
\newcommand{\kdc}{\xi}
\newcommand{\hkm}{t}

\newcommand{\ind}{\mathrm{ind}}

\newcommand{\lb}{L}
\newcommand{\blb}{N}

\newcommand{\nd}{N}
\newcommand{\nf}{\mc{N}}
\newcommand{\fpc}{z}
\newcommand{\prb}{\xi}
\newcommand{\bc}{u}
\newcommand{\bct}{v}
\newcommand{\ptbc}{U}
\newcommand{\cf}{\mc{F}}
\newcommand{\cbr}{\zeta}
\newcommand{\fdm}{n_{\mathrm{def.}}}
\newcommand{\nch}{n_{\mathrm{ch.}}} 
\newcommand{\nhy}{n_{\mathrm{hyp.}}}
\newcommand{\nte}{n_{\mathrm{ten.}}}
\newcommand{\nve}{n_{\mathrm{vec.}}}
\newcommand{\numu}{n_{\mathrm{U}(1)}}
\newcommand{\hetfib}{F}
\newcommand{\fbc}{[W]}
\newcommand{\fbl}{W}
\newcommand{\fss}{Y_3}
\newcommand{\tacx}{x}
\newcommand{\tacy}{y}
\newcommand{\tacz}{w}
\newcommand{\fbs}{\mc{B}_2}
\newcommand{\bfss}{\hat{Y}_3}
\newcommand{\bfbs}{\hat{\mc{B}}_2}
\newcommand{\fkp}{p_F}
\newcommand{\hec}{\chi_\mathrm{hol.}}
\newcommand{\wsf}{f}
\newcommand{\wsg}{g}
\newcommand{\nolb}{\mathfrak{r}}

\newcommand{\ssu}{\mc{S}}
\newcommand{\scu}{\mc{C}}

\newcommand{\gw}{\mathrm{w}}

\usepackage{hyperref}

\begin{document}

\begin{centering}
\vspace*{1.2cm}
{\Large \bf NS5-Branes and Line Bundles in Heterotic/F-Theory Duality}

\vspace{1cm}

{\large
{\bf{Andreas P. Braun$^{a,}$}\footnote{andreas.braun@maths.ox.ac.uk}},
{\bf{Callum R. Brodie}$^{b,}$\footnote{callum.brodie@physics.ox.ac.uk}},  
{\bf{Andre Lukas$^{b,}$}\footnote{lukas@physics.ox.ac.uk}},
{\bf{Fabian Ruehle$^{b,}$}\footnote{fabian.ruehle@physics.ox.ac.uk}}
\bigskip}\\[0pt]
\vspace{0.23cm}
${}^a$ {\it 
Mathematical Institute, University of Oxford \\
Woodstock Road, Oxford OX2 6GG, UK
}
\\[2ex]
${}^b$ {\it 
Rudolf Peierls Centre for Theoretical Physics, University of Oxford\\
Keble Road, Oxford OX1 3NP, UK
}

\begin{abstract}\noindent
We study F-theory duals of heterotic line bundle models on elliptically fibered Calabi-Yau threefolds. These models necessarily contain NS5-branes which are geometrised in the dual F-theory compactifications. We initiate a systematic study of the correspondence between various configurations of NS5-branes and the dual geometries in F-theory and perform several checks of the duality. Furthermore, we discuss the singular transitions between different configurations of NS5-branes.
\end{abstract}
\end{centering}

\newpage

\tableofcontents

\newpage


\section{Introduction}
Over the years, F-theory and heterotic string theory have proven particularly promising candidates for providing a UV completion to the (supersymmetric extension of the) Standard Model (MSSM). The duality between these theories has been extensively studied (see e.g.~Refs.~\cite{Friedman:1997yq,Aspinwall:1997ye,Aspinwall:1998bw,Aspinwall:1998he,Diaconescu:1999it,deBoer:2001wca,Donagi:2008ca,Beasley:2008dc,Kumar:2009ac,Kumar:2010ru,Ludeling:2014oba,Braun:2017uku}) since the discovery of F-theory \cite{Vafa:1996xn,Morrison:1996na,Morrison:1996pp} more than 20 years ago. Despite this wealth of publications, the duality has been mostly studied within the context of the rather complicated set of heterotic vector bundle models using the spectral cover construction. For the class of heterotic line bundle models \cite{Anderson:2011ns,Anderson:2012yf,Anderson:2013xka,Blaszczyk:2015zta,Nibbelink:2015vha,Nibbelink:2016wms}, where most physical examples are known and whose heterotic description is more tractable, the duality has so far not been worked out in detail.

In this paper, we initiate a study of this duality. We consider line bundles sums in $\mathrm{E}_8 \times \mathrm{E}_8$, together with NS5-branes, on the heterotic side. In order to apply the standard heterotic/F-theory duality fiberwise in the stable degeneration limit, we need to restrict the heterotic model we start with. First, we have to assume that the Calabi-Yau (CY) on the heterotic side is elliptically fibered. We will also assume that this fibration has a holomorphic section, so we can mostly focus on a Weierstrass description. For the duality to hold we furthermore construct a heterotic vector bundle, given in terms of a sum of line bundles, that is flat on the fiber. As we shall discuss, this necessitates the inclusion of heterotic NS5-branes wrapping cycles in the base, or their M5-brane analogues in the Ho\v{r}ava-Witten M-theory description \cite{Horava:1995qa,Horava:1996ma}. NS5-branes typically feature prominently in heterotic constructions \cite{Lukas:1997fg,Lukas:1998yy} and the detailed study of their duality will be the main focus of this paper. Albeit discussed in the context of line bundle models, we expect many of the results of the 5-brane duality to carry over to other bundle constructions as well, since the majority of the discussion applies to M5-branes in the bulk and thus away from the $\mathrm{E}_8$ branes. For simplicity, we will focus on the case where the line bundles are embedded into the first $\mathrm{E}_8$ only; a generalisation to an embedding into both $\mathrm{E}_8$ factors should be straightforward. 

A crucial difference between the F-theory duals of bona fide heterotic vector bundles described by a spectral cover and heterotic line bundle models is that the latter have a trivial spectral sheet and the entire information of the bundle is captured by the spectral sheaf\footnote{Line bundle sums actually have no spectral cover description in the original construction of Ref.~\cite{Friedman:1997yq,Friedman:1997ih}, as they are not `regular' on the elliptic fiber. A slightly more general construction should allow a spectral cover description, though this is not important for our purposes, as we will not use this description.}. In the former description, the vector bundle, which breaks the primordial heterotic gauge group, is mapped onto geometry in such a way that the dual F-theory ADE singularity corresponds to the commutant of the original heterotic $\mathrm{E}_8\times\mathrm{E}_8$ gauge group with the structure group of the bundle. For the latter case of line bundle models, however, the bundle is not mapped onto geometry, and the F-theory dual still comes with the full $\mathrm{E}_8\times\mathrm{E}_8$ singularity. Since the structure group of each line bundle on the heterotic side is U(1) and hence commutes, the heterotic gauge group is broken to a non-Abelian (in general) subgroup times a collection of U(1) factors, which are, however, generically Green-Schwarz massive. On the F-theory side, breaking of the $\mathrm{E}_8\times\mathrm{E}_8$ symmetry is purely accounted for by a non-trivial $G_4$ flux on the $\mathrm{E}_8$ branes, onto which the heterotic line bundle data is mapped.

The aforementioned NS5- or M5-branes also exhibit an interesting behaviour under the duality. Depending on the cycle that is wrapped by the 5-brane, they are either mapped onto D3-branes or onto geometry on the F-theory side \cite{Rajesh:1998ik,Berglund:1998ej,Diaconescu:1999it}. However, on the heterotic side these different branes are on equal footing. Since the duality with and the physics of D3-branes is very well understood, we focus in this work on the duality between NS5-branes and geometry. As the F-theory geometry dual to heterotic line bundle models is in some respects quite clean, it provides a useful arena in which to study this duality. The world-volume theories of NS5-branes are notoriously hard to study and they exhibit interesting behaviour, such as the appearance of superconformal field theories from tensionless strings \cite{Heckman:2013pva,DelZotto:2014hpa,Heckman:2015bfa,Strominger:1995ac,Ganor:1996mu,Bershadsky:1996nu}.

The rest of the paper is organised as follows: in Section~\ref{sec:prelims} we introduce our notation and review the basics of heterotic line bundle models and heterotic/F-theory duality. We then describe how to construct F-theory duals of heterotic line bundle models. We demonstrate two crucial consequences of the condition to choose a flat bundle in the fiber direction on the heterotic side in order to construct the F-theory dual: first, we necessarily need NS5-branes and second, the resulting models cannot be chiral. In Section~\ref{sec:glob_6d_models} we discuss various aspects of the duality in compactifications to six dimensions. This serves to illustrate key concepts of the duality in a simpler and cleaner setup first. Our analysis heavily relies on toric geometry. After this, we move on to the discussion of compactifications to four dimensions in Section~\ref{sec:glob_4d_models}. On top of our discussion of 5-branes, we discuss the duality map for anomaly conditions, bundle stability, and the spectrum (i.e.\ the gauge group and matter representations) on the heterotic and F-theory side. We end the section with a discussion of a subtle case where an NS5-brane wraps a curve in the heterotic base that does not intersect the anti-canonical divisor of the base. In Section~\ref{sec:coin_and_int_5branes} we discuss the duality of multiple coincident and intersecting NS5-branes. We first describe in detail the relationship between the NS5-brane configuration and the F-theory geometry. When the NS5-branes intersect or coincide, there are singularities in the F-theory base. As the description of the four-dimensional theory resulting from compactifying F-theory on the resulting singular CY fourfold is challenging and subtle, we restrict to describing the three-dimensional theory which arises upon compactification of M-theory on the fourfold, which lifts to the four-dimensional theory. In Section~\ref{sec:Conclusion}, we summarise and provide an outlook on open problems and future research directions. Appendices~\ref{app:explicit_blowups_6d} and~\ref{app:explicit_blowups_4d} contain details of toric resolutions in the six-dimensional and four-dimensional models of Sections~\ref{sec:glob_6d_models} and~\ref{sec:glob_4d_models}, respectively.


\section{F-theory duals of heterotic line bundle models}
\label{sec:prelims}

\subsection{Heterotic line bundle models}
\label{sec:het_lbs}

For the heterotic compactification space we choose a Calabi-Yau (CY) $n$-fold $\hsg$. We need to further specify a vector bundle $V=V_1\oplus V_2$ with the structure group of $V_i$ contained in $\mathrm{E}_8$. For line bundle models, the vector bundle $V$ is taken to simply be a sum of line bundles,
\be
V  = V_1 \oplus V_2 = \bigoplus_{a=1}^\nolb \lb_a \,. \label{eq:LineBundleSumModel}
\ee
Each line bundle is associated to a particular U(1) factor in the Cartan subgroup of $\mathrm{E}_8 \times \mathrm{E}_8$. The first Chern classes of $V_1$ and $V_2$ need to be trivial,
\be
c_1(V_1)=c_1(V_2)=0 \,,
\ee
in order to ensure that the structure group of $V$ can be embedded into $\mathrm{E}_8 \times \mathrm{E}_8$, which is traceless. For the sake of brevity, we will often take $V_2$ to be trivial. We will write $\mc{O}_{\hsg}(D)$ for a line bundle with first Chern class or Chern character (Poincar\'{e} dual to) $D$,
\be
\mathrm{ch}_1(\mc{O}_{\hsg}(D))=c_1(\mc{O}_{\hsg}(D))=D \,.
\ee
A divisor $D$ can be expanded in a basis $\{D_I\}$ of $H_{2n-2}(\hsg,\mbb{Z})$, so we can write $D=k^ID_I$ where $k^I\in\mbb{Z}$, and this list of integers $k^I$ characterises the line bundle. Thus, line bundle sums \eqref{eq:LineBundleSumModel} are characterised by a matrix of integers $k_a^I$. For later computations involving the anomaly condition in compactifications to four dimensions, we will need the expression for the second Chern character of a line bundle sum on a CY threefold $\hs$, which reads
\begin{align}
 \mathrm{ch}_2(\lb_a) = \frac{1}{2}d_{IJK}k_a^Ik_a^JC^K \,, \qquad
 \mathrm{ch}_2(V) = \frac{1}{2}d_{IJK}\left(\sum_{a=1}^\nolb k_a^Ik_a^J\right)C^K  \,.
\end{align}
In these expressions, $d_{IJK}\equiv D_I \cdot D_J \cdot D_K$ are the triple intersection numbers of $\hs$, and $C^K$ are a basis of curves satisfying $C^I \cdot D_J = \delta^I_J$. We also note that for a vector bundle $E$ with $c_1(E)=0$, we simply have $\mathrm{ch}_2(E)= -c_2(E)$.

The gauge and compactification background have to be chosen such that they satisfy the anomaly cancellation and supersymmetry constraints. The former is ensured via the Bianchi identity of the NS three-form field $H$,
\be
\mathrm{d} H=\mathrm{ch}_2(V_1)+\mathrm{ch}_2(V_2)-\mathrm{ch}_2(\hsg) - \fbl \,. \label{eq:het_anom_canc}
\ee
Here, $\mathrm{ch}_2(V_i)$ and $\mathrm{ch}_2(\hsg)$ are the second Chern characters of the vector and the tangent bundle, respectively. Furthermore, we have allowed for the presence of NS5-branes that span the $d$ non-compact dimensions and wrap an internal $(6-d)$ cycle with homology class $\fbc$.

Unbroken supersymmetry requires that the vector bundle satisfies the Hermitian Yang-Mills equations, which can be achieved if the flux is a $(1,1)$-form and it is poly-stable with slope zero \cite{uhlenbeck1986existence,Donaldson:1985zz}. As line bundles are slope-stable, this reduces to a slope zero condition for each line bundle. The slope $\mu_{\hs}(\lb_a)$ of a line bundle $\lb_a$ on a CY threefold $\hs$ is defined as
\be
\mu_{\hs}(\lb_a):=\int_{\hs} J \w J \w c_1(\lb)=d_{IJK}\hkm^I\hkm^Jk_a^K \,,
\ee
where $J=t^ID_I$ is the K\"ahler form expanded in a basis of divisors $D_I$ with K\"ahler parameters $t^I$.

\medskip

As we shall discuss in the next subsection, the CY manifold $\hsg$ needs to be elliptically fibered over a base $\hbg$,\footnote{While the existence of a section is not required, we will focus on Weierstrass models that have a holomorphic section, the so-called zero section.}
\be
\begin{array}{ll}
\label{eq:Heterotic_FibrationStructure}
T^2\lhook\joinrel\xrightarrow{~~~~} & \hsg\\
&\;\Big\downarrow \hp \\ 
&\hbg
\end{array}
\ee
We will write $\hsec: \hbg \to \hsg$ for the section of the elliptic fibration, and $\hetfib$ for the generic fiber. We will be discussing the duality in compactifications to six dimensions ($n=2$) and four dimensions ($n=3$).

We now collect some general details and notation on this heterotic geometry in the case of compactification to four dimensions, which we will make use of throughout. In this case we have a CY threefold $\hs$ with an elliptic fibration $\hp: \hs \to \hb$. We write $\{\hbc^i\}$ for an integral basis of curves on $\hb$, and $\{\hbc_i\}$ for a dual basis satisfying $\hbc_i \cdot \hbc^j = \delta_i^j$. We also define $\bim_{ij}:=\hbc_i\cdot\hbc_j$. For an integral basis of curves $\{C^I\}$ and divisors $\{D_I\}$ on $\hs$, where $I=(0,i)$, we can then take
\begin{align}\begin{split}
C^0 = F \,, &\quad C^i = \hsec(\hbc^i) \,, \\
D_0 = \hsec(\hb) - \kdc^i D_i \,, &\quad D_i = \hp^*(\hbc_i) \,, \label{eq:div_cur_basis}
\end{split}\end{align}
where $\kdc^i = K_{\hb} \cdot \hbc^i$ and $\sigma(\hb) \cong \hb$. Here $K_{\hb}$ is the canonical divisor on $\hb$. Note we will freely use the same symbol for the canonical divisor and the canonical bundle. We will also use the same symbol for Poincar\'{e} dual forms and divisors. This divisor basis is chosen to be orthonormal to the curve basis, $C^I \cdot D_J = \delta_I^J$, as seen from the following intersection properties.
\begin{center}
\begin{tabular}{C|C|C}
\cdot			& \hsec(\hb) 			& \hp^*(\hbc')			\\ \hline
F        			& 1					& 0					\\ \hline
\hsec(\hbc) 		& \hbc \cdot K_{\hb}	&  \hbc\cdot\hbc'		\\ 
\end{tabular}
\quad\quad\quad
\begin{tabular}{C|C|C}
\cdot			& \hsec(\hb) 			& \hp^*(\hbc')			\\ \hline
\hsec(\hb) 		& K_{\hb}				& \hsec(\hbc')				\\ \hline
\hp^*(\hbc) 	& \hsec(\hbc)			&  (\hbc\cdot\hbc')\hetfib	\\ 
\end{tabular}
\end{center}
In particular, we can identify $\hkm^0$ as the volume of the fiber in the {\kahl} form $J = \hkm^ID_I$. The triple intersection numbers in this basis are
\be
 d_{000}=\kdc_i \kdc^i \,,	\quad  d_{00i}=-\kdc_i \,, 		\quad d_{0ij}=\bim_{ij} \,,	 \quad d_{ijk}=0 \,. \label{eq:X3IntersectionRing}
\ee
Additionally, the second Chern class of the tangent bundle is (see e.g.~Ref.~\cite{Friedman:1997yq})
\begin{align}\begin{split}
c_2(\hs) = -\mathrm{ch}_2(\hs) &= \hp^*\left(c_2(\hb)\right)+11\hp^*\left(c_1(\hb)^2\right)+12\hsec(c_1(\hb)) \\
&=  \left[\int_{\hb} (c_2(\hb)+11c_1(\hb)^2)\right]\hetfib-12\hsec(K_{\hb}) \,,
\label{eq:c2X}
\end{split}\end{align}
and we will need this expression in computations involving the anomaly condition.

In compactifications to four dimensions on an elliptically fibered CY manifold $\hs$, the heterotic NS5-branes can wrap either (i) the fiber, or (ii) a curve in the base\footnote{This characterisation is slightly crude. We will be more precise in Section~\ref{sec:glob_4d_models}.}, or (iii) some combination. We refer to the first kind as `vertical' NS5-branes, and to the second as `horizontal' NS5-branes. In a Ho\v{r}ava-Witten picture, heterotic NS5-branes are M5-branes with positions in the bulk of the 11d interval $S^1/\mbb{Z}_2$, and we will often make use of this picture. At the two ends of the interval, M5-branes transition into small instantons on the respective $\mathrm{E}_8$ brane. We will occasionally refer to `horizontal' or `vertical' instantons, depending on the type of M5-brane they originate from under the small instanton transition.

\subsection{F-theory models and fiberwise duality}
\label{sec:fthmod_fibdual}

Let us now turn to F-theory duals of heterotic models. Similarly to the heterotic case, we need to specify the compactification data, which means choosing a CY $(n+1)$-fold $\fsg$ and a four-form flux $G_4$. We will discuss the $n=2$ and the $n=3$ case, the first in order to develop understanding, and the second as compactifications to four dimensions will be our primary interest. In the $n=3$ case, the F-theory analogue of the heterotic Bianchi identity is D3-brane tadpole cancellation \cite{Andreas:1997ce,Andreas:1999zv}, which reads \cite{Sethi:1996es}
\be
\nd_3-\frac{\ec(\fs)}{24}+\frac{1}{2}\int_{\fs}G_4 \w G_4 = 0 \,,
\ee
where $\nd_3$ is the number of D3-branes and $\ec(\fs)$ is the Euler number of $\fs$.

The compactification background $\fsg$ needs to be elliptically fibered and -- in case of duality to the heterotic string -- the base of the elliptic fibration needs to furthermore be $\mbb{P}^1$ fibered (so that $\fsg$ is fibered by K3 surfaces with an elliptic fibration),
\be
\label{eq:Ftheory_FibrationStructure}
\begin{array}{ll}
T^2\lhook\joinrel\xrightarrow{~~~~} & \fsg\\
&\;\Big\downarrow~\fpr\\
\mbb{P}^{1}\lhook\joinrel\xrightarrow{~~~~}&\fbg\\
&\;\Big\downarrow~\tilde{\pi}_{F}\\
&\hbg
\end{array}\,,\qquad
\begin{array}{ll}
K3\lhook\joinrel\xrightarrow{~~~~} & Y_{n+1}\\
&\;\Big\downarrow~\fkp\\
&\hbg
\end{array}\,.
\ee
Note that the base $\hbg$ in \eqref{eq:Ftheory_FibrationStructure} is identified with the base $\hbg$ in \eqref{eq:Heterotic_FibrationStructure}, which is why we use the same symbol. We will refer to the extra $\mbb{P}^1$ which appears in F-theory, i.e.\ the base of the elliptic fibration on the K3 surface, as the `F-theory~$\mbb{P}^1$'.

This duality can be understood by fiberwise application of heterotic F-theory duality in 8d. Here, heterotic strings on a $T^2$ with Wilson line moduli was proposed to be dual to F-theory on a K3 surface \cite{Vafa:1996xn},
\begin{align*}	
\text{Heterotic on }T^2 \xleftrightarrow{\text{~~dual~~}}\text{ F-theory on K3}\,,
\end{align*}
due to the equivalence of their moduli spaces.

In order to capture compactifications of heterotic string theory in the supergravity limit, the K3 fiber of the F-theory background must be in the stable degeneration limit, in which the K3 fiber degenerates to a union of two rational elliptic surfaces \cite{Friedman:1997yq,Aspinwall:1997ye,Aspinwall:1998bw}. The extension of the duality to compactification to fewer dimensions consists of fibering both sides over a common space, giving an elliptically fibered and a K3 fibered CY manifold respectively. Besides guaranteeing that the elliptic fiber on the heterotic side is large compared to the string scale, the stable degeneration limit is also crucial to ensure that the heterotic background is described by conventional geometry, i.e.\ not patched together by maps involving T-dualities, see Refs.~\cite{Braun:2013yla,Garcia-Etxebarria:2016ibz} for a discussion of the relevant geometry. 

The limit in which the heterotic string becomes weakly coupled corresponds to the adiabatic limit of the K3 fibration on the F-theory side, that is the limit in which the volume of the `F-theory $\mathbb{P}^1$' (the base of the elliptic fibration on the K3 surface) becomes small compared to the (properly normalised) volumes of $\hbg$ (the base of the K3 fibration). This is in line with thinking about the F-theory $\mbb{P}^1$ as a fibration of $S^1$ over an interval, which is identified with the interval of the M-theory lift of heterotic strings. 

Similarly, we will assume that the elliptic fibration on the heterotic side is adiabatic in the sense that the volume of the fiber $T^2_{\mathrm{Het.}}$ is small compared to volumes of the base $\hbg$. Note that in the conventional application of duality between heterotic string theory and F-theory employing vector bundles constructed by means of spectral covers, this condition ensures stability of the bundles \cite{Friedman:1997yq,Friedman:1997ih}. The required hierarchies of volumes can then be summarised as,
\be
\mathrm{vols}(\hbg) \gg \mathrm{vol}(T^2_{\mathrm{Het.}}) \to \infty \,, \quad \mathrm{vols}(\hbg) \gg \mathrm{vol}(\mbb{P}^1_{\mathrm{F-th.}}) \to 0 \,.
\ee

\medskip

We now collect some general details and notation on this F-theory geometry in the case of compactification to four dimensions. In this case we have a CY fourfold $\fs$ with a K3 fibration $\fkp:\fs\to\hb$ and an elliptic fibration $\fpr:\fs\to\fb$. We will often represent the generic fiber of the elliptic fibration in a $\mbb{P}_{123}$ ambient space whose coordinates we denote $\{\tacx,\tacy,\tacz\}$. The F-theory elliptic fibration, as on the heterotic side, is characterised by a Weierstrass equation, which describes the fiber in its ambient space over each point in the base,
\be
\tacy^2 = \tacx^3 + \wsf \tacx \tacz^4 + \wsg \tacz^6 \,,
\label{eq:wei_eq}
\ee
where the functions appearing are sections, $\wsf = \Gamma(K_{\fb}^{-4})$ and $\wsg = \Gamma(K_{\fb}^{-6})$. The elliptic curve degenerates over the locus defined by the vanishing of the discriminant
\be
\Delta = 4\wsf^3+27\wsg^2 \,,
\ee
and the vanishing orders of $(\wsf,\wsg,\Delta)$ reflect the severity of the degeneration.

If the vanishing orders of $(\wsf,\wsg,\Delta)$ are $(4,6,12)$ or higher in codimension $\geq2$, as in the cases we discuss, resolving with a blow-up which does not change the base of the elliptic fibration results in a `non-flat fibration', in which the fiber dimension can jump. On this fourfold, tensionless strings appear in the F-theory limit, and the resulting theory is expected to have no standard supergravity description. Alternatively, for such singularities, one can find crepant resolutions which change the base of the elliptic fibration, resulting in a flat fibration with an $\mc{N}=1$ supergravity description in the F-theory limit \cite{Morrison:1996pp}. We will be interested in resolutions of the latter type.

\subsection{F-theory duals of heterotic line bundle models}

Above we have discussed the heterotic and F-theory compactifications and the basic idea behind the duality between them, and this completes the general review. Now we turn to analysing the F-theory duals of heterotic line bundle models. In this section we will give a basic description of many aspects of the duality which we will cover in more detail in the remainder of the paper.

\medskip

As mentioned above, historically the spectral cover construction has been particularly useful for constructing F-theory duals of heterotic vector bundle models. In the original 8d duality, the heterotic bundle on the elliptic fiber has only Wilson line moduli. If we consider lower-dimensional dualities obtained by fibering the 8d duality, the vector bundles restricted to the generic elliptic curve should retain this property. Consequently, on each elliptic fiber of the heterotic threefold, each of the $\mathrm{E}_8$ vector bundles is determined by a set of points \cite{Friedman:1997yq}. This provides a branched covering of the heterotic base called the `spectral sheet' which describes the deformations of the $\mathrm{E}_8$ singularities on the F-theory side. In addition to the spectral sheet, the description of the heterotic vector bundles requires also a `spectral sheaf' on the spectral sheet, together giving the spectral cover. This information can then be translated into F-theory geometry and flux.

However, for duals of heterotic line bundle models the spectral cover construction becomes essentially trivial. The condition that a bundle restricts on the generic elliptic fiber to give only Wilson line moduli means that it restricts to a sum of degree zero line bundles. For a global line bundle sum $V$, each of the line bundles in $V$ has to be trivial on the fiber, and hence is a pullback bundle from the base,
\be
V = \bigoplus_a^{\nolb} \hp^* \left( \blb_a \right) \,, \label{eq:LBS_FiberFlat}
\ee
where $\blb_a$ are line bundles on the base $\hbg$. Recalling the divisor basis defined in Equation~\eqref{eq:div_cur_basis}, this condition is equivalent to requiring that the matrix of integers $k_a^I$ determining the line bundle sums satisfies
\be
k_a^0 = 0 ~~ \forall a \,.
\ee
Since the restriction of a pullback bundle to the elliptic fiber is trivial, the spectral sheet is trivial -- it is $\nolb$ copies of the zero section -- and the information about the bundle is encoded entirely in the spectral sheaf. Trivial spectral sheets correspond to two undeformed surfaces of $\mathrm{E}_8$ singularities in the F-theory dual, each of which is diffeomorphic to the heterotic base $\hbg$. We will frequently refer to these as the $\mathrm{E}_8$ branes. As the two heterotic bundles are pullbacks from $\hbg$, we naturally expect the fluxes on the two $\mathrm{E}_8$ stacks to be equal to the fluxes pulled back from $\hbg$, or equivalently that the two spectral sheaves correspond to the two bundles on $\hbg$. This constitutes a main difference to the situations usually studied in pure F-theory constructions or in heterotic/F-theory duality with bona fide vector bundles: the $\mathrm{E}_8$ symmetries on the F-theory side are broken \emph{purely by flux}, rather than by geometry.

\medskip

In order to link the flux $F$ in the heterotic theory and the flux $G_4$ in F-theory, we can use the M-theory description. In this description, we resolve the singular F-theory space $\fsg$ to a smooth space $\bfsg$. Next, we can expand $G_4$ in terms of two-forms $\gw_a^{(1)}$, $\gw_b^{(2)}$ ($a,b = 1, \ldots , \text{rk}(\mathrm{E}_8)$) dual to the exceptional divisors of the blow-up of the two $\mathrm{E}_8$ singularities,
\be
G_4 = \tilde{F}^{(1)\;a} \w \gw_a^{(1)} +\tilde{F}^{(2)\;a} \w \gw_a^{(2)}  \,, \label{eq:g4_exp}
\ee
where $\tilde{F}^{(1)\,a}$ and $\tilde{F}^{(2)\,a}$ are pullbacks of flux two-forms from the $\mathrm{E}_8$ branes. In the duality for heterotic line bundle models, from Equation~\eqref{eq:LBS_FiberFlat} the identification of flux is rather clear: the fluxes $\tilde{F}^{(1)\,a}$ and $\tilde{F}^{(2)\,a}$ correspond to the two $\mathrm{E}_8$ fluxes on the heterotic base $\hbg$ that are pulled back to the CY manifold $\hsg$. One could also phrase this in a local picture of the $\mathrm{E}_8$ branes, in which again the flux identification is clear. In Section~\ref{sec:glob_4d_models} we will discuss this duality of fluxes in much greater detail.

\medskip

The restriction \eqref{eq:LBS_FiberFlat} on the heterotic gauge flux has far-reaching consequences for compactifications to four dimensions, which will be analysed in the rest of the paper. First, in order to satisfy the Bianchi identities \eqref{eq:het_anom_canc}, we necessarily have to include horizontal NS5-branes, wrapping curves in the base $\hb$. The class of these NS5-branes is easily seen to be $-12K_{\hb}$, as we discuss in Section~\ref{sec:req_branes_and_blowups} below. The F-theory duals of NS5-branes are well-known: while vertical NS5-branes are mapped to D3-branes~\cite{Friedman:1997yq}, horizontal branes are mapped to geometry, or more precisely to blow-ups~\cite{Morrison:1996pp,Andreas:1997ce,Andreas:1999zv}, 
\begin{align*}
\begin{array}{lcl}
\text{vertical heterotic NS5/M5-branes} 		&	\xleftrightarrow{~\text{dual}~}	& 	\text{D3-branes}		\\
\text{horizontal heterotic NS5/M5-branes} 	&	\xleftrightarrow{~\text{dual}~}	&	\text{blow-ups in } \fb	\\
\end{array} \,.
\end{align*}

In particular, the blow-ups are of the following kind. On the F-theory side, the base of the elliptic fibration is in simple cases a $\mbb{P}^1$ times the heterotic base $\hb$. The blow-ups dual to the NS5-branes are then over curves in the base $\fb$ that sit at either end of the F-theory $\mbb{P}^1$. Each of these ends is diffeomorphic to the heterotic base $\hb$, and the contained curves over which blow-ups are performed correspond to the curves wrapped by the NS5-branes. These blow-ups precisely resolve the singularities of the F-theory fourfold $\fs$ where the vanishing orders of $(\wsf,\wsg,\Delta)$ are (4,6,12) or higher, as discussed in detail in Section~\ref{sec:glob_6d_models}. The {\kahl} moduli of these blow-ups correspond to the positions of the corresponding M5-branes in the Ho\v{r}ava-Witten interval. In Section~\ref{sec:glob_6d_models} we will give a detailed description of this aspect of the duality in the simpler case of compactification to six dimensions. This aspect will then form a key ingredient in the global models we construct in Section~\ref{sec:glob_4d_models}.

We will freely switch between referring to the branes as NS5- or M5-branes in this paper. The picture of heterotic M-theory, where NS5-branes become M5-branes, gives the right geometric intuition in that two NS5-branes which wrap coincident or intersecting curves in the CY manifold are not truly intersecting branes unless their positions in the interval $S^1/\mbb{Z}_2$ of the Ho\v{r}ava-Witten M-theory description coincide.

Second, in compactifications to four dimensions, line bundle sums subjected to \eqref{eq:LBS_FiberFlat} are, unfortunately, necessarily non-chiral, i.e.\
\begin{align}
\chi(\hs,V)=\int_{\hs} \text{ch}(V)\text{td}(\hs)=0\,, \label{eq:HRR}
\end{align}
where td$(\hs)$ is the Todd class,
\be
\mathrm{td}(\hs) =1+\tfrac{1}{2}c_1+\tfrac{1}{12}(c_1^2+c_2) +\tfrac{1}{24}c_1c_2+\ldots \,, \label{eq:td_class}
\ee
in which we have written $c_i(\hs) \equiv c_i$ on the right. This can be seen by expanding $\text{ch}(V)$ and $\text{td}(\hs)$ in terms of a divisor basis $D_I$. Using $c_1(V)=c_1(\hs)=0$, $k_a^0=0$ due to flatness of the flux along the fiber \eqref{eq:LBS_FiberFlat}, and the intersection ring properties \eqref{eq:X3IntersectionRing} (in particular that $d_{ijk}=0$), we find that \eqref{eq:HRR} vanishes identically. This means that pure heterotic line bundle sums that are flat on the fiber are not immediately useful for model-building purposes. It is an interesting question whether the duality could be extended to more general line bundle sums, see e.g. Ref.~\cite{Anderson:2014gla} for some progress in this direction. In this paper, however, we restrict ourselves to this class of line bundle sums\footnote{Much of what we will discuss below will apply not just to such line bundle sums but to any pullback bundle.}. 

\medskip

The duality between the heterotic line bundle flux in the two $\mathrm{E}_8$ gauge groups and the F-theory $G_4$ flux on the two $\mathrm{E}_8$ branes is rather simple. It is hence rather plausible that checks of this aspect of the duality, for example of the duality of stability conditions and of the spectrum in this sector, will work out. We will perform these checks in Section~\ref{sec:dual_checks_ind_of_ns5_config}. These are essentially independent of the duality between NS5-branes and blow-ups. However, we choose to first deal with this latter aspect of the duality, in order to have the full F-theory geometry at hand.


\section{Blow-ups and branes: review and global 6d models}
\label{sec:glob_6d_models}

We would like to build global F-theory models dual to heterotic line bundle models, in compactifications to four dimensions. As discussed in Section~\ref{sec:prelims}, the duality between horizontal NS5-branes and F-theory geometry is the most non-trivial aspect of this duality. In this section we will discuss this in detail, in the simpler case of compactification to six dimensions. This is intended to be a pedagogical discussion, to build intuition for our later construction of global models in four dimensions. We first review what is known in the literature, and flesh out ideas in toy examples. We will be particularly interested in the relevant aspects of toric descriptions. In Section~\ref{sec:glob_models_in_6d}, we build global models for compactification to six dimensions. When we then turn to the description of global models for compactification to four dimensions in Section~\ref{sec:glob_4d_models}, we will freely use the intuition developed here.

\subsection{Singularities and small instantons in 6d}
\label{sec:sing_and_inst_in_6d}

First we discuss the duality between singularities of the F-theory threefold and heterotic small instantons. We first briefly recall aspects of the six-dimensional duality. The heterotic compactification space is a K3 surface $\hsz$. We will choose to have no heterotic flux, since the purpose of our discussion of compactification to six dimensions is to develop an understanding of the duality between heterotic NS5-branes and F-theory geometry. (When we discuss the four-dimensional case we will include flux, in the form of line bundles, since four-dimensional line bundle models are our primary interest.) Since we have no heterotic flux, there must be NS5-branes: as the second Chern class of a K3 surface integrates to 24, the anomaly cancellation condition \eqref{eq:het_anom_canc} implies that there are 24 NS5-branes or, before their movement into the bulk of the Ho\v{r}ava-Witten interval, 24 small instantons. These span the 6 non-compact dimensions and are point-like in the compact space.

The dual F-theory space is a K3 fibration over a $\mbb{P}^1$ and an elliptic fibration over a twofold base $\fbs$. Hence $\fbs$ is a fibration of $\mbb{P}^1$ over $\mbb{P}^1$, and hence is a Hirzebruch surface $\mbb{F}_m$. The type $m$ of this Hirzebruch surface is matched on the heterotic side by the distribution $(12-m,12+m)$ of the 24 instantons on the two $\mathrm{E}_8$ branes. There are $\mathrm{E}_8$ singularities at either end of the F-theory $\mbb{P}^1$, since on the heterotic side there is no flux. At higher codimension there will be singularities worse than $\mathrm{E}_8$, whose resolution requires blow-ups in the base. In particular as we will see there are 24 such points. This is the same number as of NS5-branes, as expected since in six dimensions all NS5-branes are dual to blow-ups, as we will review.

\bigskip

We first recall the geometry of singularities in F-theory arising where the remaining discriminant locus intersects the $\mathrm{E}_8$ singularities. We assume a trivial fibration $\fbs = \mbb{P}^1 \times \mbb{P}^1$ for the F-theory base. The result after blow-ups will be independent of this choice; this is clear since on the heterotic side, whatever the initial instanton distribution the end result will be the same, with all M5-branes pulled into the bulk. We write the F-theory $\mbb{P}^1$ as $\mbb{P}_\fpc$ with coordinates $\{\fpc_1,\fpc_2\}$, and the heterotic base, which is the other $\mbb{P}^1$ in $\fbs$, as $\mbb{P}^1_\bc$ with coordinates $\{\bc_1,\bc_2\}$. The canonical divisor of $\fbs$ is the sum
\be
K_{\fbs} = K_{\mbb{P}^1_\fpc} + K_{\mbb{P}^1_\bc} \,,
\ee
where for simplicity we do not write the obvious pullback maps. We recall that in the Weierstrass polynomial in Equation~\eqref{eq:wei_eq}, we have $f \in \Gamma(K_{\fbs}^{-4})$, $g \in \Gamma(K_{\fbs}^{-6})$, and
\be
\Delta = \left(4\wsf^3+27\wsg^2\right) \in \Gamma(K_{\fbs}^{-12}) \,.
\ee
Since $K_{\mbb{P}^1}=\mc{O}_{\mbb{P}^1}(-2)$, $\wsf$ and $\wsg$ are respectively of bidegree (8,8) and (12,12) in the coordinates of $\mbb{P}_\fpc^1$ and $\mbb{P}_\bc^1$. Hence they can be written as
\be
\wsf = \sum_{i=0}^8 \fpc_1^i\fpc_2^{8-i} \wsf_i \,, \quad \wsg = \sum_{i=0}^{12} \fpc_1^i\fpc_2^{12-i} \wsg_i \,,
\ee
where $\wsf_i \in \Gamma(K_{\mbb{P}^1_{\bc}}^{-4})$ and $\wsg_i \in \Gamma(K_{\mbb{P}^1_{\bc}}^{-6})$. To enforce the presence of the two $\mathrm{E}_8$s at $\fpc_1=0$ and $\fpc_2=0$ requires vanishing orders $(\wsf,\wsg,\Delta)\sim(\geq4,5,10)$, so that in the above expansion very few terms can be present. This constrains $\wsf$, $\wsg$, and $\Delta$ to 
\be
\wsf = \fpc_1^4\fpc_2^4\wsf_4 \,, \quad \wsg = \fpc_1^5\fpc_2^5\left( \fpc_1^2\wsg_7+\fpc_1\fpc_2\wsg_6+\fpc_2^2\wsg_5\right) \,, \quad \Delta = \fpc_1^{10}\fpc_2^{10}\Delta_r \,,
\ee
\be
\Delta_r \equiv 4\fpc_1^2\fpc_2^2\wsf_4^3+27\left(\fpc_1^2\wsg_7+\fpc_1\fpc_2\wsg_6+\fpc_2^2\wsg_5\right)^2  \,,
\ee
where we have defined the `remaining discriminant' $\Delta_r$. We also note that $\wsf_4$ and $\wsg_6$ are identified in the duality with the $\wsf$ and $\wsg$ polynomials in the Weierstrass equation for the heterotic K3 surface \cite{Morrison:1996pp},
\be
\textrm{Heterotic K3:} \quad \tacy^2 = \tacx^3+\wsf_4\tacx\tacz^4+\wsg_6\tacz^6 \,,
\ee
where for simplicity we have used the same coordinate names for the $\mbb{P}_{123}$ ambient space of the heterotic elliptic curve as we did on the F-theory side.

The degeneration of the fiber worsens at the intersection of the $\mathrm{E}_8$ brane stacks with the rest of the D7-brane locus $\{\Delta_r=0\}$. Setting $\fpc_1=0$ or $\fpc_2=0$ in $\Delta_r$, and fixing the other coordinate to 1, we have respectively
\be
\left.\Delta_r\right|_{\fpc_1=0} = 27\wsg_5^2 \,, \quad \left.\Delta_r\right|_{\fpc_2=0} = 27\wsg_7^2 \,.
\ee
Each of these intersections consists of 12 double points, for generic $\wsg_7$ and $\wsg_5$. At these points, the vanishing orders of $(\wsf,\wsg,\Delta)$ are $(4,6,12)$, so the fiber degenerations are so severe that resolutions not changing the base lead to a non-flat fibration.

\bigskip

As we now explain, the singularity at these points is characterised in affine coordinates by an equation of form
\be
0 = \tacy^2+\tacx^3+\fpc^6+\bc^6 \,,
\label{eq:hyp_e8}
\ee
and we will refer to this singularity as an $\tilde{\mathrm{E}}_8$. As a weighted homogeneous singularity, its type depends only on the weights of the coordinates: using the scalings $\tacy\sim\lambda^{1/2}, \tacx\sim\lambda^{1/3},\ldots$, additional terms that scale as $\lambda$ are `marginal deformations'. The Weierstrass equation for the above models is of this form in the vicinity of an $\tilde{\mathrm{E}}_8$ singularity: if a zero of $\wsg_5$ is at $\bc_1=0$, then close to the singularity and in affine coordinates we have up to numerical factors
\be
0 = \tacy^2 + \tacx^3 +(\fpc_1^4+\ldots)\tacx + (\fpc_1^5\bc_1+\ldots) \,,
\ee
which expresses the $\tilde{\mathrm{E}}_8$ singularity, sitting in a locus of $\mathrm{E}_8$ singularities. Crucially this singularity admits a crepant resolution, in which an extra coordinate $\ep$ is introduced, with a new scaling relation
\be
\begin{tabular}{c c c c c}
$\tacy$ & $\tacx$ & $\fpc_1$ & $\bc_1$ & $\ep$ \\
\hline
3 & 2 & 1 & 1 & -1
\end{tabular} \,,
\label{eq:hyp_e8_res}
\ee
and the singular point is excised. The proper transform corresponds to dividing out 6 powers of $\ep$ from the equation defining the hypersurface, so that this is a crepant resolution. Conversely, $\wsf$ and $\wsg$ have to be divisible by $\ep^4$ and $\ep^6$ respectively for such a blow-up to be crepant, reflecting the requirement of vanishing orders (4,6,12). These are the resolutions which are dual to the introduction of NS5-branes and the {\kahl} moduli associated to the blow-ups correspond to positions of the M5-branes in the Ho\v{r}ava-Witten interval. We will discuss this in some depth in Section~\ref{sec:tor_desc_in_6d}. However first we must review the explicit duality between the $\tilde{\mathrm{E}}_8$ singularities and small instantons.

\bigskip

\begin{figure}[t]
\centering
\includegraphics[scale=1.5]{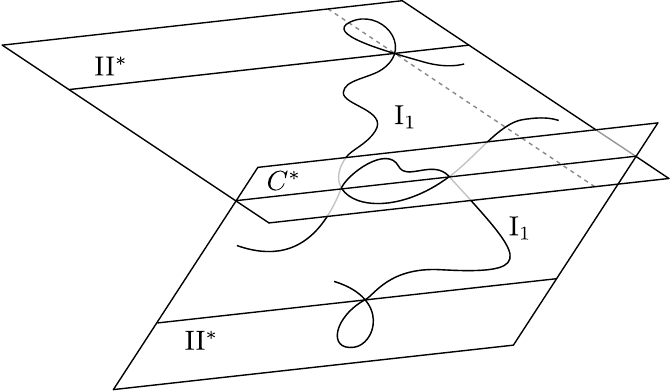}
\caption{Schematic depiction of the F-theory base after stable degeneration of a Hirzebruch base. Note this is a rendition of Figure 3 in Ref.~\cite{Aspinwall:1997ye}. In this picture there are no five-branes in the heterotic dual, only small instantons. Note $C_*$ is identified with the heterotic base.}
\label{fig:stabdegbase}
\end{figure}

The presence of (4,6,12) singularities is dual to the presence of heterotic small instantons. Beyond the match of their numbers, the explicit duality between singularities and small instantons has been discussed in Ref.~\cite{Aspinwall:1997ye}, and we briefly review this discussion. Before stable degeneration, the F-theory base is a Hirzebruch surface. In the stable degeneration limit, the F-theory $\mbb{P}^1$ splits into two $\mbb{P}^1s$ which meet at a point. Figure~\ref{fig:stabdegbase}, which is a rendition of Figure 3 in Ref.~\cite{Aspinwall:1997ye}, shows a schematic depiction of the base of the F-theory fibration after stable degeneration. This base consists of two Hirzebruch surfaces that meet along a curve $C_*$ that we identify with the $\mbb{P}^1$ base of the heterotic K3, and the elliptic fibration over $C_*$ is identified with the heterotic K3. We can note that the II$^*$ fibers are away from this intersection.

The remaining I$_1$ locus intersects the II$^*$ fibers at a set of points. Each point sits in a particular copy of a component of the F-theory $\mbb{P}^1_{\fpc}$ fiber over a point in the heterotic $\mbb{P}^1_{\bc}$. This F-theory $\mbb{P}^1_{\fpc}$, represented by the dotted line in Figure~\ref{fig:stabdegbase}, intersects $C_*$ at one point. Since $C_*$ is identified with the heterotic base, it is natural to identify this point with the position of the corresponding heterotic small instanton in the heterotic base. Additionally, the small instanton has a position in the heterotic fiber, which corresponds in the duality to Ramond-Ramond moduli (in the IIB picture) on the F-theory side. This aspect will mainly not concern us, since in four dimensions the horizontal NS5-branes/small instantons usually cannot move in the heterotic fiber, however it will be important in the discussion of Section~\ref{sec:dualchecks_ns5s_and_blowups}.

This moduli match for small instantons also holds for NS5-branes. The transition from a small instanton to an NS5-brane corresponds to a position in the 11d bulk: a position at either endpoint of the interval gives a small instanton and positions in between give an NS5-brane. Moving an instanton into the bulk corresponds on the F-theory side to resolving the singularities discussed above, where {\kahl} moduli are dual to the positions in the bulk. Throughout this transition, the position of the instanton/NS5-brane in the heterotic space and the position of the singularity/blown-up locus in the F-theory space remain unchanged, so the matching between positions of the heterotic small instantons and the positions of $(4,6,12)$ singularities in F-theory carries over to a match of NS5-branes and the loci of blow-ups. In the K3 fibration of the resolved threefold $\bfss$ on the F-theory side, these become the loci of reducible K3 fibers. In perfect agreement with the spirit of fiberwise duality, the monodromy in the K3 fibration corresponds to a shift in the B-field on the heterotic side, as appropriate for an NS5-brane. A general discussion of reducible K3 fibers, their monodromies, and the relation to NS5-branes can be found in Ref.~\cite{Braun:2016sks}.

\subsection{Toric descriptions of required resolutions}
\label{sec:tor_desc_in_6d}

In Section~\ref{sec:sing_and_inst_in_6d} we discussed the duality between (4,6,12) singularities in F-theory and small instantons in heterotic string theory. We now discuss the blow-ups of these singularities. We would like to later give a toric description for the resulting global F-theory models dual to heterotic line bundle models, as many computations are thus simplified. Hence we develop toric descriptions of the required blow-ups dual to NS5-branes. See e.g.~Refs.~\cite{fulton1993introduction,cox2011toric} for an introduction to toric geometry, Refs.~\cite{Skarke:1998yk,Knapp:2011ip} for an introduction to its applications in F-theory, and Ref.~\cite{Buchmuller:2017wpe} for a recent base-independent discussion of toric transitions.

\bigskip

\begin{figure}[t]
\centering
\includegraphics[scale=0.8]{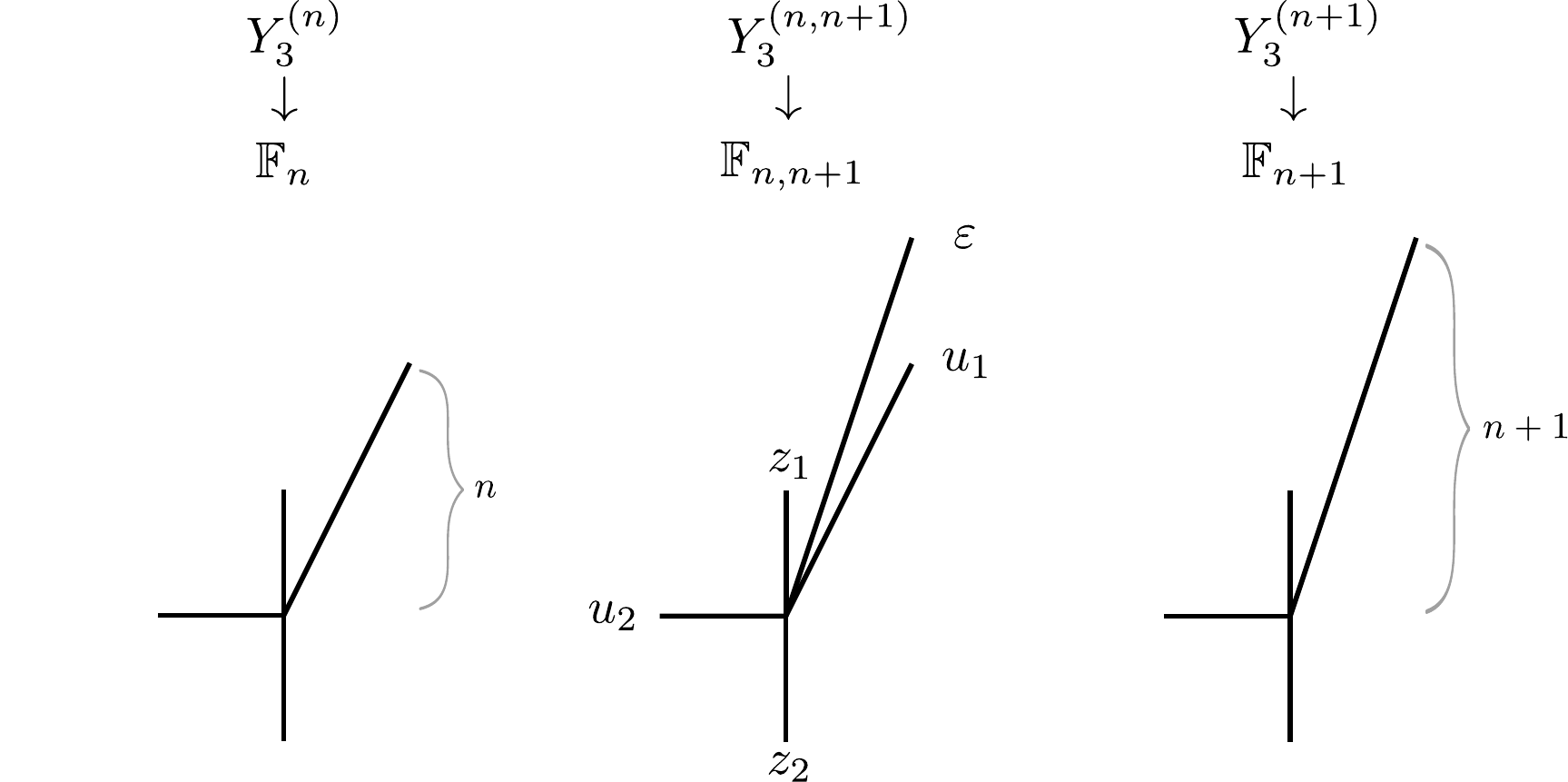}
\caption{Toric fans for the Hirzebruch surfaces $\mbb{F}_n$ and $\mbb{F}_{n+1}$, as well as the fan for the intermediate space, which we call $\mbb{F}_{n,n+1}$, in which we have blown up $\mbb{F}_n$ but have not yet blown down to $\mathbb{F}_{n+1}$. We have also indicated that we wish to consider elliptic fibrations $\fss^{(\cdot)}$ over these bases.}
\label{fig:hirzebruch_fans}
\end{figure}

We can understand the F-theory geometry dual to the presence of an NS5-brane in multiple ways. One way is to note that different heterotic instanton distributions correspond to different Hirzebruch surface bases in F-theory, so that the presence of an NS5-brane is an intermediate situation between different distributions. From this perspective, it is clear that the intermediate geometry should allow blow-downs to both geometries dual to the two instanton distributions. More precisely, these blow-downs will require flops; we discuss this below. There should also be an elliptic fibration over some intermediate base, and this base should admit blow-downs to the two Hirzebruch surfaces. An obvious candidate for an intermediate space, which we call $\mbb{F}_{n,n+1}$, is shown in Figure~\ref{fig:hirzebruch_fans} along with the two Hirzebruch spaces.

Another way to understand this is through the duality between singularities and small instantons. A small instanton is dual to an $\tilde{\mathrm{E}}_8$ singularity, which has a crepant resolution, and this resolution is dual to the transition to an NS5-brane. Letting the small instanton first sit at $\bc_1=0$, from Equation~\eqref{eq:hyp_e8_res} the resolution clearly demands the introduction of the following ray $\vec{\ep}$ in the polytope of the ambient toric fourfold,
\be
\vec{\ep} = 2\vec{\tacx} + 3\vec{\tacy} + \vec{\fpc_1} + \vec{\bc_1} \,,
\ee
where the coordinate $\ep$ describes the exceptional divisor of the blow-up. This resolution corresponds to moving to the intermediate space in Figure~\ref{fig:hirzebruch_fans}, after starting from the $\mbb{F}_n$ base. This is the toric implementation of the weighted blow-up
\be
\tacy = \hat{\tacy}\ep^3 \,, \quad \tacx = \hat{\tacx}\ep^2 \,, \quad \fpc_1 = \hat{\fpc}_1\ep \,, \quad \bc_1 = \hat{\bc}_1\ep \,,
\ee
where hats temporarily indicate coordinates after the blow-up. The proper transform of the Weierstrass equation consists of dividing by $\ep^6$ to give a new Weierstrass equation
\be
\hat{\tacy}^2 = \hat{\tacx}^3 + \hat{\wsf}\hat{\tacx}\tacz^4 + \hat{\wsg}\tacz^6 \,,
\ee
where we have defined $\hat{\wsf} = \ep^{-4}\wsf$ and $\hat{\wsg} = \ep^{-6}\wsg$. This also corresponds to having $\hat{\wsf}$, $\hat{\wsg}$ be sections of $K_{\mbb{F}_{n,n+1}}^{-4}$ and $K_{\mbb{F}_{n,n+1}}^{-6}$ respectively, as we know must be the case. It is easy to see that the resulting space is CY. As in the previous perspective, we have maintained a toric description by placing the instanton/brane at the vanishing locus of homogeneous coordinates. More general instanton/brane configurations will be considered in detail below.

The weight system of the resulting ambient toric fourfold can be written as
\be
\begin{tabular}{c c c c c c c c}
$\tacy$ & $\tacx$ & $\tacz$ & $\fpc_1$ & $\fpc_2$ & $\bc_1$ & $\bc_2$ & $\ep$ \\
\hline
3 & 2 & 1 & 0 & 0 & 0 & 0 & 0 \\
6 & 4 & 0 & 1 & 1 & 0 & 0 & 0 \\
$3n+6$ & $2n+4$ & 0 & 0 & n & 1 & 1 & 0 \\
3 & 2 & 0 & 1 & 0 & 1 & 0 & -1 \\
\end{tabular} \,.
\ee
Here we have written the final line in a form allowing us to read off the structure of the blow-up of the threefold with $\mbb{F}_n$ base. These weights reflect the resolution of a $\tilde{\mathrm{E}}_8$ singularity as described in Equation~\eqref{eq:hyp_e8_res}. Equivalently, we can rewrite the weight system as
\be
\begin{tabular}{c c c c c c c c}
$\tacy$ & $\tacx$ & $\tacz$ & $\fpc_1$ & $\fpc_2$ & $\bc_1$ & $\bc_2$ & $\ep$ \\
\hline
3 & 2 & 1 & 0 & 0 & 0 & 0 & 0 \\
3 & 2 & 0 & 0 & 1 & -1 & 0 & 1 \\
$3(n+1)+6$ & $2(n+1)+4$ & 0 & 0 & n+1 & 0 & 1 & 1 \\
6 & 4 & 0 & 1 & 1 & 0 & 0 & 0 \\
\end{tabular} \,.
\ee
The third row of this form reflects that one can reach this situation also by blowing up the threefold with $\mbb{F}_{n+1}$ base. This corresponds to the fact we can pull an M5-brane off either $\mathrm{E}_8$ brane in the 11d interval, or inversely, absorb the brane back into either $\mathrm{E}_8$ brane. The weight system above describes the 4d polytope of the intermediate situation, but not the 4d fan: a triangulation is also required. We might start with either Hirzebruch base case, which each have given fans, and include the necessary blow-up ray, giving a new specific fan. However as we now discuss, these blow-ups destroy the elliptic fibration structure, and flops are required to retrieve it. 

\bigskip

In the two Hirzebruch base space cases and the intermediate case, there exists a polytope of the ambient fourfold which may admit many triangulations. In each case, for an F-theory interpretation we are interested in those triangulations that describe an elliptic fibration for the CY hypersurface. In fact this condition uniquely determines the fan, as follows. Writing down a Weierstrass model anticipates that the fibration descends from the toric ambient space in a particular way, namely from a fibration of $\mbb{P}_{123}$ over the base. This projection is a particular instance of a toric morphism. Toric morphisms are induced from maps, in our case a projection, of the fan. In order to be a toric morphism it must not happen that cones are mapped one-to-many, but each cone must be mapped onto a unique cone (of course many cones can be mapped to the same cone). Fans of this type can be built from the `product' of cones in the base and cones in the fiber, so for example in the case of a fibration of $\mbb{P}_{123}$ over the base $\mbb{F}_n$ shown in the left of Figure~\ref{fig:hirzebruch_fans}, the top-dimensional cones are given by all pairings
\be
\{(\bc_2,\fpc_1),(\fpc_1,\bc_1),(\bc_1,\fpc_2),(\fpc_2,\bc_2)\} \times \{(\tacx,\tacy),(\tacy,\tacz),(\tacz,\tacx)\} \,,
\ee
where $\times$ is the Cartesian product and each pairing, e.g.~$(\bc_2,\fpc_1,\tacx,\tacy)$, is the cone spanned by the given four rays. 

Including blow-up rays simply refines the cone in which such a ray sits, and for this reason destroys the fibration structure. Blowing up by inserting the ray $\vec{\ep} = 2\vec{\tacx}+3\vec{\tacy}+\vec{\bc}_1+\vec{\fpc}_1$, the resulting cones are as before except one of the old cones is replaced by four,
\be
(\tacx,\tacy,\bc_1,\fpc_1) \to \{(\ep,\tacy,\bc_1,\fpc_1),(\tacx,\ep,\bc_1,\fpc_1),(\tacx,\tacy,\ep,\fpc_1),(\tacx,\tacy,\bc_1,\ep)\} \,.
\ee
The fibration of the original threefold over the base is inherited from one on the ambient space, specifically the map that collapses the directions of the $\mbb{P}_{123}$, and after introducing the blow-up ray, this map is no longer a toric morphism. We can see this by the fact that under the projection map, the map of cones includes for example
\be
(\ep,\tacy,\bc_1,\fpc_1) \to (\bc_1,\ep) \cup (\ep,\fpc_1) \,,
\ee
so that this cone is split in half by the projection and the map between cones is one-to-many. This means that there is no toric morphism, and hence no projection of the ambient space for the threefold to inherit, so that the fibration has been destroyed.

However, we would like to preserve the fibration structure of the threefold in order to have an F-theory interpretation. The blown-up space has a flop\footnote{Before the flops, there does not appear to be an obvious F-theory interpretation and it appears F-theory does not see the intermediate process \cite{Morrison:1996pp}.} from this `naively' blown-up fan to that with the product cone structure. We will henceforth assume the required flops have been performed, giving an elliptic fibration in situations with NS5-branes. Practically speaking, this means we should imagine blowing up the base, and then adding the elliptic fibration, and we will often speak in this way.

\subsection{Toric global models in 6d}
\label{sec:glob_models_in_6d}

We have discussed the explicit duality between heterotic NS5-branes and F-theory blow-ups in compactification to six dimensions, and we can use this knowledge to build global F-theory models dual to six-dimensional heterotic models with only NS5-branes. This will be useful to study their generalisation to duals of four-dimensional line bundle sum models. 

\begin{figure}[t]
\centering
\includegraphics[scale=0.7]{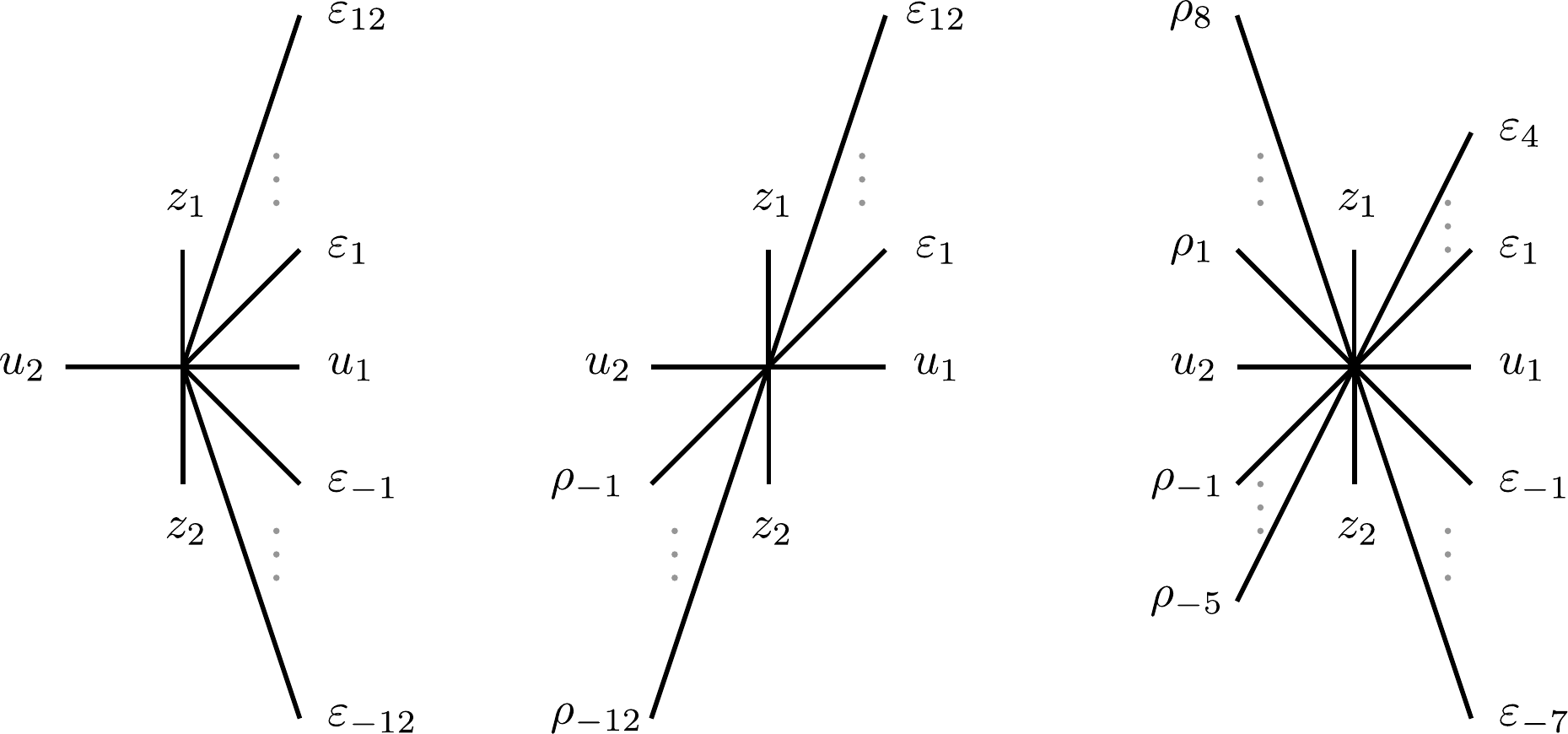}
\caption{Possible toric bases of the F-theory threefold after blowing up all intersections of the $\mathrm{E}_8$ stacks with the remaining brane locus.}
\label{fig:fullyblownup_fan}
\end{figure}

We have discussed the toric description of a blow-up dual to a single NS5-brane, however anomaly cancellation requires 24 heterotic NS5-branes, dual to the blow-ups of all 24 $\tilde{\mathrm{E}}_8$ singularities in the F-theory threefold. We would like to give a toric description of the geometry in F-theory after all of these blow-ups. We first note that before the transition of instantons into NS5-branes, the precise Hirzebruch surface occurring as the F-theory base reflects the initial distribution of heterotic instantons. One can take as a starting point a $\mbb{P}^1 \times \mbb{P}^1$ base, since all instantons will be pulled into M5-branes in the bulk, washing out dependence on initial distribution. Then 12 blow-ups will be associated to each $\mathrm{E}_8$ stack. From the discussion in Section~\ref{sec:tor_desc_in_6d}, the base spaces shown in Figure~\ref{fig:fullyblownup_fan} are obvious proposals for the resulting geometry. The different distributions of blow-up rays correspond to different configurations of small instantons/NS5-branes: blow-up rays associated to $\bc_1$ ($\bc_2$) are dual to NS5-branes at $\bc_1=0$ ($\bc_2=0$). These bases are those with simple toric descriptions, and this corresponds to these restricted positions for NS5-branes in the heterotic space. More general brane configurations require more involved descriptions, but as these are only general configurations of points they are not very interesting. We will discuss arbitrary brane configurations in the more interesting four-dimensional context in Section~\ref{sec:more_gen_glob_models}.

We take the left possibility in Figure~\ref{fig:fullyblownup_fan} as a concrete example in the following. This base space is a $\mbb{P}^1 \times \mbb{P}^1$ with two series of blow-ups. In order for these blow-up rays to reduce all singularities to mere $\mathrm{E}_8$s, the intersection of the remaining brane locus with the $\mathrm{E}_8$ stacks must be tuned to sit at the vanishing of toric coordinates. After this tuning, in each series of blow-ups at each stage a $\mbb{P}^1$ is blown-up on a point of the previous exceptional $\mbb{P}^1$. In Appendix \ref{app:explicit_blowups_6d} we verify explicitly that this toric prescription indeed appropriately reduces the severity of the singularities, leaving only two curves of $\mathrm{E}_8$ singularities which do not worsen further over points in the base. These can then be resolved using standard techniques preserving the flatness of the elliptic fibration.
In particular, these crepant resolutions are achieved torically by including new rays at
\be\begin{aligned}
(0,\pm i,2,3)  ~&\mathrm{for}~ i=1,2,3,4,5,6 \,, \hspace{.3cm}
&(0,\pm i,1,2)  ~&\mathrm{for}~ i=1,2,3,4 \,, \hspace{.3cm}
& (0,\pm i,1,1)  ~\mathrm{for}~ i=1,2,3 \,, \hspace{.3cm} \\
(0,\pm i,0,1)  ~&\mathrm{for}~ i=1,2    \,, \hspace{.3cm}
&(0,\pm i,0,0)  ~&\mathrm{for}~ i=1 &
\label{eq:newrayse86D}
\end{aligned}\ee
where $+$ and $-$ are for the $\mathrm{E}_8$ branes at $\fpc_1=0$ and $\fpc_2=0$ respectively, and the first two values are coordinates in the ray diagrams in Figure~\ref{fig:fullyblownup_fan}, while the final two coordinates are in the $\mbb{P}_{123}$ in which $\tacx=(-1,0)$, $\tacy=(0,-1)$, $\tacz=(2,3)$. Note that working with a description of $X$ in terms of reflexive polytopes, the requirement of convexity of the $N$-lattice polytope $\Delta^*$ already forces the introduction of the rays \eqref{eq:newrayse86D}.

\bigskip

In order to perform the blow-ups in the base using toric methods, we had to tune the intersections of the remaining brane locus with each $\mathrm{E}_8$ stack to be at the zeroes of toric coordinates. Deformations of these intersection positions, or correspondingly of the positions of the NS5-branes in the heterotic base, cannot be seen directly in this toric description. We hence expect these to appear as non-polynomial deformations, as we now verify. We recall Batyrev's formula \cite{Batyrev:1994hm} for complex structure moduli: for a CY hypersurface $X$ defined by an $m$-dimensional reflexive $M$-lattice polytope $\Delta$, we have
\be
h^{m-2,1}(X) = l(\Delta)-m-1-\sum_{\mathrm{codim}\;\Theta=1}l^*(\Theta) + \sum_{\mathrm{codim}\;\Theta=2}l^*(\Theta)l^*(\Theta^*) \,,
\ee
where $\Theta$ is a face of $\Delta$, and $\Theta^*$ the dual face in the dual reflexive polyhedron, and where $l(\cdot)$ denotes the number of integral points while $l^*(\cdot)$ denotes the number of integral points in the interior. The various terms can be identified with the number of polynomial deformations minus the number of automorphisms of the ambient space together with the number of non-polynomial deformations, which correspond to the last term. Such non-polynomial deformation are complex structure deformations of the CY hypersurface which are frozen in the embedding into a toric variety under consideration. In the example under discussion, one can easily verify that the number of complex non-polynomial deformations is 23. These clearly correspond to the moduli used up in stacking 24 NS5-branes in the heterotic base. The middle and right diagrams of Figure~\ref{fig:fullyblownup_fan} both have 22 complex non-polynomial deformations, as expected since on the heterotic side there are now two stacks of NS5-branes.

\bigskip

Finally, before turning to the construction of global four-dimensional models, we perform a multiplet match in the simple above example of the six-dimensional duality, as the four-dimensional multiplet matches will be more involved but somewhat similar. We recall that in compactification to six dimensions we have chosen to have no flux on the heterotic side, so there is no corresponding flux on the $\mathrm{E}_8$ brane stacks. In the following we write $\bfss$ for the F-theory threefold and $\bfbs$ for the base, where both are the result upon performing all required resolutions, including of the $\mathrm{E}_8$ singularities.

On the F-theory side, we have the following expressions for the number of tensor multiplets $\nte$ and the number of hypermultiplets\footnote{We also have $\mathrm{rk}(V) = h^{1,1}(\bfss)-h^{1,1}(\bfbs)-1$ for the rank $\mathrm{rk}(V)$ of the gauge group. From this we can conclude that $\mathrm{rk}(V)=16$, but since we have an unbroken $\mathrm{E}_8 \times \mathrm{E}_8$ we already know the number of vectors is $\nve=496$.} $\nhy$ \cite{Morrison:1996pp}
\be
\nte = h^{1,1}(\bfbs)-1 \,, \quad \nhy = h^{2,1}(\bfss)+1 \,, 
\ee
For our specific threefold in F-theory, we find that $h^{2,1}(\bfss) = 43$, $h^{1,1}(\bfss) = 43$, and $h^{1,1}(\bfbs) = 26$, so that $\nte = 25$ and $\nhy = 44$. We recall that a hypermultiplet contains four real scalars, so that altogether we have 201 real scalars.

On the heterotic side, we have first 24 real parameters for the positions of the NS5-branes in the interval of the 11d Ho\v{r}ava-Witten picture, and these 24 real scalars sit in tensor multiplets. Together with the tensor multiplet containing the dilaton, we have $\nte = 25$ as on the F-theory side. We also have $24 \cdot 4 = 96$ real scalars from the positions in the K3 of the 24 five-branes. Additionally K3 has $b_2 = 22$, giving rise to 22 real scalars. Finally the moduli space of Ricci-flat metrics on K3 has 58 real moduli, and so together we get $96+22+58=176$ real scalars. Since these all sit in hypermultiplets this gives $\nhy = 44$ as on the F-theory side. It is clear the anomaly condition $29\nte+\nhy-\nve=273$ is satisfied.


\section{Global 4d models}
\label{sec:glob_4d_models}

We now turn to a description of global F-theory models dual to heterotic line bundle models in compactifications to four dimensions. After briefly determining the required branes and blow-ups, we will build our first global models in analogy with the toric six-dimensional constructions of Section~\ref{sec:glob_models_in_6d}. We then give a construction that allows for more general NS5-brane configurations. Afterwards, we verify various aspects of the proposed duality between these models and heterotic line bundle models. Note we will freely use the ideas reviewed and developed above in the six-dimensional case, as many ideas required in the four-dimensional case are analogous.

We briefly remind ourselves of some notation in the four-dimensional case. On the heterotic side we have a CY threefold $\hs$ which is an elliptic fibration $\hp: \hs \to \hb$. Since the heterotic space is now a threefold, there are multiple choices for the twofold base $\hb$, rather than just $\mbb{P}^1$ as was the case in compactification to six dimensions. We will mainly treat toric bases, for which there are already many possibilities. On the F-theory side, there is a CY fourfold $\fs$ which is a K3 fibration $\fkp: \fs \to \hb$ and an elliptic fibration $\fpr: \fs \to \fb$. We write $\bfs$ and $\bfb$ for the spaces after any required blow-ups have been performed.

\subsection{Required branes and blow-ups}
\label{sec:req_branes_and_blowups}

We first determine the required number of heterotic NS5-branes and F-theory blow-ups, for heterotic line bundle sums. We recall that in compactification to four dimensions, NS5-branes in $\hs$ can wrap the fiber or curves in the base; these are `vertical' and `horizontal' branes respectively. The number and type of NS5-branes required for line bundle sum models with F-theory duals follows from anomaly cancellation. From the triple intersection numbers of $\hs$ and the fact that $k_a^0=0$ for each line bundle $\lb_a$ in one of the line bundle sums $V_{1,2}$ (as reviewed in Section~\ref{sec:prelims}), we have that
\be
\mathrm{ch}_2(V_{1,2}) = \sum_{a=1}^\nolb \frac{1}{2}d_{IJK}k_a^Ik_a^JC^K = \left(\sum_{a=1}^\nolb \frac{1}{2}k_a^ik_a^j\bim_{ij}\right) F \,,
\ee
where $\hetfib$ is the heterotic fiber. Then recalling the expression for $\mathrm{ch}_2(\hs)$ in Equation~\eqref{eq:c2X}, we see from the anomaly cancellation condition in Equation~\eqref{eq:het_anom_canc} that we require NS5-branes wrapping curves in the base $\hb$ with total class ${\fbc=12\hsec(-K_{\hb})}$. This is the required horizontal NS5-brane content. We also have a choice in how to cancel the part of ch$_2(\hs)$ proportional to $\hetfib$: we can use vertical NS5-branes or the line bundle sums. We will choose to have no vertical NS5-branes, so line bundle sums make up the remainder of the anomaly condition. Summarising,
\be
\fbc_{\mathrm{hor.}} = 12\hsec(-K_{\hb}) \,, \quad \fbc_{\mathrm{ver.}}=0 \,.
\ee
Note that as vertical NS5-branes are dual to D3-branes, there are no D3-branes on the F-theory side.

The F-theory dual of a horizontal NS5-brane is a blow-up in the F-theory base $\fb$, and these blow-ups are the obvious generalisations of the six-dimensional case discussed in Sections \ref{sec:sing_and_inst_in_6d} and \ref{sec:tor_desc_in_6d}. The F-theory $\mathrm{E}_8$ symmetries are broken purely by flux, not geometry. The loci of $\mathrm{E}_8$ singularities, which are now twofolds, intersect the remaining brane locus to give curves of $\tilde{\mathrm{E}}_8$ singularities, which are (4,6,12) curves, whose resolution is dual to the introduction of NS5-branes. For simplicity we can take the initial instanton distribution to be symmetric, so that $\fb = \hb \times \mbb{P}^1$. Then the two intersection curves have classes
\be
[\fpc_{1}]\cdot[\Delta_r] = [\wsg_5]=-6K_{\hb}|_{\fpc_{1}=0} \,, \quad [\fpc_{2}]\cdot[\Delta_r] = [\wsg_7]=-6K_{\hb}|_{\fpc_{2}=0} \,,
\ee
together determining a point locus in the class $-12K_{\hb}$ that is dual to the NS5-brane locus. The precise correspondence between positions of blow-ups and branes is the generalisation of the six-dimensional case in Figure~\ref{fig:stabdegbase}.

\bigskip

The characterisation of horizontal branes as sitting in the base is slightly too crude, so let us take a moment to be more precise. A curve $\mc{C}$ in the heterotic base $\hb$ can always be embedded holomorphically into the heterotic threefold $\hs$ by the zero section, ${C=\hsec(\mc{C})}$. However in addition, over a given curve $\mc{C}$ in $\hb$ there is an elliptic surface $S_\mc{C}$ in $\hs$ which may have more than one section. In particular, if the elliptic surface $S_\mc{C}$ is a $dP_9$ there will be (countably) infinitely many sections, and if the elliptic fibration on $S_\mc{C}$ is trivial there is a continuous family of sections. A nice discussion on such situations is found in Ref.~\cite{Donagi:1999jp}. For simplicity, we will tend to consider all horizontal branes to be embedded by the zero section, unless explicitly stated otherwise. As in the case of compactification to six dimensions, the positions of NS5-branes in the fiber are expected to be dual to Ramond-Ramond moduli (in the IIB picture) on the F-theory side, hence equivalently we assume particular expectation values for these fields. We will return to a discussion of these extra parameters in Section~\ref{sec:dualchecks_ns5s_and_blowups}, where we will be particularly interested in the case of trivially fibered elliptic surfaces.

\subsection{Toric global models}
\label{sec:toric_global_models}

In Section~\ref{sec:glob_models_in_6d} above we have built global F-theory models dual to six-dimensional heterotic line bundle sum models, which had particularly simple toric descriptions. Examples of the F-theory base space were shown in Figure~\ref{fig:fullyblownup_fan}. It is easy to build global four-dimensional models in direct analogy with those simple constructions, so we will discuss these models first before moving on to more general constructions.

\bigskip

\begin{figure}[t]
\centering
\includegraphics[scale=0.8]{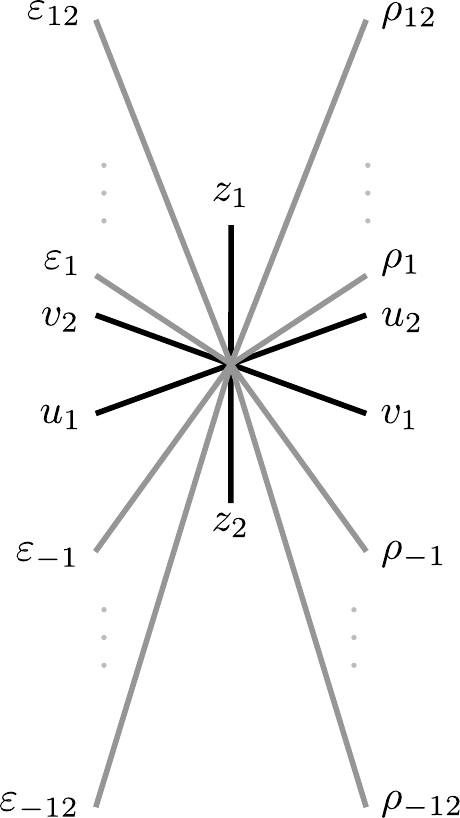}
\caption{Ray diagram for the result of blowing up the F-theory base $\left(\mbb{P}^1\right)^3$ to remove intersections of the remaining brane locus with the two $\mathrm{E}_8$ stacks. Blow-up rays are a different colour only for clarity. The $\ep_i$ and $\rho_i$ rays are blow-up rays, in the $u-z$ and $v-z$ planes respectively.}
\label{fig:fullyblownup_P1P1P1_fan}
\end{figure}

\begin{figure}[t]
\centering
\includegraphics[scale=0.8]{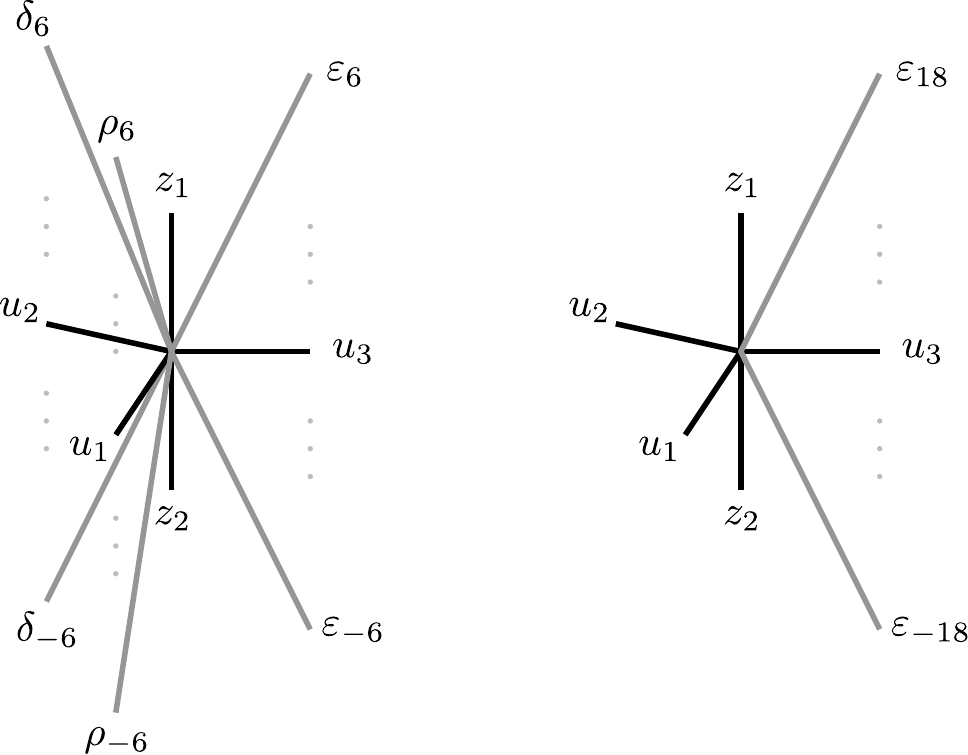}
\caption{Examples of blow-up distributions in the case of a heterotic base space $\mbb{P}^2$. Blow-up rays are all above/below one of the rays in the base ($\bc_1,\bc_2,\bc_3$). In the left diagram the blow-ups are distributed democratically, while on the right they are all associated to one base ray.}
\label{fig:fullyblownup_examples_4d}
\end{figure}

Let us assume the F-theory base space is initially a trivial fibration $\fb = \mbb{P}^1 \times \hb$. The locus of $\tilde{\mathrm{E}}_8$ singularities, whose resolution is dual to the introduction of the required NS5-branes, consists of curves. As in the case of compactification to six dimensions, these loci on the two $\mathrm{E}_8$ surfaces are determined by the vanishing of $\wsg_7$ and $\wsg_5$. In order to perform toric blow-ups over these curve loci, we must tune these polynomials to give zeroes only over toric curves, which are those at the vanishing of the toric coordinates. For any such tuning, the natural expectation for how to then build the resolved base space is as follows. We introduce toric blow-up rays into the base ray diagram, in towers above/below the rays of $\hb$, such that the projection of these onto the $\hb$ part of the ray diagram gives a set of rays which, when counted with multiplicities, have divisor class $-12K_{\hb}$. That is, the blow-up rays are at
\be
\vec{\ep}_{(a),i} = (\vec{\bc}_a,i) ~~ \mathrm{for}~~ i = -m_a,-m_a+1,\ldots,m_a \,,
\ee
\be
\mathrm{with} \quad \sum_a m_a = \sum_a n_a = 6 \cdot c_2(\hb) \,, \nonumber
\ee
where $\vec{\bc}_a$ are the rays of $\hb$, the third coordinate is along the $\vec{\fpc}_1$ direction, and $c_2(\hb)$ is simply the number of rays in $\hb$. We show in Figure~\ref{fig:fullyblownup_P1P1P1_fan} an example ray diagram for the F-theory base, where the heterotic base is a $\mbb{P}^1 \times \mbb{P}^1$ and we have chosen a particular distribution of the blow-up rays. We also show in Figure~\ref{fig:fullyblownup_examples_4d} some possible distributions of blow-ups for the heterotic base choice $\hb = \mbb{P}^2$, as this base choice allows reasonably clear diagrams for multiple blow-up distributions.\footnote{Note that, unlike in six dimensions, the ray diagram of the base $\bfb$ is not enough to specify its fan. While the extra rays are understood to be blow-ups of the original base, this still leaves freedom for the triangulation. For the purposes of this section, the particular triangulation is not important, and we leave a discussion of triangulations to Section~\ref{sec:global_dual_fth_geom}.}

The distribution of blow-up rays reflects the tuning of $\wsg_7$ and $\wsg_5$. In the example of Figure~\ref{fig:fullyblownup_P1P1P1_fan}, the choice of tuning is
\be
\wsg_7 = \alpha \bc_1^{12}\bct_1^{12} \,, \quad \wsg_5 = \beta \bc_1^{12}\bct_1^{12} \,,
\ee
with $\alpha$ and $\beta$ arbitrary complex numbers. With this choice, the remaining discriminant locus initially intersects each $\mathrm{E}_8$ stack at $\{\bc_1=0\} \cup \{\bct_1=0\}$. This gives four curves along which blow-ups are required. These blow-ups are performed by the introduction of the rays in Figure~\ref{fig:fullyblownup_P1P1P1_fan}. In Appendix \ref{app:explicit_blowups_4d}, we explicitly work through the required tuning and blow-up procedure for the example in Figure~\ref{fig:fullyblownup_P1P1P1_fan}, showing that this ray diagram is indeed the result of resolving the singularities. We also verify that the loci $\{\fpc_1=0\}$, $\{\fpc_2=0\}$ in the base, are each diffeomorphic to $\hb$, so that the GUT surfaces have the same topology as the heterotic base. Similarly we check that the remaining brane locus does not intersect the $\mathrm{E}_8$ stacks, so that the singularities worse than $\mathrm{E}_8$ have all been removed. We also show that the $\mathrm{E}_8$ singularities themselves are now geometrically non-Higgsable.

\bigskip

To retain a simple toric description, $\wsg_7$ and $\wsg_5$ are tuned to give zeroes at the vanishing of toric coordinates. This choice of tuning is dual to the choice of NS5-brane configuration on the heterotic side. As in six dimensions, the number of non-polynomial deformations of the toric hypersurface reflects the choice of NS5-brane configuration, analogously to the discussion in Section~\ref{sec:glob_models_in_6d}. For example, in the case of Figure~\ref{fig:fullyblownup_P1P1P1_fan}, the heterotic dual of this specific F-theory base has two stacks of 24 horizontal NS5-branes, one at $\bc_1=0$ and one at $\bct_1=0$; each brane wraps a $\mbb{P}^1$. We find from Batyrev's formula that the number of non-polynomial deformations is 46, as expected as this is the number of moduli used up in forcing two stacks of 24 $\mbb{P}^1$ NS5-branes in $\mbb{P}^1 \times \mbb{P}^1$. For different distributions of the blow-ups, the count of non-polynomial deformations will be different, reflecting the number of parameters involved in specifying such a brane configuration.

\subsection{More general global models}
\label{sec:more_gen_glob_models}

The four-dimensional F-theory models constructed in Section~\ref{sec:toric_global_models} had particularly simple toric descriptions, but also were rather restrictive. In particular the dual horizontal NS5-branes were forced to wrap $\mbb{P}^1$s in stacks at loci defined by the vanishing of toric coordinates. Clearly these are not the most general brane configurations; in general on each $\mathrm{E}_8$ brane stack there is an arbitrary curve locus of (4,6,12) singularities with class $-6K_{\hb}$, and these singularities are removed by blowing up over this locus. While the blow-ups in the toric hypersurface models were over toric curves, giving up the hypersurface description allows arbitrary blow-ups. Often the complete intersection CY manifold can be described by a nef partition \cite{Batyrev:1994pg}, which allows for straightforward calculation of the Hodge numbers. (However at high codimension, this can become computationally prohibitive.)

\bigskip

A blow-up over a non-toric locus can be achieved torically by increasing the dimension of the toric ambient space by one. First a new coordinate $\prb$ is introduced with the addition of a ray perpendicular to the old fan. An extra equation then expresses that the zero locus of $\prb$ is to be determined by a function $F$ of the other coordinates. Along with any toric constraints, this gives the locus we wish to blow up. Writing $c_a$ for these other coordinates, we have for the extra ray and the extra equation,
\be
\vec{\prb} = (\vec{0},1) \,, \quad \mathrm{with} \quad \prb = F(c_a) \,.
\ee
For the extra equation to be well-defined under all scalings, the positions of the rays $\vec{c}_a$ in the new direction must be such that $\prb$ and $F(c_a)$ have the same weights. The blow-up is then performed by adding a ray
\be
\vec{\cbr} = \vec{\prb} + \ldots \,,
\ee
where the dots represent other rays that correspond to the toric constraints on the blow-up.

We take as an example the case where the heterotic base is $\hb = \mbb{P}^2$. We begin with the situation where the blow-ups have all been performed in the simple toric manner in the diagram on the right of Figure~\ref{fig:fullyblownup_examples_4d}, except we remove the top $n$ rays in the tower over $\bc_3$. The highest ray appearing is then $\ep_{18-n}$. This leaves a singularity on the $\fpc_1=0$ surface, which we tune into a more general curve defined by a polynomial of degree $n$ in the $\mbb{P}^2$ scaling. We will then blow up on this locus by going to a toric complete intersection. This polynomial is represented in the ambient space by a function of coordinates on the pullbacks of curves in $\hb$ under the $\mbb{P}^1$ projection. That is, we can use the products of the rays associated to each base ray,
\be
\bc_1 \,, ~~ \bc_2 \,, ~~ \ptbc_3 := \bc_3 \prod_{a=-18}^{18-n} \ep_a\,.
\ee
These `coordinates' have the same behaviour under all scalings, i.e.\ they correspond to the same divisor class. We define $\cf_n$ to be a degree $n$ polynomial in $\bc_1,\bc_2,\ptbc_3$; this is a well-defined function up to an overall scaling. This equation defines a curve on the GUT surface upon intersection with $\{\fpc_1=0\}$. 

We then perform the blow-up over the curve. In addition to the rays in the original fan,
\begin{align}
\vec{\bc}_1 &= (-1,0,0,2,3) \,,  &\vec{\bc}_2 &= (0,-1,0,2,3) \,, &\vec{\bc}_3 = (1,1,0,2,3) \,, \nonumber \\ 
\vec{\fpc}_1 &= (0,0,1,2,3) \,,  &\vec{\fpc}_2 &= (0,0,-1,2,3) \,, &\vec{\ep}_i = (1,1,i,2,3) \,, \\
\vec{\tacx} &= (0,0,0,-1,0) \,,  &\vec{\tacy} &= (0,0,0,0,-1) \,, &\vec{\tacz} = (0,0,0,2,3) \,, \nonumber
\end{align}
we then introduce a new coordinate $\prb$, as well as the additional equation
\be
\vec{\prb} = (0,0,0,0,0,1) \,, \quad \prb = \cf_n(\bc_1,\bc_2,\ptbc_3) \,.
\ee
We also give all the old rays the value zero in the new direction, except for $\bc_2$ (for example) which we give the value $-n$, i.e.
\be
\vec{\bc}_2 = (0,-1,0,2,3,-n) \,.
\ee
This ensures that the equation $\prb = \cf_n$ is well-defined under the new scaling relation that has been introduced. Next we perform a blow-up by introducing an extra ray
\be
\vec{\cbr} = 2\vec{\tacx}+3\vec{\tacy}+\vec{\prb}+\vec{\fpc_1} = (0,0,1,0,0,1)\,.
\ee
(We then flop to the required fan, as discussed in Section~\ref{sec:tor_desc_in_6d} above.) The coefficients of $\tacx$ and $\tacy$ reflect that we are resolving a $\tilde{\mathrm{E}}_8$ singularity. The F-theory space $\bfs$ is then the CY fourfold defined as a complete intersection, in the 6d toric ambient space, by two equations
\be
\mc{W} = 0 \,, \quad \prb\cbr = \cf_n(\bc_1,\bc_2,\ptbc_3) \,,
\ee
where $\mc{W}$ is the Weierstrass polynomial. We also have to take the proper transform of $\mc{W}$, as discussed above. The base $\bfb$ of the F-theory fibration can be described as a hypersurface in a 4d toric ambient space by the second of the above equations.

\bigskip

\begin{figure}[t]
\centering
\includegraphics[scale=1.7]{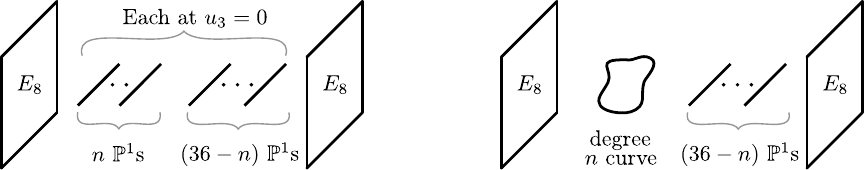}
\caption{Horizontal NS5-brane configurations on the heterotic side for a choice of a $\mbb{P}^2$ heterotic base, in the case of (i) a configuration that has a dual geometry described by a toric hypersurface, and (ii) a configuration whose dual geometrical description requires a codimension two complete intersection.}
\label{fig:different_brane_configs}
\end{figure}

In this example we have replaced $n$ blow-ups over a $\mbb{P}^1$ by a single blow-up over a degree $n$ curve on the $\mathrm{E}_8$ surface. In the heterotic dual, we have replaced $n$ $\mbb{P}^1$ NS5-branes with a single NS5-brane with the topology of a degree $n$ curve in $\mbb{P}^2$. For example if $n=3$, the resulting NS5-brane wraps a torus. Figure~\ref{fig:different_brane_configs} shows the dual NS5-brane configuration, in the toric hypersurface case and in the complete intersection case we have just discussed\footnote{Note that we have not discussed the transition between these situations -- we will return to this question in Section~\ref{sec:coin_and_int_5branes} below.}. This example was particularly simple. In the slightly more complicated example of a $\mbb{P}^1 \times \mbb{P}^1$ base in Figure~\ref{fig:fullyblownup_P1P1P1_fan} above, we could for example blow down $n$ of the top rays from the $\ep$ tower and $m$ of the top rays from the $\rho$ tower, and then perform a blow-up over a general curve of degrees $n$, $m$, analogously to what we have just done for the $\mbb{P}^2$ case. The new brane will have curve class $n[\bc_1]+m[\bct_1]$.

One can note that the procedure we have used replaces the $n$ top or bottom $\mbb{P}^1$s, and thus creates a general brane configurations `near' a heterotic $\mathrm{E}_8$ brane, i.e.\ where there are no branes between that M5 and the $\mathrm{E}_8$ brane. This is sufficient for the purposes of this section, as multiplet counting and other checks will not depend on the ordering of branes in the bulk. We will discuss arbitrary configurations of branes with arbitrary orderings in Section~\ref{sec:global_dual_fth_geom} below.

\subsection{Duality checks independent of NS5-brane configuration}
\label{sec:dual_checks_ind_of_ns5_config}

We have now constructed global F-theory models dual to heterotic line bundle sums in four dimensions and spent some time discussing the duality between geometry of the F-theory side and horizontal NS5-brane configurations on the heterotic side. In addition to these ingredients, there is the duality of flux on the two sides: on the heterotic side the two sets of $\mathrm{E}_8$ flux on the base, and on the F-theory side the $G_4$ flux on the two $\mathrm{E}_8$ surfaces, were naturally proposed to be the same in Section~\ref{sec:prelims}. With the flux identification as well as the geometry/NS5-brane identification, we are now in a position to check various aspects of this duality.

We begin with those aspects that do not depend on the specific NS5-brane configuration. First we match the heterotic anomaly, which depends only on the class of the NS5-branes and not their configuration, to the D3-charge anomaly. Next we discuss a match of the stability conditions for the flux on the two sides, and then we match the charged matter content as well as the number of massless U(1) vector bosons (these matches only concern the $\mathrm{E}_8$s and not the branes/blow-ups). In the next section we turn to the match of matter content arising from NS5-branes and blow-ups.

\bigskip
\noindent\underline{\textbf{\large{Anomalies}}}
\smallskip

We begin with matching anomaly conditions on the two sides. In the process we will also show that the Euler characteristic of the F-theory fourfold is independent of which precise blow-ups are performed, which is dual to independence of the NS5-brane configuration. We recall that on the heterotic side both bundles $V_1$ and $V_2$ are taken to be line bundle sums. We use the notation
\be
c_1(V_1) = \sum_{a=1}^{\nolb_1} c_1(\lb_a) = k_a^iD_i \,, \quad c_1(V_2) = \sum_{a=1}^{\nolb_2} c_1(\tilde{\lb}_a) = \tilde{k}_a^iD_i \,,
\ee
where we have recalled that $k_a^0=\tilde{k}_a^0=0$ $\forall a$ or equivalently $\lb_a = \hp^*\blb_a$ and $\tilde{\lb}_a = \hp^*\tilde{\blb}_a$, as found in Section~\ref{sec:prelims}. The heterotic anomaly condition is
\be
\mathrm{ch}_2(V_1) + \mathrm{ch}_2(V_2) - \mathrm{ch}_2(\hs)= \fbc \,,
\ee
where $\fbc$ is the NS5-brane class. We recall the expressions for the second Chern characters,
\begin{align}
-\mathrm{ch}_2(\hs) &=  \left[\int_{\hb}(c_2(\hb)+11c_1(\hb)^2)\right] C^0 - 12\kdc_iC^i \,, \\
\mathrm{ch}_2(V_1) &= \frac{1}{2}d_{IJK}k_a^Ik_a^JC^K = \frac{1}{2}\bim_{ij}k_a^ik_a^jC^0\,, \\
\mathrm{ch}_2(V_2) &= \frac{1}{2}d_{IJK}\tilde{k}_a^I\tilde{k}_a^JC^K = \frac{1}{2}\bim_{ij}\tilde{k}_a^i\tilde{k}_a^jC^0\,,
\end{align}
where we use the notation defined in Section~\ref{sec:prelims}, and where we recalled the triple intersection numbers from Section~\ref{sec:prelims}. Hence
\be
\mathrm{ch}_2(V_1) + \mathrm{ch}_2(V_2) - \mathrm{ch}_2(\hs) =  \left( \int_{\hb} \left(c_2(\hb) + 11c_1(\hb)^2\right) + \frac{1}{2}\bim_{ij}k_a^ik_a^k + \frac{1}{2}\bim_{ij}\tilde{k}_a^i\tilde{k}_a^k\right)C^0 - 12\kdc_iC^i \,.
\ee
In our setup, we have chosen the NS5-brane content so that $\fbc = - 12\kdc_iC^i$, i.e.\ we chose not to include vertical NS5-branes, and hence we have 
\be\begin{aligned}
0 &=  \int_{\hb} \left(c_2(\hb) + 11c_1(\hb)^2\right) + \frac{1}{2}\bim_{ij}k_a^ik_a^j + \frac{1}{2}\bim_{ij}\tilde{k}_a^i\tilde{k}_a^k  \\
&=  \int_{\hb} \left(c_2(\hb) + 11c_1(\hb)^2 + \frac{1}{2}\left(\sum_a c_1(\blb_a)^2 + \sum_a c_1(\tilde{\blb}_a)^2\right) \right) \,,
\end{aligned}\ee
as the remaining part of the anomaly condition.

On the F-theory side, the corresponding anomaly condition is that of the cancellation of D3-brane charge, which reads \cite{Sethi:1996es}
\be
\nd_3-\frac{\ec(\bfs)}{24}+\frac{1}{2}\int_{\bfs}G_4 \w G_4 = 0 \,,
\ee
where $\nd_3$ is the number of D3-branes. In our setup there are no D3-branes, so $\nd_3=0$.\footnote{Note that the inclusion of D3-branes gives a contribution to the anomaly condition which trivially matches that of the corresponding vertical NS5-branes.} For the flux contribution, we can see from the expansion of $G_4$ in two-forms dual to the blow-ups of the $\mathrm{E}_8$ singularities that
\be
\int_{\bfs} G_4 \w G_4 = - \mathrm{Tr}\int_{\idb_1} F_1 \w F_1 - \mathrm{Tr}\int_{\idb_2} F_2 \w F_2 \,,
\ee
where $\idb_i$ are the two $\mathrm{E}_8$ surfaces. Since in our identification the D7-brane flux $F_i$ is equal to that of the corresponding heterotic bundle $V_i$ on the heterotic base space, we have
\be
\mathrm{Tr}\int_{\idb} F_1 \w F_1 + \mathrm{Tr}\int_{\idb} F_2 \w F_2 = \sum_a c_1(\blb_a)^2 +\sum_a c_1(\tilde{\blb}_a)^2 \,.
\ee
Hence, the only non-trivial part of the matching of the two anomaly conditions is
\be
 \int_{\hb} \left(c_2(\hb) + 11c_1(\hb)^2\right) \stackrel{?}{=} \frac{\ec(\bfs)}{24} \,.
\ee

We now show this last match holds. Recall that in addition to the blow-ups of the F-theory fourfold corresponding to horizontal NS5-branes, the geometrically non-Higgsable $\mathrm{E}_8$ singularities must be resolved. These blow-ups enter in the computation of the Euler characteristic. The Euler characteristic for a smooth elliptically fibered CY fourfold that is the result of resolving $\mathrm{E}_8$ singularities over a divisor class $S$ is known to be \cite{Andreas:1997ce, Esole:2017kyr}
\be
\ec(\bfs) = \int_{\bfb} 12(c_1c_2+30c_1^3-80c_1^2S+70c_1S^2-20S^3) \,,
\ee
where for brevity we write $c_i \equiv c_i(\bfb)$ and do not write wedge products. In our case, $S=[\fpc_1]+[\fpc_2]$. Additionally, before the blow-ups dual to horizontal NS5-branes, the first Chern class of $\fb$ is a sum $c_1(\fb) = c_1(\hb) + c_1(\mbb{P}^1)$. Under the blow-ups, the exceptional divisors are added to this expression. Since the class $S$ is equal to $c_1(\mbb{P}^1)$, we see that we can write $c_1(\bfb) = S + \Sigma$ where $\Sigma$ is a sum over pullbacks from $\hb$ to $\bfb$. To evaluate $\ec(\bfs)$, we first note that
\begin{align}
\ec(\bfs) &= \int_{\bfb} 12\left[c_1c_2+30(S^3+3S^2\Sigma+3S\Sigma^2+\Sigma^3)-80(S^3+2S^2\Sigma+S\Sigma^2)+70(S^3+S^2\Sigma)-20S^3\right] \nonumber \\
&= \int_{\bfb} 12\left[c_1c_2+10S\Sigma^2+30\Sigma^3\right] \,,
\end{align}
where many terms cancelled. The remaining terms are straightforward to evaluate. It is easy to see that the third term vanishes, since three pullbacks will not intersect. For the second, it is straightforward to see that $ \int_{\bfb} S \cdot \Sigma \cdot \Sigma =  \int_{\hb} 2c_1(\hb)^2$\,: since $\Sigma$ is a sum of pullbacks, the intersection $\Sigma \cdot [\fpc_{1,2}]$ clearly has curve class $c_1(\hb)$ in $\{\fpc_{1,2}=0\}$, and hence also clearly $ \int_{\bfb} [\fpc_{1,2}] \cdot \Sigma \cdot \Sigma =  \int_{\hb} c_1(\hb)^2$. In the toric hypersurface cases of Figures \ref{fig:fullyblownup_P1P1P1_fan} and \ref{fig:fullyblownup_examples_4d}, this is the observation that the presence of $[\fpc_1]$ restricts to the cones one sees from `above', which join $[\fpc_1]$ to the `highest' ray in each tower, and these cones form a copy of $\hb$; similarly for the term with $[\fpc_2]$.

It only remains to establish that $ \int_{\bfb} \tfrac{1}{2}c_1(\bfb)c_2(\bfb)= \int_{\hb} \left(c_1(\hb)^2+c_2(\hb)\right)$. In fact both expressions are equal to 12. This follows from the Hirzebruch-Riemann-Roch theorem, which states that for a holomorphic vector bundle $E$ on a compact complex manifold $X$,
\be
\ind(X,E) = \int_X \mathrm{ch}(E)\mathrm{td}(X) \,,
\ee
where $\ind(X,E)$ is the bundle index, and ch$(\cdot)$ and td$(\cdot)$ are the Chern character and Todd class respectively. We apply this to the case of the trivial line bundle, $E=\mc{O}_X$, for which we have ch$(E)=1$. We also recall the expression for the Todd class in Equation~\eqref{eq:td_class}. Now, both $\hb$ and $\bfb$ have $h^{i,0}=0$ for $i>0$ since they form the bases of CY manifolds, hence $\ind(X,\mc{O}_X)=1$ in both cases, and so we have
\begin{align}
\mathrm{\hb:} \quad&  \int_{\hb} \left(c_1(\hb)^2+c_2(\hb)\right) = 12 \,, \\
\mathrm{\bfb:} \quad&  \int_{\bfb} c_1(\bfb)c_2(\bfb) = 24 \,.
\end{align}
This establishes $ \int_{\bfb} \tfrac{1}{2}c_1(\bfb)c_2(\bfb)= \int_{\hb} \left(c_1(\hb)^2+c_2(\hb)\right)$, which was the final equality that we needed. Hence we have established the equivalence of the two anomaly conditions. We note this shows that $\ec(\bfs)$ is independent of the NS5-brane configuration, as we only used the fact that the $\mathrm{E}_8$ singularities are geometrically non-Higgsable, dual to the fact the horizontal NS5-brane class is $-12K_{\hb}$.

\bigskip
\noindent\underline{\textbf{\large{Stability condition}}}
\smallskip

Next we would like to match the flux stability conditions on the two sides of the duality. We write $V$ for either of the heterotic bundles $V_i$, whose flux is identified with that on one of the $\mathrm{E}_8$ surfaces in F-theory. We recall we have a sum of line bundles that are pullbacks,
\be
V = \bigoplus_a \lb_a = \bigoplus_i \hp^*(\blb_a) \,.
\ee
As reviewed in Section~\ref{sec:prelims}, for preservation of supersymmetry we require that the slopes of all the line bundles vanish simultaneously,
\be
0 \stackrel{!}{=} \int_{\hs} J^2 \wedge c_1(\lb_a) = \int_{\hs} J^2 \wedge \hp^*\left(c_1(\blb_a)\right) \quad \forall a \,.
\ee
We can rewrite this expression using Poincar\'{e} dual divisors and a knowledge of the triple intersection numbers from Section~\ref{sec:prelims}. We have
\be\begin{aligned}
\int_{\hs} J^2 \w c_1(\lb_a) 
&= \hkm^I\hkm^Jk_a^k D_I \cdot D_J \cdot D_k  =  \hkm^I\hkm^Jk_a^k d_{IJk} \\
&= -(\hkm^0)^2\kdc_k k_a^k + 2 \hkm^0\hkm^ik_a^k\bim_{ik} \\
&= 2\hkm^0\left(J_{\hb} - \tfrac{1}{2}\hkm^0K_{\hb} \right) \cdot \hbc \stackrel{!}{=} 0 \,,
\end{aligned}\ee
where $J_{\hb} \equiv \hkm^i\hbc_i$ and $\hbc \equiv c_1(\blb)$, and where we have recalled the intersection numbers from Section~\ref{sec:prelims}. We note that $\hkm^0\neq0$ as it corresponds to the volume of the fiber, so the $\hkm^0$ factor drops out of the condition. As we are working in the adiabatic limit of $\hs$, for which the volume of the elliptic fiber is small compared to volumes in the base, we can drop the second term and recover the usual D-term stability condition on a fluxed 7-brane
\begin{equation}
\int J \wedge F_a \stackrel{!}{=} 0 \,,
\end{equation}
where the $F_a$ are the U(1) fluxes on the 7-brane corresponding in the duality to the heterotic U(1) fluxes of the line bundles $L_a$. One can see that this is the stability condition for a fluxed 7-brane from a local analysis, or from a global F-theory analysis where it corresponds to the condition $J \wedge G_4 = 0$. See for example Refs.~\cite{Cvetic:2012xn,Grimm:2011tb,Grimm:2010ks} for discussions of this condition.

\bigskip
\noindent\underline{\textbf{\large{Matter multiplets}}}
\smallskip

Next we would like to match the counts of matter multiplets charged under the $\mathrm{E}_8$ gauge groups. We begin with the heterotic side. We have two vector bundles $V_i$, corresponding to the two $\mathrm{E}_8$ factors, each given by a sum of line bundles that are pulled back from the base. We write $V= \bigoplus_a \lb_a = \bigoplus_i \hp^*(\blb_a)$ for either of these two bundles, as above. We are interested in the cohomologies of the $\lb_a$, as well as of tensor products and duals, since these determine the charged matter content. As the heterotic threefold $\hs$ is elliptically fibered, we will use the Leray spectral sequence to determine these cohomologies. See e.g.~Refs.~\cite{Friedman:1997yq,Andreas:2007ev,Donagi:2004ia} for details and examples of the application of this sequence in related contexts.

First we recall the direct image and higher direct image of the trivial bundle on the heterotic threefold,
\be
(\hp)_*\mc{O}_{\hs} = \mc{O}_{\hb} \,, \quad R^1(\hp)_{*}\mc{O}_{\hs} =  K_{\hb} \,,
\ee
as well as the projection formulae $(\hp)_*(\lb_a) = (\hp)_* \mc{O}_{\hs} \otimes \blb_a$ and $R^1 (\hp)_*(\lb_a) = R^1 (\hp)_* \mc{O}_{\hs} \otimes \blb_a$. Next we recall from the Leray spectral sequence that
\be
H^0(\lb_a) = E_2^{0,0} \,, \quad H^3(\lb_a) = E_2^{2,1} \,,
\ee
\be
0 \to E_2^{1,0} \to H^1(\lb_a) \to E_2^{0,1} \to E_2^{2,0} \to H^2(\lb_a) \to E_2^{1,1} \to 0 \,,
\ee
where $E_2^{p,q}\equiv H^p(R^q(\hp)_*\lb_a)$ and where the sequence is exact. If for a given $\lb_a$ one of $E_2^{0,1}$ or $E_2^{2,0}$ vanishes then the exact sequence splits. In fact we will show in a moment that $h^2(\blb_a)=0$, so the exact sequence indeed always splits as the $E_2^{2,0}$ term is zero. Hence we have
\be\begin{aligned}
H^0(\lb_a) &= H^0(\blb_a) \,, \\
H^1(\lb_a) &= H^0(K_{\hb} \otimes \blb_a )\oplus H^1(\blb_a) = H^2(\blb_a^* )^*\oplus H^1(\blb_a) \,, \\
H^2(\lb_a) &= H^1(K_{\hb} \otimes \blb_a)\oplus H^2(\blb_a) = H^1(\blb_a^*)^*\oplus H^2(\blb_a)\,, \\
H^3(\lb_a) &= H^2(K_{\hb} \otimes \blb_a) = H^0(\blb_a^*) \,.
\end{aligned}\ee
where we have also included the results after using Serre duality, which eliminates occurrences of $K_{\hb}$. 

We are also interested in the cohomologies of tensor products and duals of the line bundles, since these enter in the counts of multiplets in various representations of the gauge group. We note that the pullback commutes with taking duals or tensor products, so that the line bundles we need to consider will always be pullbacks, and hence the above analysis holds also in these cases. Hence quite generally, writing $\tau$ for a representation corresponding to a particular vector bundle $\tilde{V}=\hp^*\tilde{\mc{V}}$ built from taking tensor product and dual operations on $V$, we then have for the number $n_{\tau}$ of multiplets in this representation and the number $n_{\tau^*}$ in the conjugate representation,
\be\begin{aligned}
n_{\tau} &= h^1(\hb,\tilde{\mc{V}}) + h^2(\hb,\tilde{\mc{V}}^*) \,, \\
n_{\tau^*} &= h^1(\hb,\tilde{\mc{V}}^*) + h^2(\hb,\tilde{\mc{V}}) \,,
\label{eq:chg_mult_count}
\end{aligned}\ee
which completes the computation of the numbers of charged multiplets. 

Finally we also note the following results on vanishing cohomologies. From the supersymmetry conditions on the vector bundle $V$, it follows \cite{Anderson:2012yf} for a sum of non-trivial line bundles that $h^0=h^3=0$ for $V$ and $V^*$. Since $V$ is a line bundle sum, it follows that each line bundle in $V$ must have $h^0=h^3=0$. Hence it also follows that $h^0=h^3=0$ for $\tilde{V}$ and $\tilde{V}^*$, since these are built from tensor products and duals of the line bundles. Additionally, it is then straightforward to see that $h^2(\hb,\tilde{\mc{V}})=h^2(\hb,\tilde{\mc{V}}^*)=0$ if $-K_{\hb}$ is effective (which is required for the existence of the elliptic fibration of $\hs$) as follows. We note the inclusions
\be\begin{aligned}
H^2(\hb,\tilde{\mc{V}}^{\phantom{*}}) = H^0(\hb,\tilde{\mc{V}}^* \otimes K_{\hb})^* &\subseteq H^0(\hb,\tilde{\mc{V}}^*)^* \\
H^2(\hb,\tilde{\mc{V}}^*) = H^0(\hb,\tilde{\mc{V}}^{\phantom{*}} \otimes K_{\hb})^* &\subseteq H^0(\hb,\tilde{\mc{V}}^{\phantom{*}})^* \,,
\end{aligned}\ee
where in the equalities we have used Serre duality. Since $h^0(\hs,\tilde{V})=0$ and $h^0(\hb,\tilde{\mc{V}})=h^0(\hs,\tilde{V})$, and similarly for $\tilde{V}^*$ and $\tilde{\mc{V}}^*$, we see from the inclusions that indeed $h^2(\hb,\tilde{\mc{V}})=h^2(\hb,\tilde{\mc{V}}^*)=0$. This result shows the exact sequence in the Leray spectral sequence always splits, as assumed above. It also reduces Equation~\eqref{eq:chg_mult_count} to the $h^1$ terms.

\bigskip

On the F-theory side, the proposed dual geometry contains two $\mathrm{E}_8$ brane stacks, and it is the matter coming from flux on these brane stacks that we expect to match the above heterotic matter. We can consider each $\mathrm{E}_8$ stack separately, so we will write $\idb$ for either surface. The background flux is described by a vector bundle on this surface, and in the proposal in Section~\ref{sec:prelims}, this vector bundle is the same as appears on the base of the heterotic threefold for the corresponding $\mathrm{E}_8$ gauge group. The count of the charged multiplets for a given background flux on the surface has been computed in Ref.~\cite{Beasley:2008dc}. In particular the result is that the matter content in a representation $\tau$ corresponding to a vector bundle $\mc{T}$ is given by
\be
H^0(\idb,\mc{T}^*)^* \oplus H^1(\idb,\mc{T}) \oplus H^2(\idb,\mc{T}^*)^* \,,
\ee
so the number $n_{\tau}$ of multiplets in the $\tau$ representation and the number $n_{\tau^*}$ of multiplets in the conjugate representation $\tau^*$ are given by
\be\begin{aligned}
n_{\tau} &= h^0(S,\mc{T}^*) + h^1(S,\mc{T}) + h^2(S,\mc{T}^*) \,, \\
n_{\tau^*} &= h^0(S,\mc{T}) + h^1(S,\mc{T}^*) + h^2(S,\mc{T}) \,.
\end{aligned}\ee
However, a non-zero $h^0(\idb,\mc{T})$ or $h^0(\idb,\mc{T}^*)$ would mean the F-theory compactification is inconsistent \cite{Donagi:2008ca}. We note that these were also zero on the heterotic side. Rewriting the number of multiplets with this taken into account, we have
\be\begin{aligned}
n_{\tau} &= h^1(\idb,\mc{T}) + h^2(\idb,\mc{T}^*) \,, \\
n_{\tau^*} &= h^1(\idb,\mc{T}^*) + h^2(\idb,\mc{T}) \,.
\end{aligned}\ee
We see that this result precisely matches the heterotic side, so that we have established the matching of the two charged matter spectra. We also note that in Ref.~\cite{Beasley:2008dc} they find that $h^2(\idb,\mc{T})=h^2(\idb,\mc{T}^*)=0$ if both $-K_{\idb}$ is effective and $h^{2,0}(\idb)=0$. Since $\idb$ is diffeomorphic to the heterotic base $\hb$, both of these conditions hold since $\hb$ forms the base of a CY elliptic fibration. Hence these second cohomologies vanish for the $\idb$ we consider, which matches what was found on the heterotic side. 

\newpage
\bigskip
\noindent\underline{\textbf{\large{Massless U(1)s}}}
\smallskip

Finally we wish to match the count of extra massless U(1) vector bosons, which are a well-known possibility in line bundle sum models. For example, if a line bundle sum is used to break one of the $\mathrm{E}_8$ gauge groups to a group containing SU(5), the commutant also necessarily contains four additional U(1) factors. These extra U(1) vector bosons tend to be massive by the Green-Schwarz mechanism, see e.g.~Refs.~\cite{Nibbelink:2009sp,Anderson:2012yf}. We will count on both sides of the duality the number of the extra massless U(1) vector bosons. On the heterotic side, the mass matrix for the extra U(1) vector bosons in line bundle models is found to be at lowest order
\be
M_{ab} = k_a^IG_{IJ}k_b^J \,,
\ee
where $G$ is the {\kahl} metric, $k_a^I$ are the integers specifying the line bundle sum as discussed in Section~\ref{sec:prelims}, and as also discussed there, in our models $k_a^0=0 ~\forall a$, so that only $k_a^i$ appears. Since $G$ is invertible, the number of massless U(1) vector bosons is determined by the rank of $k_a^i$.

On the F-theory side, we recall the discussion from an M-theory perspective \cite{Grimm:2010ks}. We write $\omega_{\alpha}$ for two-forms pulled back from the base $\bfb$ by the projection map of the elliptic fibration, and $\gw_a$, $a=1,\ldots,\mathrm{rk}(\mathrm{E}_8)$, for two-forms dual to the exceptional divisors in the blow-up of the $\mathrm{E}_8$ singularity, whose intersections give Cartan matrix factors $C_{ab}$. The $G_4$ flux is expanded in the $\gw_a$, which are the natural geometric objects,
\be
G_4 = \tilde{F}^a \w \gw_a \,,
\ee
where the $\tilde{F}^a$ are pullbacks of flux two-forms on the $\mathrm{E}_8$ surface $\idb$. This corresponds to a choice of generators $T_a$ of the adjoint representation of the Lie algebra such that Tr$(T_aT_b)=C_{ab}$. The relevant St\"{u}ckelberg mass term appears in a gauging of fields $T_\alpha$, which are {\kahl} moduli of the base $\bfb$,
\be
\dD T_\alpha = \dd T_\alpha + iX_{\alpha a}\tilde{A}^a \,, \quad \mathrm{where} \quad X_{\alpha i}:= \tfrac{1}{2} \int_{\bfs}\omega_\alpha \w  \gw_a \w G_4 \,,
\ee
and it follows straightforwardly that the St\"{u}ckelberg mass term is
\be
X_{\alpha a}\tilde{A}^a = -\frac{1}{2}C_{ab}\tilde{A}^a\int_{\idb} \omega_\alpha \w \tilde{F}^b \,.
\ee
The $C_{ab}$ factor reflects the geometrically convenient basis choice in the Lie algebra. More generally we clearly have
\be
-\frac{1}{2}\mathrm{Tr}\left(A\int_{\idb} \omega_\alpha \w F\right) \,.
\ee
We write $\{\omega_i\}$ for the subset of $\{\omega_{\alpha}\}$ that are pullbacks in $\bfb$ from curves in $\idb$ under the projection that collapses the $\mbb{P}^1$. These survive in the above integral. Additionally as we have a line bundle sum on $\idb$, we can write $F = F^a k_a^i \omega_i$, where the $k_a^i$ are identified in the duality with those of the heterotic line bundle sum. Then as well as factors of the intersection matrix $\bim_{ij} \equiv \int_{\idb} \omega_i \w \omega_j$ on $\idb$ and factors of the moduli metric, which are both invertible, the mass matrix of U(1) vector bosons contains only occurrences of $k_a^i$. Hence the number of massless U(1) vector bosons is determined by the rank of $(k_a^i)$ exactly as on the heterotic side.

\bigskip

Before turning to duality checks involving the NS5-brane/blow-up configuration, we make two final comments. First, there is an additional condition on the F-theory flux $G_4$ which we have not yet discussed. The flux $G_4$ is subject to the Freed-Witten quantisation condition \cite{Witten:1996md},
\be
G_4 + \frac{1}{2}c_2(\bfs) \in H^4(\bfs,\mbb{Z}) \,.
\ee
In the F-theory models that we have constructed, $G_4$ is manifestly integrally quantised since the heterotic flux is integrally quantised, c.f.~Equation~\eqref{eq:g4_exp}. Hence we expect that $c_2(\bfs)$ is always even. This is however difficult to fully verify, due to the difficulty in computing an integral basis for $H^4$. The fact that in our examples the Euler characteristic is divisible by 24 is a necessary check, since this is implied by evenness of $c_2(\bfs)$ \cite{Witten:1996md,Klemm:1996ts}. Additionally, in the D3-brane charge cancellation condition, we know the number of D3-branes from the heterotic dual, and including also the $\ec(\bfs)/24$ contribution and the flux contribution from the $\mathrm{E}_8$ branes this condition is satisfied, which seems to leave no room for additional $G_4$ flux. Finally, half-integrally quantised $G_4$ fluxes are localised on stacks of 7-branes \cite{Collinucci:2010gz}, and in our situation the only 7-brane stacks are the $\mathrm{E}_8$ brane stacks, whose flux seems to be matched with the heterotic line bundle flux. For further discussion of the Freed-Witten quantisation condition in F-theory see for example Refs.~\cite{Collinucci:2010gz,Collinucci:2012as}.

Second, F-theory has an attractive mechanism \cite{Donagi:2008kj,Beasley:2008dc} for breaking a GUT group to the gauge group of the Standard Model leaving the hypercharge massless. This requires the existence of a curve class in the GUT surface $\idb$ that maps to a trivial curve class in the embedding of $\idb$ into the base $\bfb$. In our models, the GUT surfaces form the two sections of a $\mbb{P}^1$ fibration. Curves in the GUT surfaces pull back to non-trivial divisors on $\bfb$, whose intersection with the sections of the fibration return the original curves. Hence this F-theory mechanism is not possible in these models. This is expected, as it is well-known that this is not possible for models with a heterotic dual. See also Ref.~\cite{Anderson:2014hia} for more detail on the situation in heterotic string theory.

\subsection{Duality checks concerning NS5-branes and blow-ups}
\label{sec:dualchecks_ns5s_and_blowups}

In Section~\ref{sec:dual_checks_ind_of_ns5_config} we have performed checks of the duality between heterotic line bundle models and their proposed dual F-theory models, which did not depend on the precise NS5-brane configuration. We now turn to a match of multiplets whose number does depend on the precise configuration of NS5-branes. As explained in Sections \ref{sec:sing_and_inst_in_6d} and \ref{sec:tor_desc_in_6d}, this corresponds to the choice of blow-ups in the F-theory base that resolve the $\tilde{\mathrm{E}}_8$ singularities appearing where the $\mathrm{E}_8$ stacks intersect the remaining D7-brane locus. We will write $\bfs$ for the F-theory CY fourfold after all resolutions (both those that alter the base and those that do not), and $\bfb$ for the blown-up base. On the heterotic side we have a CY threefold $\hs$ with a base $\hb$.

\bigskip

We first collect general expressions for the number of multiplets on each side, for a general NS5-brane configuration. We begin with the heterotic side, and we will ignore the vector multiplets of the $\mathrm{E}_8 \times \mathrm{E}_8$. There are $h^{1,1}(\hs)+h^{2,1}(\hs)$ chiral multiplets from the geometry and one from the heterotic dilaton. Second, we have the moduli from the NS5-branes, which have been counted in Ref.~\cite{Lukas:1998hk}, and we briefly recall this count. For a single NS5-brane wrapping a genus $g$ curve in the CY manifold, there is a single universal chiral multiplet and $g$ vector multiplets. Additionally there are chiral multiplets from the deformation moduli.\footnote{The number of deformation moduli in our cases will be rather clear, since we consider only vertical and horizontal NS5-branes, rather than branes that are a combination. For a detailed discussion of the moduli space of general NS5-branes on elliptic CY threefolds see Ref.~\cite{Donagi:1999jp}.} We write $\nf_5$ for the number of NS5-branes, and we write $M_i$ for the $i^{\mathrm{th}}$ brane. We will also write $\fdm(M_i)$ for the number of (complex) deformation moduli of the $i^{\mathrm{th}}$ brane, and $g(M_i)$ for its genus. The number $\nch$ of chiral multiplets and the number $\nve$ of vector multiplets (excluding those from the $\mathrm{E}_8 \times \mathrm{E}_8$ sector) are then,
\begin{align}
\nch &= h^{1,1}(\hs) + h^{2,1}(\hs) + 1 + \nf_5 + \sum_{M_i} \fdm(M_i) \,, \\
\nve &= \sum_{M_i} g(M_i) \,.
\end{align}
Note that in our models we have chosen not to include vertical branes: as discussed in Section~\ref{sec:req_branes_and_blowups}, the anomaly condition can be satisfied using only horizontal branes and line bundle flux.

On the F-theory side we have the following. We first note that in our models there is no $G_4$ flux away from the $\mathrm{E}_8$ stacks, which could have lifted moduli. The number of massless neutral chiral multiplets $\nch$ and the rank of the resulting gauge group $\mathrm{rk}_V$ are then \cite{Mohri:1997uk,Andreas:1997pd,Curio:1997rn} (see also Refs.~\cite{Grimm:2010ks,Grimm:2012yq} for a detailed discussion)
\begin{align}
\nch &= h^{3,1}(\bfs) + h^{1,1}(\bfb) + \left( h^{2,1}(\bfs) - h^{2,1}(\bfb) \right) + 3\nd_3 \,, \label{eq:numch} \\ 
\mathrm{rk}_V &= h^{1,1}(\bfs) + h^{2,1}(\bfb) - h^{1,1}(\bfb) - 1 + \nd_3 \,, \label{eq:numu}
\end{align}
where $\nd_3$ is the number of D3-branes, which is zero in our models\footnote{Note however that the match between D3-branes and vertical NS5-branes is very simple. Each vertical NS5-brane contributes 1 to $\nf_5$, and has 2 complex deformation parameters as it sits over a point in a twofold base, giving 3 chiral multiplets. These branes wrap genus 1 curves, giving a single vector multiplet. These match the D3-brane contributions.}, corresponding to the choice to not include vertical NS5-branes. Additionally, we have not broken the two $\mathrm{E}_8$ gauge symmetries by geometry, rather only by flux, so that in the rank computation we will find a contribution of 16 corresponding to these, which we have already discussed. The mass of the vector multiplets from these $\mathrm{E}_8$ gauge groups are determined by the background $G_4$ flux pulled back from the brane, as reviewed in Section~\ref{sec:dual_checks_ind_of_ns5_config}. We then have 
\be
\nve = h^{1,1}(\bfs) + h^{2,1}(\bfb) - h^{1,1}(\bfb) - 1 -16
\ee
for the number of vector multiplets $\nve$ from the remaining sector.

\bigskip

We now match the vector multiplet counts in general. First we note that $h^{1,1}(\bfs) - h^{1,1}(\bfb) - 1 - 16 $ counts extra sections of the F-theory elliptic fibration, and these are not present in our F-theory models. Hence for the models we consider the matching condition for vector multiplets is
\be
\sum_{M_i} g(M_i) \stackrel{?}{=} h^{2,1}(\bfb) \label{eq:gen_h21_rel} \,.
\ee
Before the blow-ups dual to horizontal NS5-branes, $\fb = \hb \times \mbb{P}^1$, so that $h^{2,1}(\fb)=0$ (using the Kunneth formula and that $h^{1,0}(\hb)=0$ since otherwise $h^{1,0}(\hs)\neq0$). In each blow-up, a curve $C$ of genus $g$, diffeomorphic to the NS5-brane curve, is replaced by a $\mbb{P}^1$ fiber bundle $E$ over the curve, so by the additivity of the Euler characteristic,
\be
\Delta \ec(\fb) = \ec(E)-\ec(C) = 2(2-2g)-(2-2g) = 2-2g \,,
\ee
and noting that $\ec(\fb)=2+2(h^{1,1}(\fb)-h^{2,1}(\fb))$, which follows since $h^{1,0}(\fb)=h^{2,0}(\fb)=0$ for $\fs$ to be CY, this means $\Delta(h^{1,1}(\fb)-h^{2,1}(\fb))=1-g$. Since $C$ is irreducible we know $\Delta h^{1,1}(\fb)=1$, so we have shown that $\Delta h^{2,1}(\fb)=g$. This proves the matching condition.

Next we look at matching the chiral multiplet counts. The required match is
\be
h^{1,1}(\hs) + h^{2,1}(\hs) + 1 + \nf_5 + \sum_{M_i} \fdm(M_i) \stackrel{?}{=} h^{3,1}(\bfs) + h^{1,1}(\bfb) + \left( h^{2,1}(\bfs) - h^{2,1}(\bfb) \right) \,.
\ee
It will be useful for our purposes to reduce this equation to one involving fewer Hodge numbers. One uninterested in the derivation can skip directly to Equation~\eqref{eq:red_chmult_match}. We first note that the heterotic CY threefold $\hs$ is an elliptic fibration with a single section, from which we have
\begin{align}
h^{1,1}(\hs) &= h^{1,1}(\hb) + 1 \,, \\
\ec(\hs) &\equiv 2\left(h^{1,1}(\hs)-h^{2,1}(\hs)\right) =  -60\int_{\hb}c_1(\hb)^2 \,, \\
\Rightarrow \quad h^{2,1}(\hs) &= h^{1,1}(\hb)+1+ 30 \int_{\hb} c_1(\hb)^2 \,.
\end{align}
The expression for the Euler characteristic $\ec(\hs)$ of a smooth elliptic threefold $\hs$ described by a Weierstrass equation is a simple consequence of adjunction. Furthermore
\be
\int_{\hb} c_2(\hb) = h^{1,1}(\hb) + 2 \,.
\ee
On the F-theory side, there is a relation between Hodge numbers of the CY fourfold \cite{Klemm:1996ts}, so that the Euler characteristic can be written in terms of any three,
\begin{align}\begin{split}
h^{2,2}(\bfs) &= 44 + 4h^{1,1}(\bfs) + 4h^{3,1}(\bfs) -2h^{2,1}(\bfs) \,, \\
\ec(\bfs) &= 4 +2h^{1,1}(\bfs) - 4h^{2,1}(\bfs) + h^{2,2}(\bfs) + 2h^{3,1}(\bfs) \,, \\
\Rightarrow \quad \ec(\bfs) &= 6\left(8 +h^{1,1}(\bfs) - h^{2,1}(\bfs) + h^{3,1}(\bfs)\right) \,, \\
\Rightarrow \quad h^{3,1}(\bfs) &= \frac{1}{6}\ec(\bfs) - 8 - h^{1,1}(\bfs) + h^{2,1}(\bfs) \,, \\
\mathrm{where} \quad \frac{1}{6}\ec(\bfs) &= 48 + 40 \int_{\hb} c_1(\hb)^2 \,,
\end{split}\end{align}
in which the expression for $\ec(\bfs)$ was given in the anomaly cancellation match in Section~\ref{sec:dual_checks_ind_of_ns5_config}. We also recall from that discussion that
\be
 \int_{\hb} \left(c_1(\hb)^2+c_2(\hb)\right) = 12 =  \int_{\bfb} \frac{1}{2}c_1(\bfb)c_2(\bfb) \,.
\ee
There is a single section of the F-theory elliptic fibration as noted in the vector multiplet match, and additionally from knowledge of the duality between F-theory base blow-ups and NS5-branes we have together
\begin{align}
h^{1,1}(\bfs) =& h^{1,1}(\bfb) + 1 + 16 \,, \\
h^{1,1}(\bfb) =& h^{1,1}(\hb) + 1 + \nf_5 \,,
\end{align}
and we recall the relation in Equation~\eqref{eq:gen_h21_rel}. Using all of these expressions, it is straightforward to reduce the chiral multiplet match to
\be
\sum_{M_i} \fdm(M_i) +\nf_5 + \sum_{M_i} g(M_i) -12c_1(\hb)^2 - 2h^{2,1}(\bfs) \stackrel{?}{=} 0 \,.
\label{eq:red_chmult_match}
\ee

\bigskip

\newcommand{\nb}{N_{C/\hb}}
\newcommand{\degr}{\mathrm{deg}}

This is clearly a more complicated match than in the case of vector multiplets. We will show now that it holds under a particular assumption, and then we will discuss the general case in Section~\ref{sec:subtle_curves} below. The assumption we impose is that 
\be
-K_{\hb} \cdot C \neq 0 ~~ \forall C\quad  \mathrm{and} \quad h^{2,1}(\bfs)=h^{2,1}(\bfb)  \,,
\label{eq:assump_for_chmult_proof}
\ee
where $C$ is any curve in the heterotic base $\hb$ wrapped by an NS5-brane. We will argue in Section~\ref{sec:subtle_curves} that the first condition is expected to enforce the second, so we will refer to this as a single assumption. We emphasise that there are many situations satisfying this. For example, this holds for any of the toric hypersurface cases with heterotic base $\mbb{P}^1 \times \mbb{P}^1$ or $\mbb{P}^2$, examples of which were shown in Figures~\ref{fig:fullyblownup_P1P1P1_fan} and \ref{fig:fullyblownup_examples_4d}. Using this assumption, the match is straightforward to show. From the first equation it follows that for a brane $M_i$ wrapping a curve $C$,
\be
\fdm(M_i) = -K_{\hb} \cdot C -1 + g(C) \,,
\ee
where $g(C)$ is the genus of $C$. (We will prove this in a moment.) Additionally, the second equation allows us to write ${h^{2,1}(\bfs) = \sum_{M_i} g(M_i)}$, as shown in the proof of Equation~\eqref{eq:gen_h21_rel}. Substituting these two expressions into the chiral multiplet match, we find that it is satisfied. Hence the match holds under the assumption above.

To derive the expression for $\fdm(M_i)$, let a horizontal NS5-brane wrap a curve $C$ of genus $g$ in the heterotic base $\hb$. As we will discuss below in Section~\ref{sec:subtle_curves}, the condition $-K_{\hb} \cdot C \neq 0$ ensures that the NS5-brane can only be deformed within $\hb$. Hence the number of holomorphic deformations is counted by
\be
h^0(C,N_{C/\hs}) = h^0(C,\nb) \,,
\ee
where $N_{C/\hs}$ and $\nb$ are the normal bundles of $C$ within $\hs$ and $\hb$, the latter being a line bundle. We can use the Riemann-Roch formula
\be
h^0(C,\nb)-h^1(C,\nb) = c_1(\nb)+1-g \,,
\ee
to rewrite $h^0(C,\nb)$, since as we now show, $h^1(C,\nb)=0$. Note by Serre duality on $C$ that
\be
h^1(C,\nb)=h^0(C,\nb^* \otimes K_C) \,.
\label{eq:h1_normbund}
\ee
The bundle $\nb^* \otimes K_C$ has degree
\be
\degr(\nb^* \otimes K_C) = -c_1(\nb)-c_1(T_C) = -c_1(T_{\hb}|_C) = K_{\hb} \cdot C \,,
\ee
and we can note that $K_{\hb} \cdot C \leq 0$ as $\hb$ is weak Fano, with the equality case ruled out by the assumption on $C$. Hence the degree is negative, giving $h^1(C,\nb)=0$ from Equation~\eqref{eq:h1_normbund}. So the Riemann-Roch formula gives
\be
h^0(C,\nb) = c_1(\nb)+1-g = -K_{\hb} \cdot C -1 + g \,,
\ee
where we noted $c_1(\nb) = c_1(T_{\hb}|_C) - c_1(T_C) = -K_{\hb} \cdot C - \ec(C)$. This derivation completes the multiplet match above, under the specified assumption.

\bigskip

Before moving on to cases where the assumption in Equation~\eqref{eq:assump_for_chmult_proof} has been relaxed, we now give some examples that illustrate the matches of vector and chiral multiplets in cases obeying the assumption. We first consider the example in Figure~\ref{fig:fullyblownup_P1P1P1_fan}, in which the F-theory fourfold is a toric hypersurface. Here the heterotic base is $\hb = \mbb{P}^1 \times \mbb{P}^1$, and the NS5-branes/blow-ups are distributed equally between $\bc_1=0$ and $\bct_1=0$. For this choice of heterotic base space, we find the Hodge numbers
\be
h^{1,1}(\hs) = 3 \,, \quad h^{2,1}(\hs) = 243 \,.
\ee
Additionally we have 48 NS5-branes wrapping $\mbb{P}^1$s. As these branes wrap genus zero curves they do not give vector multiplets. Each however has a single complex deformation modulus, so that the branes contribute 96 chiral multiplets. Altogether this gives $\nch=343$ chiral multiplets. On the F-theory side we find the following Hodge numbers,
\be\begin{aligned}
\bfb: \quad & h^{1,1}(\bfb) = 51 \,, \quad h^{2,1}(\bfb) = 0  \,, \\
\bfs: \quad & h^{1,1}(\bfs) = 68 \,, \quad h^{2,1}(\bfs) = 0 \,, \quad h^{3,1}(\bfs) = 292 \,, \quad h^{2,2}(\bfs) = 1484 \,, \\
\mathrm{with} ~~& \ec(\bfs) = 2208 \,.
\end{aligned}\ee
We see from the above expression for the multiplet count in F-theory that we have $\nch = 343$, as well as the $\mathrm{E}_8$ vector multiplets, so that the multiplet content matches.

Next we illustrate the multiplet count for some models with more complicated NS5-brane configurations, whose F-theory dual geometries are described in general by complete intersections as described in Section~\ref{sec:more_gen_glob_models}. We take as an example the case of a heterotic base $\mbb{P}^2$, where we keep some NS5-branes at the vanishing loci of toric coordinates, but the rest of the brane class will be in a single general genus $g$ NS5-brane. For this choice of heterotic base, we have the Hodge numbers
\be
h^{1,1}(\hs)=2 \,, \quad h^{2,1}(\hs)=272 \,.
\ee
The NS5-brane configuration was shown in Figure~\ref{fig:different_brane_configs}. The toric description of the F-theory geometry in this situation was given in Section~\ref{sec:more_gen_glob_models}. The complete intersection can be described by a nef partition, and hence the Hodge number computation is straightforward. In Table \ref{tab:p2_ns5braneconfig_multiplet_counts} we record on the two sides of the duality the number of chiral multiplets $\nch$, and the number of U(1) vector bosons $\numu$ associated to this sector, in the case of a single NS5-brane wrapping a degree $n$ curve in the heterotic base, for several choices of $n$. Since the heterotic base is $\mbb{P}^2$, the genus of the Riemann surface wrapped by the brane is $g=\frac{1}{2}(n-1)(n-2)$.

\begin{table}[h]
\begin{center}
\begin{tabular}{|c||c|c||c|c|c|c|c||c|c|}
\hline
$n$ & 	$\nf_5$  & 	$\fdm$ &			$h^{3,1}(\bfs)$ & 	$h^{2,1}(\bfs)$ & 	$h^{1,1}(\bfs)$ & 	$h^{1,1}(\bfb)$ & 	$h^{2,1}(\bfb)$ &	$\nch$ &		$\numu$ 	\\ \hline
2 & 		34+1 &		$34\times2+5$& 	346 & 			0 & 				54&				37& 				0& 				383 &		0			\\ \hline
3 & 		33+1 &		$33\times2+9$& 	348 & 			1 & 				53&				36& 				1&				384 &		1			\\ \hline
6 & 		30+1 &		$30\times2+27$& 	360 & 			10 & 			50&				33& 				10&				393 &		10			\\ \hline
9 & 		27+1 &		$27\times2+54$& 	381 & 			28 & 			47&				30& 				28&				411 &		28			\\ \hline
\end{tabular}
\caption{Multiplet matching in the case of $\hb=\mbb{P}^2$, when one NS5-brane wraps a degree $n$ curve in $\hb$ and the other branes wrap $\mbb{P}^1$s. Note that $h^{1,1}(\hs)=2$ and $h^{2,1}(\hs)=272$. Here $\numu$ is only the number of U(1) vector bosons associated to this sector -- it does not include contributions from the $\mathrm{E}_8$ sectors.}
\label{tab:p2_ns5braneconfig_multiplet_counts}
\end{center}
\end{table}

\subsection[Subtleties concerning NS5-branes not intersecting \texorpdfstring{$-K_{\hb}$}{-K\_B2}]{Subtleties concerning NS5-branes not intersecting \texorpdfstring{$\boldsymbol{-K_{\hb}}$}{-K\_B2}}
\label{sec:subtle_curves}

In the restricted proof of the chiral multiplet match in Section~\ref{sec:dualchecks_ns5s_and_blowups}, we imposed that no NS5-brane wraps a curve $C$ in $\hb$ that does not intersect the anti-canonical divisor, i.e.\ ${-K_{\hb}\cdot C \neq 0}$ for all $C$ wrapped by an NS5-brane. Such branes are special, since the heterotic elliptic fibration is trivial over $C$, which follows as the discriminant locus is in the class $-K_{\hb}$ and an elliptic surface $S_C$ with no singular fibers is trivially fibered. Hence the brane sits in a subspace $S_C = {T^2 \times C}$. This fact is important because precisely these branes can move in the heterotic fiber direction, since any point in the $T^2$ can be specified holomorphically across the whole base region $C$. That is,
\be
-K_{\hb}\cdot C = 0 \quad \Rightarrow \quad S_C = T^2 \times C \quad \Rightarrow \quad \textrm{`vertical' deformation possible} \,.
\ee
Such branes then have an additional single complex deformation modulus, in addition to any deformations within the base. (We note that if $C$ has genus zero, then by adjunction $C^2=-2$ so the curve is rigid in $\hb$, however for higher genus, ${C^2\geq0}$.)

As in the case of point-like NS5-branes in compactification to six dimensions, the fiber position of an NS5-brane is expected to be dual to a Ramond-Ramond modulus (in the IIB picture) on the F-theory side. In compactifications to four dimensions, the relevant moduli are counted by\footnote{This difference of Hodge numbers can also include Wilson line moduli on the D7-branes, arising from $(1,0)$-forms on the brane $\idb$. However these are not present in our cases, since $\idb$ is diffeomorphic to $\hb$ and $b_1(\hb)=0$.} \cite{Donagi:2008ca,Grimm:2010ks}
\be
\textrm{No. of Ramond-Ramond moduli}  ~~=~~  h^{2,1}(\bfs)-h^{2,1}(\bfb) \,.
\label{eq:ramram_mod_count}
\ee
This will be important for the discussion below. It is also why in Equation~\eqref{eq:assump_for_chmult_proof} we claimed the first condition is expected to imply the second.

\bigskip

We have shown in Section~\ref{sec:dualchecks_ns5s_and_blowups} that the chiral multiplet match in Equation~\eqref{eq:red_chmult_match} holds under the assumption that there are no branes wrapping curves $C$ that do not intersect $-K_{\hb}$. However, in cases where this assumption does not hold, we have not given a proof of the match and in fact it turns out that such a match fails in general. That is, we find there is an apparent chiral multiplet mismatch
\be
\nch(\textrm{Het.}) \neq \nch(\textrm{F-th.}) \quad \mathrm{when} \quad -K_{\hb} \cdot C = 0 \,,
\ee
for some curve $C$ wrapped by a horizontal NS5-brane. We will first explain how precisely it fails, and then outline the expected reasons for the failure.

\begin{figure}[t]
\centering
\includegraphics[scale=0.7]{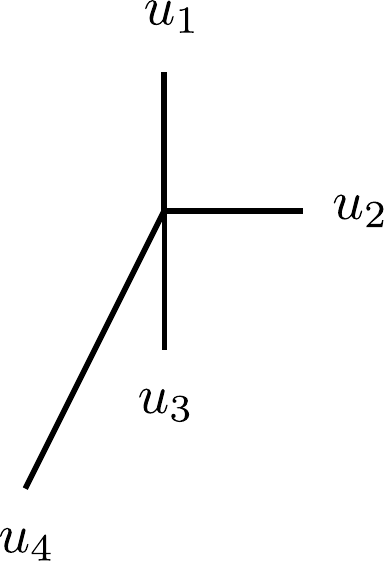}
\caption{The fan of the Hirzebruch surface $\mbb{F}_2$.}
\label{fig:f2_fan}
\end{figure}

The simplest heterotic base in which the assumption ${-K_{\hb}\cdot C\neq0}$ fails for some horizontal NS5-brane $C$ is the Hirzebruch surface, $\hb=\mbb{F}_2$, when there is a non-zero number of NS5-branes wrapping the unique holomorphic curve of self-intersection $-2$. One finds in this case that the number of chiral multiplets predicted on the heterotic side is always larger by two than the predicted number on the F-theory side. We emphasise this difference is not proportional to the number of NS5-branes wrapping the $(-2)$-curve, rather it is a constant discrepancy. For example, in the notation of Figure~\ref{fig:f2_fan}, when 2 NS5-branes are wrapped on each of $\bc_2=0$, $\bc_3=0$, $\bc_4=0$, with 22 branes wrapping $\bc_1=0$, we find the following apparent mismatch of chiral multiplets,
\begin{align}
h^{1,1}(\hs) + h^{2,1}(\hs) + 1 + \nf_5 + \sum_{M_i} \fdm(M_i) &\stackrel{?}{=} h^{3,1}(\bfs) + h^{1,1}(\bfb) + \left( h^{2,1}(\bfs) - h^{2,1}(\bfb) \right) \nonumber \\
347 = 3+243+1+28+72 &\neq 313+31+(1-0) = 345 \,. \label{eq:ch_mis_exmpl}
\end{align}

We can identify which heterotic chiral multiplets correspond to these two `extra' multiplets. We now argue that specifically these are (i) one of the brane deformation moduli of the $(-2)$-branes in the fiber direction, and (ii) one of the complex structure deformations of the heterotic threefold $\hs$. After establishing that these are the moduli involved in the apparent mismatch, we will discuss possible explanations.

We first discuss the brane deformation modulus. Each vertical deformation modulus is expected to be dual to a Ramond-Ramond modulus, and in Equation~\eqref{eq:ramram_mod_count} above we recalled the F-theory count of the latter. However, when the heterotic base is $\mbb{F}_2$ with a non-zero number of NS5-branes wrapping the $(-2)$-curve, the F-theory space is found to be missing an expected Ramond-Ramond modulus, that is
\be
h^{2,1}(\bfs)-h^{2,1}(\bfb) = \left(\textrm{No. of branes on the $(-2)$-curve}\right) -1 \,.
\ee
We saw this already in the example just given. Hence one of the moduli involved in the apparent mismatch is a modulus for the movement in the fiber direction of a horizontal NS5-brane wrapping a $(-2)$-curve.

Next we discuss the complex structure deformation. First we note that all even Hirzebruch surfaces differ only by complex structure \cite{Hirzebruch1951/52}. In particular, a Hirzebruch surface $\mbb{F}_n$ is a fiber product of $\mbb{P}^1$ over $\mbb{P}^1$, for which there are only two topologies. The $\mbb{F}_n$ then fall into two classes: for even $n$ the fiber product is trivial, while for odd $n$ it is non-trivial. The remaining distinction is in complex structure: all $\mbb{F}_{2n}$ are related by complex structure deformation, and similarly for $\mbb{F}_{2n+1}$. The deformations change the set of effective curves.

In fact, $\mbb{F}_2$ exists on a complex codimension one locus in the complex structure moduli space of ${\mbb{F}_0 = \mbb{P}^1 \times \mbb{P}^1}$. This can be seen in a description of the deformation by an embedding. Consider the family of hypersurfaces $S_t,$ in the ambient space $\mc{A} = \mbb{P}^1 \times \mbb{P}^2$,
\be
S_t: ~~ \{0 = x_0^2y_1-x_1^2y_0+tx_0x_1y_2\} \subset \mc{A} \,, \quad S_t \in \left[ \begin{array}{c|c} \mbb{P}^1 & 2 \\ \mbb{P}^2 & 1\end{array} \right] \begin{array}{l} \left[x_0,x_1\right] \\ \left[y_0,y_1,y_2\right] \end{array} \,,
\ee
where $t \in \mbb{C}$. It is not difficult to show \cite{barth2015compact} that this describes a $\mbb{P}^1 \times \mbb{P}^1$ for $t\neq0$, while for $t=0$ it describes an $\mbb{F}_2$, and the family of hypersurfaces describes the complex structure deformation. The $(-2)$-curve in the $\mbb{F}_2$ is given by $\{y_0=y_1=0\} \subset S_0$.

The heterotic threefold $\hs$ has an $\mbb{F}_2$ base, so one available complex structure deformation of $\hs$ leads to an elliptic fibration over a different base, $\mbb{F}_0$. This is seen as a non-polynomial deformation in the toric description. This deformation is also present in the F-theory fourfold $\bfs$, which is K3 fibered over $\mbb{F}_2$. Or rather, this is so until the base includes blow-ups over the $(-2)$-curve, corresponding to the presence of NS5-branes wrapping the $(-2)$-curve. In fact, these blow-ups remove this non-polynomial deformation from $h^{3,1}(\bfs)$,
\be
\textrm{NS5-branes on $(-2)$-curve} \quad \Rightarrow \quad \textrm{F-theory deformation $\mbb{F}_2 \to \mbb{F}_0$ lost from $h^{3,1}(\bfs)$} \,,
\ee
as can be seen from a count of the non-polynomial deformations. Hence one of the heterotic moduli involved in the chiral multiplet mismatch is the analogous non-polynomial complex structure deformation of the heterotic space $\hs$, taking the base $\hb$ from $\mbb{F}_2$ to $\mbb{F}_0$. (Since on both sides this deformation concerns the space $\hb$, we expect these deformation moduli to be simply matched.)

\bigskip

We have now identified which moduli are involved in the apparent chiral multiplet mismatch. The next question is the reason for the discrepancy. We will now argue that in fact the complex structure deformation of $\hs$ that takes $\hb$ from $\mbb{F}_2$ to $\mbb{F}_0$ is massive, which will reduce the discrepancy from two chiral multiplets to one. This deformation takes the heterotic base from an $\mbb{F}_2$ to an $\mbb{F}_0$. In this deformation, the $(-2)$-curve ceases to be holomorphic. This is clear since in $\mbb{P}^1 \times \mbb{P}^1$, the Mori cone is spanned by the hyperplane classes $H_1$ and $H_2$, and a $(-2)$-curve must be in a class $\pm\left(H_1-H_2\right)$, which is not effective. If an NS5-brane wraps the $(-2)$-curve, then in the deformation its embedding cannot remain holomorphic, breaking supersymmetry. Hence we can conclude that the complex structure deformation taking $\mbb{F}_2$ to $\mbb{P}^1 \times \mbb{P}^1$ is obstructed if a brane wraps the $(-2)$-curve, as in the situation under discussion.

The other chiral multiplet corresponds to the deformation in the fiber direction of a horizontal NS5-brane wrapping a $(-2)$-curve in $\hb$, or rather some combination of all such deformations. Unlike in the case of the complex structure deformation, we have not been able to identify a source for a mass of this multiplet. This appears to be a collective effect, rather than an effect for each brane, despite these branes being generically separated in the bulk. While this is reminiscent of the decoupling of a centre-of-mass mode, we do not believe it can be explained this way. In particular, we cannot mod out by isometries along the constant fiber over the $(-2)$-curve in $\hb$, as these cannot lift to isometries of $\hs$. Hence we expect the physical spectrum to depend on the `distance' of the NS5-branes from the zero section and to contain states which are sensitive to the positions of the NS5-branes.

Another possible explanation for the discrepancy is that the corresponding modulus in F-theory becomes massless in the stable degeneration limit. As discussed in Section~\ref{sec:fthmod_fibdual}, only in this limit do we have control over the various possible corrections to the simple supergravity theories that we have been primarily working with on both sides of the duality. If an F-theory modulus does become massless in this limit, then the discrepancy is resolved in the region of moduli space where our description is valid. Further, we may expect that the corresponding heterotic modulus is lifted by one of these corrections away from the stable degeneration limit.

\begin{figure}[t]
\centering
\includegraphics[scale=2.1]{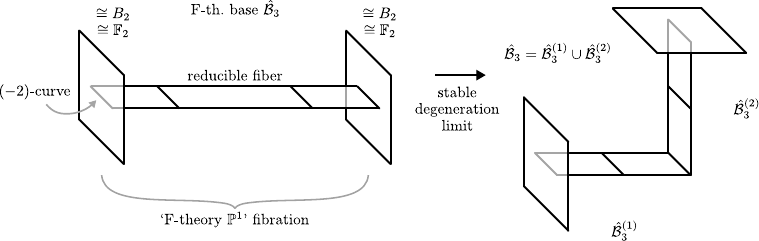}
\caption{Base $\bfb$ of the F-theory fourfold $\bfs$ in the stable degeneration limit, in the case where blow-ups have been performed over a $(-2)$-curve in the $\mathrm{E}_8$ surfaces, which corresponds to NS5-branes wrapping the $(-2)$-curve in the heterotic base $\hb=\mbb{F}_2$. In the $\mbb{P}^1$ fibration, only the reducible fibers over the $(-2)$-curve have been shown, for simplicity.}
\label{fig:stab_deg_extra_h21}
\end{figure}

We will now give a plausibility argument that indeed an F-theory modulus appears to become massless in the stable degeneration limit. First we consider the situation before we have taken the limit, and where on the heterotic side $N\geq2$ NS5-branes wrap the $(-2)$-curve in the base $\hb=\mbb{F}_2$. In the F-theory dual, $N$ blow-ups have been performed in the F-theory base $\bfb$, over either of the $(-2)$-curves sitting in the two $\mathrm{E}_8$ surfaces. This gives a reducible fiber of the $\mbb{P}^1$ fibration, with $N+1$ $\mbb{P}^1$ components, over each point on these $(-2)$-curves. These form $N+1$ exceptional divisors. This situation is depicted on the left of Figure~\ref{fig:stab_deg_extra_h21}. While the divisors formed by the two `end' components of the fiber (times the $(-2)$-curve) touch the $\mathrm{E}_8$ surfaces and hence intersect the anti-canonical divisor $-K_{\bfb}$, one can check that the other $N-1$ exceptional divisors do not. Hence the elliptic fibration of $\bfs$ is trivial over these divisors in the base $\bfb$. By taking the pullback to $\bfs$ of the (1,1)-form corresponding to one of these exceptional divisors and wedging with a (1,0)- or (0,1)-form in the elliptic curve over the divisor, one obtains contributions to $h^{2,1}(\bfs)$ and $h^{1,2}(\bfs)$. These are the Ramond-Ramond moduli contributions. We can immediately see why this contribution to $h^{2,1}(\bfs)$ does not match the number required to match the moduli of the NS5-branes, since we have introduced $N$ NS5-branes but only $N-1$ Ramond-Ramond moduli.

In the stable degeneration limit, the F-theory base $\bfb$ splits into two components, $\bfb = \bfb^{(1)} \cup \bfb^{(2)}$. The cohomology of the degenerated space can be computed from those of the two components by using the Clemens-Schmid exact sequence. (See Ref.~\cite{morrison1984clemens} for more information on this exact sequence, and for example Ref.~\cite{Aspinwall:1998bw} for a description of its use in the stable degeneration limit of heterotic/F-theory duality.) A full computation is beyond our present scope; instead we confine ourselves to a plausibility argument for the existence of a single extra contribution to $h^{2,1}(\bfs)$ in this limit. As the base has split into two components, exactly one of the $N$ exceptional divisors over the $(-2)$-curve in the base $\bfb$ is separated into two pieces in the degeneration. This situation is depicted on the right of Figure~\ref{fig:stab_deg_extra_h21}. Hence in the sum of cohomologies of $\bfb^{(1)}$ or $\bfb^{(2)}$, there exists an extra $(2,1)$-form constructed as above from the trivial elliptic fibration over the exceptional divisors being discussed. This brings the total number of  such contributions to $h^{2,1}(\bfb)$ up to $N$, matching those required for the corresponding NS5-brane moduli. We note that in the Clemens-Schmid exact sequence it is possible that this differential form will not contribute to $h^{2,1}(\bfb)$ because of cohomological equivalences, and one would have to perform the full computation to check this.

\bigskip

This completes our discussion of the apparent chiral multiplet mismatch in Equation~\eqref{eq:ch_mis_exmpl} for the example of a heterotic base $\hb=\mbb{F}_2$. We argued that one of the extra heterotic multiplets is actually massive. For the other, we were not able to find a source for the mass, however we have argued that it is plausible that an extra F-theory chiral multiplet becomes massless in the stable degeneration limit. We expect that all cases involving horizontal NS5-branes wrapping curves $C$ with $-K_{\hb} \cdot C=0$ can be treated in this way, not only the Hirzebruch $\mbb{F}_2$ case used as an example here. We also emphasise that these cases are in some sense a higher order effect, with a large set of examples already within the remit of the proof in Section~\ref{sec:dualchecks_ns5s_and_blowups}.


\section{Coincident and intersecting NS5-/M5-branes}
\label{sec:coin_and_int_5branes}

In Section~\ref{sec:glob_4d_models} above, we have constructed global F-theory models dual to heterotic line bundle models. In this context we also discussed various configurations of horizontal NS5-branes. One can consider transitions between these configurations, passing through situations with intersecting branes. It is interesting to translate the various stages of these transitions into their dual descriptions as global F-theory models. This seems particularly tractable in the present context of heterotic line bundle models, since the F-theory geometries before transition are relatively simple. We also note that the effective four-dimensional theory of coincident or intersecting NS5-branes is not so well-understood. The F-theory duals provide another perspective on these theories and their various branches, so it may prove useful to present the details of these dual models.

We will first review what is known about the local geometry dual to coincident or intersecting horizontal NS5-branes, before considering their embedding into the global models we have discussed, where we will detail the correspondence between aspects of the horizontal NS5-brane configurations and the toric descriptions of the F-theory base. Finally we will make some comments on the effective field theories through the transitions. The theory on further compactification to three dimensions is most tractable, and we review the recent literature on such M-theory compactifications, connecting this to the brane transition picture.

Before turning to this discussion, we make three preliminary remarks. First, we assume that all horizontal NS5-branes are embedded by the unique section in the elliptic fibration of $\hs$, as discussed in Section~\ref{sec:req_branes_and_blowups}. Second, we will not discuss the blow-ups of the remaining $\mathrm{E}_8$ singularities, since these are away from the blow-ups dual to the horizontal NS5-branes, and hence do not play a role here. Third, we will simply say `NS5-brane' when we mean horizontal NS5-brane, as we will not discuss vertical NS5-branes\footnote{We recall the dual of a vertical NS5-brane is a D3-brane, so that the dual of a set of coincident vertical NS5-branes is a stack of D3-branes. One can also consider the intersection of a vertical and a horizontal NS5-brane, which is an interesting question, discussed in e.g.~Ref.~\cite{Diaconescu:1999it}, but one that we do not attempt to discuss here. We hope to discuss this question in future work.}.

\subsection{Branches in NS5-brane transitions}
\label{sec:branches_in_5brane_trans}

We first briefly discuss the situation in heterotic string theory, in particular possible sets of NS5-brane configurations connected by a transition. We consider a set of M5-branes at the same position in the 11d bulk, so that they sit in the same CY threefold. This set of branes can have a complicated configuration, perhaps with point-like or curve-like intersections. We can imagine the collection of nearby possible configurations, some of which eliminate intersections. We will consider two particular situations for simplicity: (i) a configuration with a set of point-like intersections, and (ii) a configuration with a stack of completely coincident branes. In principle both can occur in the same configuration.

\begin{figure}[t]
\centering
\includegraphics[scale=1.9]{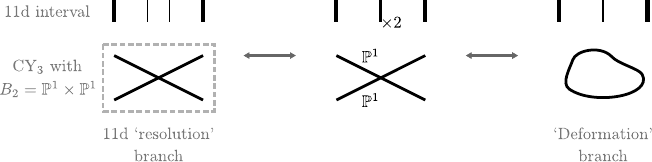}
\caption{Schematic depiction of transversely-intersecting M5-branes and the `branches' that remove the intersection, as discussed in the main text. In this example, the heterotic base is $\hb = \mbb{P}^1 \times \mbb{P}^1$, and two $\mbb{P}^1$ M5-branes sit at for example $\bc_1=0$ and $\bct_1=0$ in $\hb$.}
\label{fig:inter_5branes_branches_on_p1xp1}
\end{figure}

We first consider the collection of configurations around a transverse-intersection case. There exists an 11d `resolution' branch, in which the M5-branes are separated in the 11d bulk, and often also a `deformation' branch in which the equations describing the branes in the heterotic base $\hb$ are deformed to remove the intersection. These situations are shown schematically in an illustrative example in Figure~\ref{fig:inter_5branes_branches_on_p1xp1}. In this example the heterotic base is $\hb = \mbb{P}^1 \times \mbb{P}^1$, and we have considered bringing together just two of the M5-branes, each of which wraps a $\mbb{P}^1$ in the $\mbb{P}^1 \times \mbb{P}^1$. We note in passing that these branes cannot be made to coincide, so this does not form a possible configuration in a deformation of coincident branes.

\begin{figure}[t]
\centering
\includegraphics[scale=1.9]{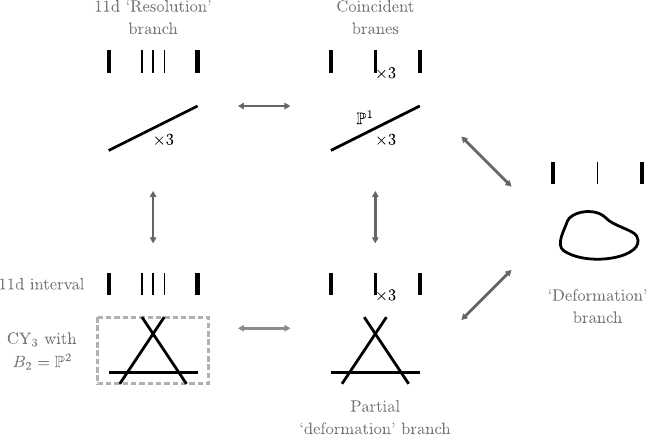}
\caption{Schematic depiction of coincident M5-branes and the `branches' that remove the coincidence, as discussed in the main text. In this example, the heterotic base is $\hb = \mbb{P}^2_u$. In the coincident situation, three $\mbb{P}^1$ M5-branes sit all at for example $\bc_1=0$ in $\hb$, while in the transverse-intersection case they are at for example $\bc_1=0$, $\bc_2=0$, and $\bc_3=0$ in $\hb$.}
\label{fig:coin_5branes_branches_on_p2}
\end{figure}

We next consider the collection of configurations related to a stack of completely coincident branes. Again there is an 11d resolution branch, as well as a deformation branch. These possibilities are shown schematically in an illustrative example in Figure~\ref{fig:coin_5branes_branches_on_p2}. In this example the heterotic base is $\hb = \mbb{P}^2$, and we have considered bringing together just three of the M5-branes, each of which wraps a $\mbb{P}^1$ in the $\mbb{P}^2$. In this example, shifting the branes in the CY manifold, giving a `partial' deformation, leaves transverse-intersections, and there also exists a full deformation branch where there is a single connected brane.

\begin{figure}[t]
\centering
\includegraphics[scale=1.9]{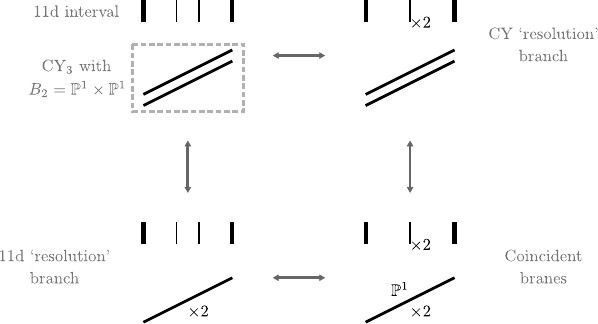}
\caption{Schematic depiction of coincident M5-branes and the `branches' that remove the coincidence, as discussed in the main text. In this example, the heterotic base is $\hb = \mbb{P}^1_u \times \mbb{P}^1_v$. In the coincident situation, two $\mbb{P}^1$ M5-branes sit at for example $\bc_1=0$ in $\hb$. Deformation to a single smooth brane is impossible, with general deformations giving separated branes.}
\label{fig:coin_5branes_branches_on_p1xp1}
\end{figure}

There may be many interesting possibilities differing in some way from these examples. One such possibility is shown in Figure~\ref{fig:coin_5branes_branches_on_p1xp1}. Here shifting the branes in the CY manifold removes all intersections, giving a set of disconnected branes. However there exists no deformation to a connected smooth brane. We do not claim to be exhaustive by the possibilities we have discussed. We also note that, as seen in Figures \ref{fig:inter_5branes_branches_on_p1xp1} and \ref{fig:coin_5branes_branches_on_p1xp1}, not all coincident cases and not all transverse-intersection cases occur in the collection of configurations connected to the other case. We see that when the two do occur in the same collection, as in Figure~\ref{fig:coin_5branes_branches_on_p2}, it is possible to move between the 11d resolution and the deformation branch by passing through either one of the coincident or transverse-intersection configurations.

\bigskip

In addition to M5-branes that can be deformed only within the base $\hb$, there exist a special set of branes wrapping curves $C$ in $\hb$ such that $-K_{\hb} \cdot C = 0$, which can also move in the heterotic fiber direction as discussed in Section~\ref{sec:subtle_curves}. These fiber positions correspond on the F-theory side to Ramond-Ramond moduli (in the IIB picture), rather than to geometry as is the case for base positions. We will not treat this part of moduli space in this section, instead restricting ourselves to deformations of M5-branes in the base $\hb$, and the corresponding F-theory geometry.

We note that upon restricting to base deformations alone, some intersecting or coincident brane situations involving curves $C$ with $-K_{\hb} \cdot C = 0$ cannot be deformed to remove the intersection/coincidence. Recall that for supersymmetric branes the embedding is holomorphic, so only algebraic deformations preserve supersymmetry. In the Hirzebruch space $\mbb{F}_2$ with weight system
\be
\mbb{F}_2: ~~
\begin{tabular}{c c c c}
$x$ & $y$ & $z$ & $w$ \\
\hline
1 & 1 & 0 & 0 \\
2 & 0 & 1 & 1 
\end{tabular} \,,
\label{eq:f2_weights}
\ee
the transverse-intersection given by $yw=0$ cannot be deformed away: the general curve in this divisor class is $0=y(az+bw)$ which preserves the intersection. Similarly the coincident brane situation given by $y^2=0$ has no deformations.

\floatsetup[figure]{style=plain,subcapbesideposition=center}
\begin{figure}[t]
\begin{tabular}{ll}
\sidesubfloat[]{\includegraphics[scale=2]{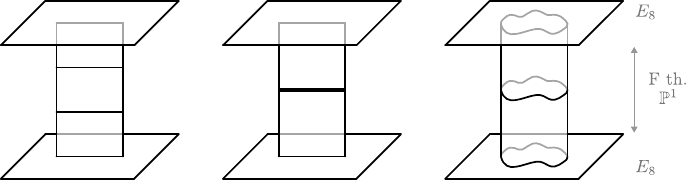}\label{fig:localgeom_dualto_coin_and_int_sub1}}\vspace{15pt}\\
\sidesubfloat[]{\includegraphics[scale=2]{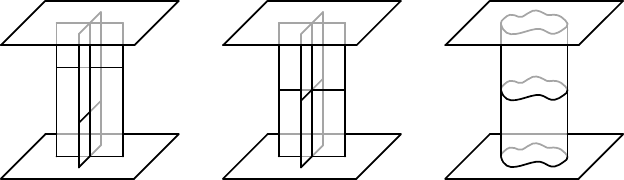}\label{fig:localgeom_dualto_coin_and_int_sub2}}
\end{tabular}
\caption{A schematic depiction of the geometry in the base $\bfb$ of the F-theory fourfold $\bfs$, when in the dual geometry we pass through a situation with (a) coincident horizontal M5-branes and (b) transversely-intersecting horizontal M5-branes. In each case on the left is the resolved side, in the middle the singular situation, and on the right the deformed side. We have shown only those fibers that have multiple components; the other fibers remain $\mbb{P}^1$s.}
\label{fig:localgeom_dualto_coin_and_int}
\end{figure}

\subsection{Review of local dual F-theory geometry}
\label{sec:loc_dual_fth_geom}

We now review the known local F-theory geometries dual to coincident or transversely-intersecting NS5-branes. We first review the case of coincident NS5-branes. These are M5-branes in the Ho\v{r}ava-Witten picture wrapping the same curve in the base $\hb$ of the CY manifold $\hs$, and which are also at the same position in the 11d bulk. We consider the case of two coincident M5-branes, as the story with $N$ branes is a straightforward generalisation. It is known \cite{Diaconescu:1999it,Rajesh:1998ik,Morrison:1996pp,Morrison:1996na} that in the F-theory dual of this situation, a singular locus appears in the fourfold $\bfs$: this locus consists of a complex surface in the fourfold $\bfs$, which arises due to the appearance of a complex curve of singularities in the F-theory base $\bfb$.

The appearance of this curve of singularities in $\bfb$ can be understood by bringing together branes with distinct positions in the 11d bulk. We recall that the blow-ups of the F-theory base correspond to the introduction of M5-branes as we pull instantons off the $\mathrm{E}_8$ branes in the Ho\v{r}ava-Witten picture. Additionally, {\kahl} moduli associated to the blow-ups correspond to the positions in the 11d bulk of the M5-branes. As we bring two M5-branes together in the bulk, in the F-theory geometry we move to an edge of the {\kahl} cone\footnote{Actually, in order to blow-down the divisor we will have to first flop the fourfold, passing into another {\kahl} cone on whose boundary the desired divisor can be blown-down. This is analogous to the discussion at the end of Section~\ref{sec:tor_desc_in_6d}.}, and a divisor collapses to zero volume. The curve of singularities in the base is the image of the base divisor under the blow-down, and corresponds to the coincident M5-branes. The situation before the blow-down is shown schematically in the middle of Figure~\ref{fig:localgeom_dualto_coin_and_int_sub1}. Here all the exceptional divisors have finite size, and as the M5-branes are brought together the {\kahl} moduli are tuned so that one divisor shrinks to zero size. As a result a complex curve of $A_1$ singularities appears in the base, which corresponds to a complex surface of $A_1$ singularities in the fourfold. In the case of $N$ coincident M5-branes, the singularity is $A_{N-1}$.

Often there will exist deformations of these singularities that are dual to deformations of the stack of NS5-branes in the heterotic base $\hb$ into a single connected NS5-brane. This branch is depicted schematically in the right of Figure~\ref{fig:localgeom_dualto_coin_and_int_sub1}. The resulting structure of reducible fibers in this branch reflects the fact that we now have on the heterotic side a single brane, usually of higher genus, with some position in the 11d interval. As noted above in Section~\ref{sec:branches_in_5brane_trans}, in the M5-brane configuration it can happen that (i) base deformation results in multiple branes, or (ii) base deformation is not possible. The former case occurs in Figure~\ref{fig:coin_5branes_branches_on_p1xp1}; the dual F-theory deformation branch will simply shift the two exceptional divisors rather than combine them. In the latter case, there must not exist deformations of the singularity.

\bigskip

Next we review the F-theory dual of transversely-intersecting horizontal NS5-branes. These are M5-branes at the same position in the 11d bulk and wrapping transversely-intersecting curves in the base $\hb$ of the CY manifold $\hs$. We review the case of two transversely-intersecting M5-branes, with a single intersection point, as the story with $N$ branes and/or multiple intersection points is a straightforward generalisation. As also discussed in Ref.~\cite{Diaconescu:1999it}, the F-theory dual of this situation contains conifold singularities in the base $\bfb$, giving rise to complex curves of singularities in the fourfold $\bfs$. 

We can imagine reaching this situation by bringing together branes with distinct positions in the 11d bulk. This discussion is very analogous to that for coincident M5-branes above. However in this case, as the F-theory {\kahl} moduli are tuned to the edge of the {\kahl} cone, a curve in the base collapses to zero volume, rather than a divisor. This situation is shown schematically in the left and middle of Figure~\ref{fig:localgeom_dualto_coin_and_int_sub2}. The result is a point-like $A_1$ singularity in the base, giving a curve of $A_1$ singularities in the fourfold. In the case of $N$ coincident M5-branes, the singularity is $A_{N-1}$. If there are multiple transverse-intersections then there are multiple conifold singularities.

As in the case of coincident branes, in the situation of transversely-intersecting branes there will often exist deformations of the singularity in the F-theory base $\bfb$ that are dual to deformations of the NS5-branes in the heterotic base $\hb$ that give rise to a single connected NS5-brane. This branch is depicted schematically in the right of Figure~\ref{fig:localgeom_dualto_coin_and_int_sub2}. Again the resulting structure of reducible fibers reflects the fact that we now have on the heterotic side a single brane. As noted above in Section~\ref{sec:branches_in_5brane_trans}, sometimes it is not possible to deform the branes in the base $\hb$ at all. In this case, the corresponding F-theory singularity must be non-deformable.

\subsection{Global dual F-theory geometry}
\label{sec:global_dual_fth_geom}

We now discuss how the local singular geometry dual to intersecting or coincident NS5-branes fits into the global F-theory models that we have constructed. The result will be a toric description of the F-theory fourfolds dual to heterotic line bundle models with intersecting or coincident NS5-branes and the surrounding brane configurations. Note that we will mainly discuss the F-theory base rather than the fourfold itself, but the F-theory fourfold is constructed by taking an elliptic fibration, and hence the fan of the ambient space for the fourfold is implicitly specified at each stage\footnote{Note when we discuss blowing down, this requires flops of the fourfold. After flopping but before blowing down the elliptic fibration is lost. (See the discussion in Section~\ref{sec:tor_desc_in_6d}.)}.

We first discuss the simplest global models, where the F-theory base is described by a toric fan. Here the fourfold is a toric hypersurface in a 5d toric ambient space, as opposed to a higher codimension complete intersection. Examples of these situations were shown in Figures \ref{fig:fullyblownup_P1P1P1_fan} and \ref{fig:fullyblownup_examples_4d}. In the corresponding heterotic situation, there are a set of NS5-branes multiply wrapping $\mbb{P}^1$s, and each brane has a distinct position in the bulk. We first discuss bringing two or more of these branes together by tuning their bulk positions. This may give rise to point-like or curve-like intersections between the branes. It will be simplest to first discuss the resolution branch and singular situation together, and then turn to the deformation branch.

\bigskip

As reviewed in Section~\ref{sec:loc_dual_fth_geom} above, transversely-intersecting NS5-branes are dual to point-like singularities in the F-theory base, while coincident NS5-branes are dual to curve-like singularities in the F-theory base. If the threefold base is described away from the singular situation by a three-dimensional fan, then point-like singularities can arise when a 1d face is deleted in a 2d face, while curve-like singularities can arise when we blow down a ray. The resulting fan will describe the global singular F-theory geometry.

\begin{figure}[t]
\centering
\includegraphics[scale=.4]{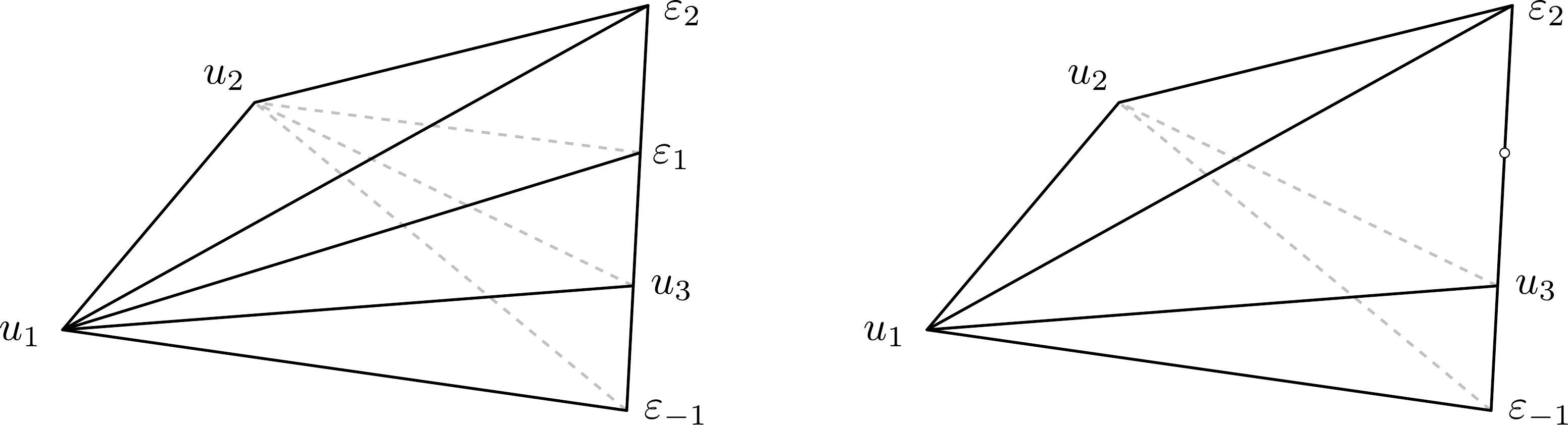}
\caption{Part of the triangulation of the ray system on the right of Figure~\ref{fig:fullyblownup_examples_4d} for the F-theory base $\bfb$ when the heterotic base $\hb$ is a $\mbb{P}^2$ with a particular NS5-brane configuration, in the situation before and after a removal of one of the blow-up rays. The result corresponds to a situation with two coincident NS5-branes, and gives rise to a curve of singularities.}
\label{fig:p2_ray_removal}
\end{figure}

We consider as a first example the F-theory base shown on the right of Figure~\ref{fig:fullyblownup_examples_4d}. In this example, the heterotic base $\hb$ is a $\mbb{P}^2$, with all NS5-branes wrapping $\mbb{P}^1$s at $\bc_3=0$. Correspondingly all blow-ups of $\bfb$ are at $\bc_3=0$. The volumes of these exceptional divisors correspond to distances between M5-branes in the bulk. Hence we can think of the gaps between the blow-up rays $\ep_i$ as corresponding to the NS5-branes, and removing one of the blow-up rays $\ep_i$ joins two gaps reflecting the fact that two M5-branes have been brought together. This creates a curve of singularities at the intersection of the divisors $\ep_{i-1}$ and $\ep_{i+1}$, where $\ep_0 \equiv \bc_3$. An illustration of the triangulations before and after removal of a ray is shown in Figure~\ref{fig:p2_ray_removal}. In general one can remove $N$ internal blow-up rays to bring together $N+1$ NS5-branes, giving rise to an $A_N$ singularity. We note that blowing down the rays at the top and bottom of the tower does not correspond to bringing NS5-branes together; rather this corresponds to putting NS5-branes back onto one of the $\mathrm{E}_8$ branes at the ends of the 11d interval.

\begin{figure}[t]
\centering
{\includegraphics[scale=.3]{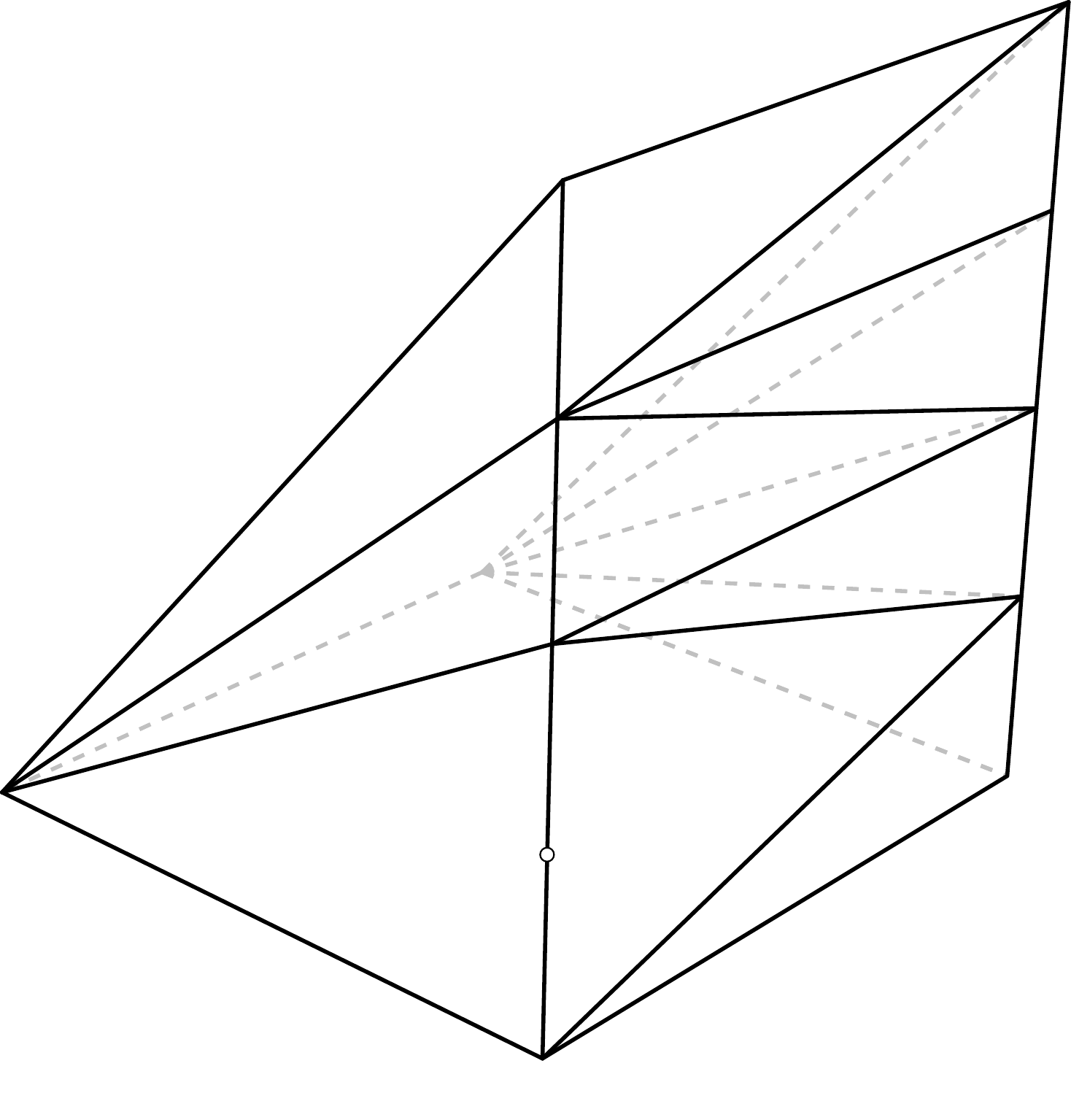}}\hspace{25pt}
{\includegraphics[scale=.3]{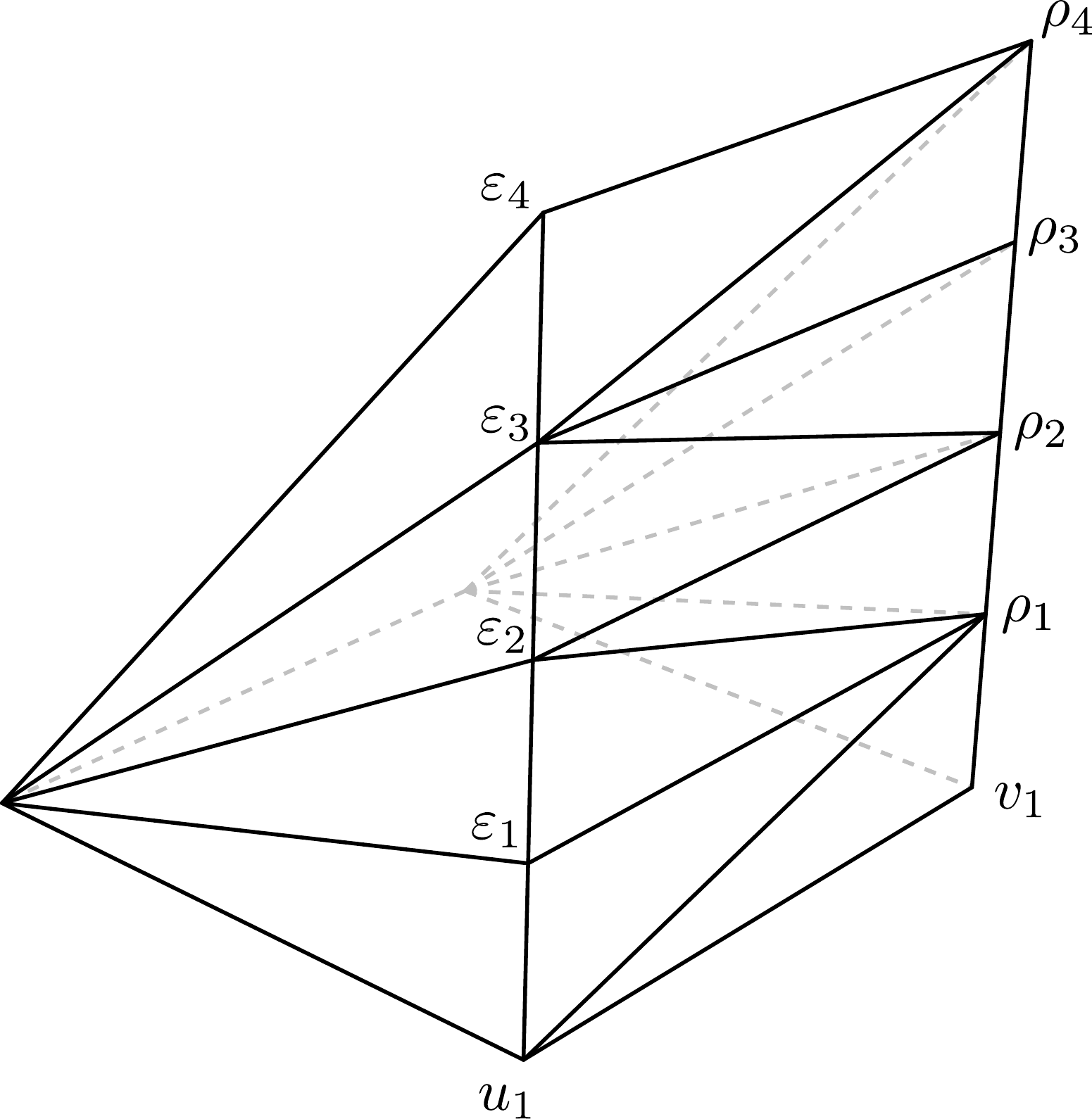}}\hspace{25pt}
{\includegraphics[scale=.3]{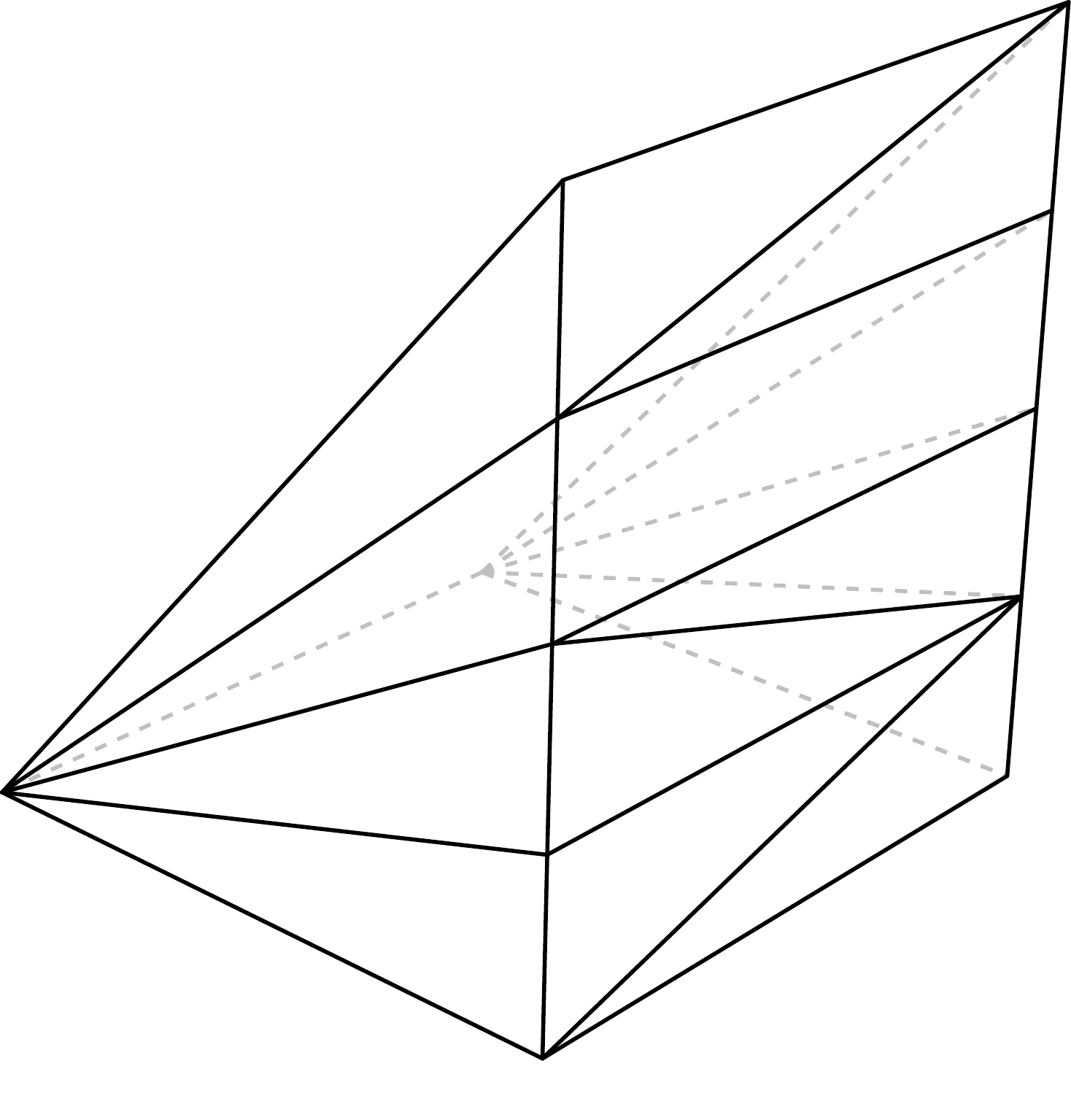}}
\caption{Part of the triangulation of the ray system in Figure~\ref{fig:fullyblownup_P1P1P1_fan} for the F-theory base $\bfb$ when the heterotic base $\hb$ is a $\mbb{P}^1 \times \mbb{P}^1$ with a particular NS5-brane configuration, in (centre) the non-singular situation, (left) after a removal of one of the blow-up rays to give coincident NS5-branes, and (right) after removing a 1d face to give transversely-intersecting NS5-branes.}
\label{fig:p1xp1_3d_triangs}
\end{figure}

As a second example, consider the F-theory base shown in Figure~\ref{fig:fullyblownup_P1P1P1_fan}. In this example the heterotic base $\hb$ is a $\mbb{P}^1 \times \mbb{P}^1$, and the NS5-branes sit in two stacks at $\bc_1=0$ and $\bct_1=0$. The gaps between the blow-up rays over $\bc_1$ and $\bct_1$ correspond to the NS5-branes at $\bc_1=0$ and $\bct_1=0$ respectively. Removing a ray corresponds to bringing branes into coincidence, creating a curve of singularities in $\bfb$. However there is also the possibility of transversely-intersecting branes, in contrast to the previous example. Thinking of the gaps between the blow-up rays as the M5-branes, we can bring a $\bc_1$ and a $\bct_1$ brane together in the bulk by removing a 1d face that separates a gap in one blow-up tower with a gap in the other. This creates a point-like singularity as expected from the dual picture. Figure~\ref{fig:p1xp1_3d_triangs} illustrates the triangulations after a ray has been removed and after a 1d face has been removed. 

\begin{figure}[t]
\centering
\includegraphics[scale=1.3]{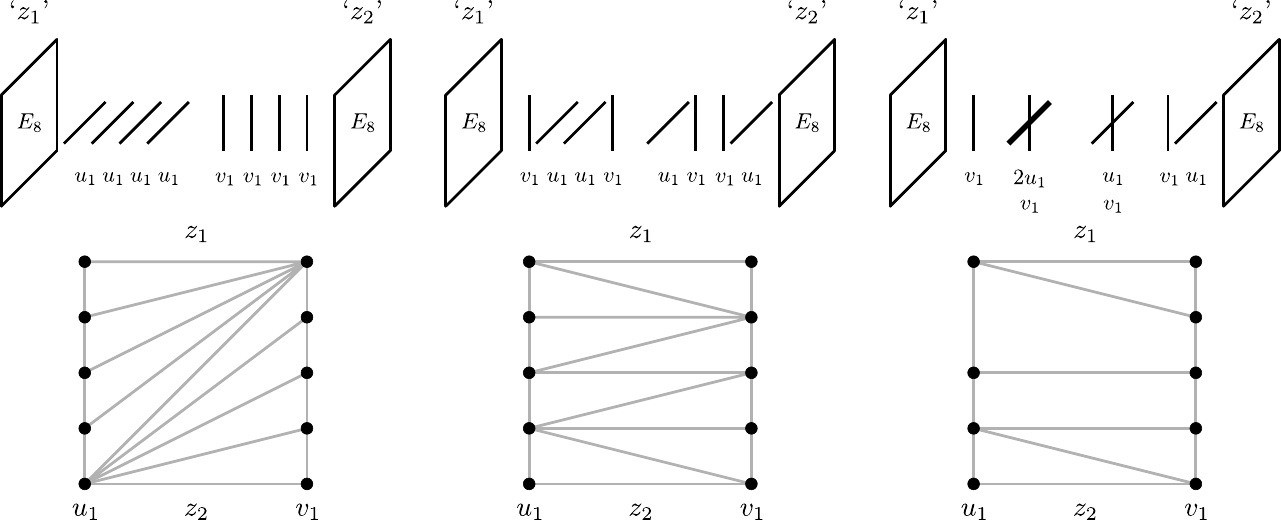}
\caption{Examples of the correspondence between the positions of the horizontal M5-branes in the bulk and triangulations of the `$\bc_1-\bct_1$' face of the ray system in Figure~\ref{fig:fullyblownup_P1P1P1_fan} for the F-theory base $\bfb$ when $\hb = \mbb{P}^1 \times \mbb{P}^1$. In the example on the right, there are intersecting and coincident branes. Here we have included only 8 blow-ups rather than the full 24 for simplicity of the diagrams.}
\label{fig:bulk_and_triangulations}
\end{figure}

Choosing a different triangulation of the $N$-lattice polytope $\Delta^*$, which translates to a different choice of fan for the base employing the same rays, corresponds to a different ordering of the M5-branes in the bulk. It is clear that only particular gaps in the two towers of blow-up rays can be connected to one another by removing a single 1d face. It is also clear that only particular rays can be removed while leaving a triangulation of the remaining polytope: for example, in the initial triangulation in Figure~\ref{fig:p1xp1_3d_triangs}, we were able to blow down $\ep_1$, but we could not have blown-down for example $\rho_1$, as the result would not be a triangulation. Hence it is clear that the triangulation reflects the ordering of the M5-branes in the 11d interval, since it reflects which branes can be brought into contact with one another. Figure~\ref{fig:bulk_and_triangulations} shows some examples of the correspondence between orderings of M5-branes in the bulk and triangulations of the F-theory base space, for the example of a heterotic base $\mbb{P}^1 \times \mbb{P}^1$ with a particular blow-up pattern. For clarity we only show a subset of the branes, or equivalently only part of the triangulation. For any toric hypersurface example, there is a more complicated but analogous relation between triangulation and ordering of branes in the bulk.

\bigskip

We have so far discussed singularities in the F-theory fourfold from the perspective of the resolution branch, as we collapse divisors and curves in $\bfb$. We now discuss the deformation branch, in particular its toric description, which then will complete the picture of the global F-theory geometry dual to intersecting or coincident NS5-brane situations in heterotic line bundle models. We note the fourfolds we have so far considered are toric hypersurfaces, which give the simplest global models from the perspective of toric geometry, and correspondingly are dual to particularly simple dual NS5-brane configurations. Once we have described the deformation branches of singularities in these cases, we have described the dual to complicated brane configurations anywhere in the 11d interval. One can then discuss singularities arising from bringing these branes together, and their deformation branches.

In the deformation branch, the singularity is removed by complex structure deformation, dual to deformations of the NS5-branes in the heterotic base $\hb$. We recall that if the $N$ M5-branes, which are to be combined and deformed, are `nearest' to an $\mathrm{E}_8$ brane in the bulk, i.e.\ there are no branes between them and an end of the 11d interval, the deformation branch can be described torically as in Section~\ref{sec:more_gen_glob_models}. There we blew down $n_i$ of the rays at the top of each $i^{\mathrm{th}}$ `tower' of blow-up rays, over a ray $\bc_i$ of $\hb$, before blowing up over a general curve in the curve class $\sum_in_i[\bc_i]$ on $\{\fpc_1=0\}$. On the heterotic side this corresponds to transforming the $N$ branes back into small instantons on an $\mathrm{E}_8$ brane, and then pulling out a single smooth M5-brane. This roundabout way of getting to the deformation branch misses the intermediate stage of coincident branes in the bulk, but must give the correct final description. More generally, if M5-branes have been brought together away from an $\mathrm{E}_8$ brane, i.e.\ there are branes between them and the $\mathrm{E}_8$ brane, then we could blow down also the rays corresponding to the intervening branes. We would then blow up first on the curve of the desired deformed brane, then sequentially on the curves of the branes that intervened between the brane stack and the $\mathrm{E}_8$. To put this another way, since we have seen that changing the order of branes in the bulk corresponds to flops, we could also reach this situation by building the general brane configuration near an $\mathrm{E}_8$ brane and then flopping until it is in the right position.

\begin{figure}[t]
\centering
\includegraphics[scale=1.7]{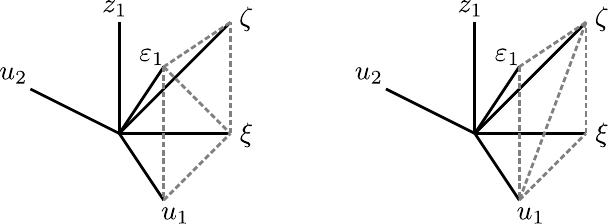}
\caption{Part of the triangulations giving the toric ambient space for the twofold F-theory base in compactification to 6d for a situation in which we have included arbitrarily placed M5-branes through the blow-up $\zeta$. In the triangulation on the left, the blow-up $\zeta$ is performed after $\ep_1$, while on the right $\zeta$ is included first and then $\ep_1$. The two situations are related by a flop, and correspond to different orderings of M5-branes in the 11d bulk.}
\label{fig:6d_deformed_triangs}
\end{figure}

This is easier to visualise in compactification to six dimensions. We recall the toric bases in Figure~\ref{fig:fullyblownup_fan}. To describe the deformation branch when $n$ of the point-like M5-branes come together in the bulk near an $\mathrm{E}_8$ brane, we can blow down the top $n$ rays over $\bc_1=0$ for example. We then extend the 2d fan of the base into 3d by adding an auxiliary ray $\vec{\xi}$, putting $\vec{\bc}_2$ at $-n$ in the new direction. We then add a blow-up ray at
\be
\vec{\zeta} =\vec{\xi}+\vec{\fpc}_1 \,.
\ee
(More precisely, at $\vec{\zeta}=\vec{\xi}+\vec{\fpc}_1+2\vec{\tacx}+3\vec{\tacy}$ in the 5d fan.) The twofold base is then a hypersurface in a 3d ambient space, whose fan is shown schematically in the left of Figure~\ref{fig:6d_deformed_triangs}. In this example we have only included one blow-up ray $\ep_1$ for simplicity. 

Clearly we may flop the triangulation of the $(\ep_1,\bc_1,\xi,\zeta)$ face of the polytope. This corresponds to altering the ordering of M5-branes in the bulk, specifically bringing the `deformed'\footnote{The `deformed' brane is a set of points, so that deformations are not very interesting in compactification to six dimensions.} brane away from an $\mathrm{E}_8$ brane, leaving an $\ep_1$ brane in between. This is clear as in each triangulation, precisely one of $\ep_1$ and $\zeta$ can be blown-down, reflecting which is nearest the $\mathrm{E}_8$ brane.\footnote{Of course, the singularity appearing during the flop should generically miss the hypersurface, intersecting it only when the hypersurface is tuned in a way that corresponds to the $\ep_1$ and $\zeta$ branes being at the same position in the heterotic base.} The situation with multiple blow-up rays $\ep_i$ is an obvious generalisation. The situation in compactification to four dimensions is clearly analogous, though more difficult to visualise. To include multiple arbitrary M5-branes will require going to higher codimension in the toric descriptions. The details of the fan associated with a choice of triangulation again reflect which of these branes can come into contact in the 11d bulk.

\bigskip

Finally we note one other global aspect of these singular transitions: the Euler characteristic is unchanged. This follows from the computation of the Euler characteristic in Section~\ref{sec:dual_checks_ind_of_ns5_config}. This computation was independent of the precise NS5-brane configuration, and hence the Euler characteristic is unchanged under a singular transition dual to a transition between NS5-brane configurations. We note this means that after such a transition, there is no need to include extra $G_4$ flux to meet the anomaly condition. All of this is clear from the heterotic side as the overall class of NS5-branes remains unchanged. 

\subsection{Comments on effective theory through transition}

In Section~\ref{sec:glob_4d_models}, we have described the field content in the various branches of a transition between NS5-brane configurations, but we have not attempted a description of the theory in the situation with coincident or transversely-intersecting branes. The coincident case is particularly interesting, and historically has been challenging to describe. We note that on the resolved side we have a gauge group U$(1)^{g\cdot N}$, and on the deformed side a gauge group U$(1)^{g_{\mathrm{def}}}$, where $g$ and $g_{\mathrm{def}}$ are the genera of the branes in the stack and the deformed brane respectively. It is difficult to see how to understand in gauge field theory terms an enhancement to SU$(N)$ at the coincident point. This is however not problematic since the intermediate theory is not expected to have a description in terms of a gauge theory. Instead, the four-dimensional theory will be a dimensional-reduction of a six-dimensional theory in which there will be tensionless strings arising from M2-branes stretching between two NS5-branes \cite{Strominger:1995ac} or an NS5-brane and an $\mathrm{E}_8$ brane \cite{Ganor:1996mu,Bershadsky:1996nu}. These strings are known as M- and E-strings, respectively.

It is beyond the scope of the present work to give a description of the four-dimensional theory obtained from compactifying F-theory on a fourfold with a curve of singularities in the base. Instead, we will confine ourselves to a study of the three-dimensional theory obtained from compactifying M-theory on the same fourfold. The advantage is that in the three-dimensional theory we have a bona fide gauge theory and avoid the subtleties coming with tensionless strings. This three-dimensional theory is then expected to lift to the four-dimensional theory. The three-dimensional resolution and deformation branches of M-theory on the smooth fourfolds on either side of the singular transition have a straightforward description. Again, at the transition point, the M-theory fourfold develops a surface of singularities, whose field theory description is more difficult. However, this has been investigated recently in Ref.~\cite{Jockers:2016bwi} and we can use their results to study the theory. The resulting correspondence may provide a useful perspective on the four-dimensional theories.

\bigskip

We first recall the field content of M-theory on a non-singular CY fourfold, and its lift to F-theory. This three-dimensional theory has $\mc{N}=2$ supersymmetry, so there is a chiral multiplet and a vector multiplet, with all multiplets equivalent on shell, containing two real scalar fields and a Majorana spinor. The numbers of three-dimensional chiral multiplets $\nch$ and vector multiplets $\nve$ are
\be
\nch = h^{2,1}(\bfs)+h^{3,1}(\bfs) \,, \quad \nve =  h^{1,1}(\bfs) \,.
\ee
When the fourfold is K3 fibered this theory is related by a `lift' to the four-dimensional theory from F-theory, in which roughly the elliptic fibers are collapsed resulting in an effective extra dimension after T-duality. In this lift the multiplets are mapped as follows \cite{Grimm:2010ks},
\begin{align}
h^{1,1}(\bfs) ~\textrm{3d vector multiplets} &\to 
\left\{
\begin{array}{l}
h^{1,1}(\bfs)-h^{1,1}(\bfb)-1 ~\textrm{4d vector multiplets} \\
h^{1,1}(\bfb) ~\textrm{4d chiral multiplets} \\
\textrm{metric of new non-compact direction} \\
\end{array}
\right. \,, \\
h^{2,1}(\bfs)+h^{3,1}(\bfs) ~\textrm{3d chiral multiplets} &\to 
\left\{
\begin{array}{l}
h^{2,1}(\bfb) ~\textrm{4d vector multiplets} \\
h^{2,1}(\bfs)-h^{2,1}(\bfb) ~\textrm{4d chiral multiplets} \\
h^{3,1}(\bfs) ~\textrm{4d chiral multiplets} \\
\end{array}
\right. \,.
\end{align}
In the smooth resolution\footnote{Note that resolution of the fourfold tends to destroy the elliptic fibration, so there is no obvious F-theory interpretation. However, this is related to an elliptic fibration by flops, which we can then perform. These leave multiplet counts unchanged: it is known that birationally equivalent smooth CY manifolds have identical Hodge numbers (see \cite{wang1998}). Hence F-theory multiplet counts can be performed before flopping.} or deformation branch of a singularity, these results give the lift of the three-dimensional matter content to four dimensions. We can note that the $h^{2,1}(\bfb)$ four-dimensional vector multiplets correspond to three-dimensional chiral multiplets. It is these that correspond to the U(1) gauge symmetries from NS5-branes in the four-dimensional duality, c.f.\ Equation~\eqref{eq:gen_h21_rel}, hence the Abelian NS5-brane gauge groups and their possible enhancement corresponds in three dimensions to the chiral multiplet sector.

\bigskip

\newcommand{\irr}{q}
\newcommand{\arge}{p_g}

We next review the three-dimensional theory resulting from compactification of M-theory on a fourfold $\bfs$ with a complex surface $\ssu$ of $A_{N-1}$ singularities, as developed in Ref.~\cite{Jockers:2016bwi}. Following their notation we will write $\irr := h^{1,0}(\ssu)$ and $\arge :=h^{2,0}(\ssu)$, which are respectively the irregularity and arithmetic genus of $\ssu$. In the singular situation, there is an SU$(N)$ symmetry and the field content includes 1 vector multiplet and $\irr+\arge$ chiral multiplets, all in the adjoint. The singularity has a resolution branch and often also a deformation branch, which correspond respectively to the Coulomb and Higgs branches of the field theory. In the Coulomb branch, the non-zero vevs of scalars in the vector multiplets measure the {\kahl} moduli of the resolution, generically breaking SU$(N)$ to U$(1)^{N-1}$. The gauge enhancement in three dimensions is straightforward, as a result of U(1) gauge symmetries being now associated to $h^{1,1}$ rather than $h^{2,1}$. In the Higgs branch the scalars of the chiral multiplets acquire vevs, measuring the complex structure deformations of the singularity, and breaking the gauge symmetry entirely.

The singularities that develop in our models are of a particular kind. They arise when we shrink some exceptional divisor(s) in the base $\bfb$, whose volumes correspond in the duality to distances between M5-branes in the Ho\v{r}ava-Witten interval. This leaves a curve of singularities diffeomorphic to the curve $\scu$ wrapped by the coincident heterotic NS5-branes. Hence the complex surface $\ssu$ of singularities in the fourfold is an elliptic surface with base curve $\scu$. Note that this means $c_1(\ssu)^2=0$, since the Poincar\'{e} dual of the first Chern class of an elliptic surface is proportional to the fiber class, and so also $12\hec=\ec$, where $\hec$ is the holomorphic Euler characteristic. The elliptic fibration of $\ssu$ is characterised by the number of singular fibers, equal to the Euler characteristic of $\ssu$. The Hodge diamond is one of two types, depending on whether the fibration is trivial (first case) or not (second case) (see e.g.~Ref.~\cite{schutt2009elliptic})
\be
\begin{tabular}{C C C C C}
   	&    		&1 		& 		& 	\\[4pt]
	&g+1	& 		&g+1	& 	\\[4pt]
g\phantom{+1} 	& 		&2g+2	&		&\phantom{1+}g	\\[4pt]
	&g+1	& 		&g+1	& 	\\[4pt]
   	&    		&1 		& 		& 	\\
\end{tabular} \,, \quad\quad
\begin{tabular}{C C C C C}
   		&    	&1 			& 	& 			\\[4pt]
		&g 	& 			&g 	& 			\\[4pt]
\arge\phantom{+1} 	& 	&10\hec+2g 	& 	&\phantom{1+}\arge		\\[4pt]
		&g 	& 			&g 	& 			\\[4pt]
   		&    	&1 			& 	& 			\\
\end{tabular} \,,
\ee
where in the second diamond $\arge = \hec+g-1=\frac{1}{12}\ec+g-1$. The number of singular fibers is counted by the intersection $-12K_{\bfb}\cdot\scu$. Note $\bfb$ has only orbifold singularities, so this intersection is well-defined. To compute this intersection, note there is by construction a projection collapsing the F-theory $\mbb{P}^1$, that maps the singular $\bfb$ to the smooth $\hb$ and under which the curve $\scu$ is mapped one-to-one to a copy in $\hb$. The intersection, hence the number of singular fibers and also the Euler characteristic of $\ssu$, must then be counted by $-12K_{\hb}\cdot\scu$ where here $\scu$ is the image in $\hb$.

One can note that $G_4$ flux may lift massless directions, altering any massless multiplet counts. The offending flux may thread cycles involved in the transition. If there were no such flux on one side, one may be forced to include it on the other if the Euler characteristic changes, to preserve anomaly cancellation. In Ref.~\cite{Jockers:2016bwi} a computation of the change in Euler characteristic in going from resolved to deformed side yields
\be
\Delta \left(\ec(\bfs)\right) = N(N-1)(N+1)K_{\ssu}^2 \,.
\ee
We see that as $K_{\ssu}^2=0$ in the singular transitions we consider, the Euler characteristic is then predicted to be unchanged, as expected since we proved in Section~\ref{sec:dual_checks_ind_of_ns5_config} that the Euler characteristic is independent of the NS5-brane configuration.

\bigskip

In Ref.~\cite{Jockers:2016bwi} the spectrum of the compactified M-theory is computed at each stage of the singular transition. However, their analysis is restricted to cases where $\irr=0$, $\arge\geq1$. This restriction is imposed to avoid non-perturbative corrections from M5-branes wrapping the shrinking divisors in the fourfold, which can contribute only if $\arge-\irr=0$. The induced non-perturbative superpotential depends on the volume $v$ of a shrinking divisor as $e^{-v}$, and hence blows up. We see from the above Hodge diamonds that the singular transitions we consider may not obey the particular restrictions imposed in Ref.~\cite{Jockers:2016bwi}, and in particular may have $\arge-\irr=0$. In the latter case, this physical obstruction to the singular transition should appear dual to world-sheet instantons in the NS5-brane transition. We see from the above Hodge diamonds that the restriction $\irr=0$, $\arge\geq1$ implies that $g=0$, so the matter spectra computations in Ref.~\cite{Jockers:2016bwi} only apply when in the heterotic dual the NS5-branes multiply wrap a $\mbb{P}^1$.

We now review the spectrum computations of Ref.~\cite{Jockers:2016bwi}, keeping in mind the above restrictions. At non-generic points in the Coulomb branch, vector multiplet scalar vevs coincide, corresponding to partial resolution of the singularity. If the $i^{\mathrm{th}}$ value occurs $k_i$ times, the gauge symmetry at this point is
\be
\mathrm{SU}(k_1) \times \ldots \mathrm{SU}(k_m) \times \mathrm{U}(1)^{m-1} \,,
\label{eq:partres_gaugesym}
\ee
where $\sum_{i=1}^m k_i = N$. At this point the Higgs branch has complex dimension
\be
(\arge-1)\sum_{i=1}^m(k_i^2-1)+\arge(m-1) \,,
\label{eq:partres_higgsbranch}
\ee
corresponding to deformations of the remaining singularities. The expressions for generic points are the specialisations with $m=N$ and $k_i=1$, $\forall i$. Additionally, one can note an alternative path from Coulomb to Higgs branch as follows. At generic points in the Coulomb branch there are $\arge(N-1)$ neutral chiral multiplets. Giving these vevs before sending the vector multiplet scalar vevs to zero results in a U$(1)^{N-1}$ gauge symmetry, and leaves $(\arge-1)N(N-1)$ charged chiral fields. The geometry then has curve-like singularities, and these can be deformed away by giving vevs to the charged fields, resulting in the full Higgs branch dimension.

Recalling the discussion above in Sections \ref{sec:loc_dual_fth_geom} and \ref{sec:global_dual_fth_geom}, it is clear how this M-theory geometry corresponds to dual NS5-brane configurations. The situation of an $A_{N-1}$ singularity on $\ssu$ corresponds to a stack of $N$ NS5-branes wrapping $\scu$ in the heterotic base $\hb$. Resolution of the singularity (Coulomb branch) corresponds to moving M5-branes apart in the bulk, and deformation (Higgs branch) to deformation of the brane stack in the base of the CY manifold. The $N-1$ three-dimensional U(1) vector bosons on the Coulomb branch come from the four-dimensional chiral multiplets parametrising distances between M5-branes in the 11d bulk. The three-dimensional chiral fields parametrising the Higgs branch come from the four-dimensional chiral multiplets parametrising brane configurations in the CY manifold. Partial resolution gives multiple stacks of branes, corresponding to the gauge group in Equation~\eqref{eq:partres_gaugesym}, and leaving available deformations of each stack, corresponding to the branch in Equation~\eqref{eq:partres_higgsbranch}. Turning on vevs of the $\arge(N-1)$ neutral chiral multiplets at generic points in the Coulomb branch corresponds to moving the branes in the CY while at different 11d bulk positions. The alternate path between Coulomb and Higgs branch corresponds to first such a shift, followed by bringing the branes together in the bulk, and finally deforming the resulting transversely-intersecting configuration, as when following the lower path in the example of Figure~\ref{fig:coin_5branes_branches_on_p2}, from resolution to deformation.

Since the four-dimensional field theories of the resolution and deformation branches are best-understood, the three-dimensional theory of the singular situation is of most interest. Given the above correspondence, the three-dimensional gauge enhancement at the brane stack appears to correspond to enhancement of four-dimensional chiral multiplets. This appears to be distinct from any gauge enhancement of four-dimensional vector multiplets, since those correspond to three-dimensional chiral multiplets. This is an interesting role reversal. We hope to investigate the lift to four dimensions further in future work.


\section{Summary and outlook}
\label{sec:Conclusion}


We set out to describe F-theory duals of heterotic line bundle models in compactification to four dimensions. As we have argued, the spectral cover does not provide a useful description for these models, as all of the heterotic bundle information is in the spectral sheaf with the spectral sheet being trivial. The dual F-theory geometry correspondingly contains two $\mathrm{E}_8$ singularities with $G_4$ flux on these loci that is naturally dual to the heterotic line bundles. The requirement of having a standard F-theory dual restricts the possible line bundle models, in particular they are necessarily non-chiral. These models furthermore require `horizontal' NS5-branes wrapping curves in the base of the elliptic threefold for anomaly cancellation. The remainder of the anomaly is cancelled by either line bundle flux or `vertical' NS5-branes on the fiber.

The horizontal NS5-branes are dual to blow-ups in the base of the F-theory fourfold, and we treat this aspect of the duality in detail. We reviewed the local F-theory geometry dual to the inclusion of a horizontal NS5-brane, and described the global structure of F-theory fourfolds dual to horizontal line bundle models with various choices of NS5-brane content. We studied situations for which the F-theory fourfold is described by a toric hypersurface, as well as more general situations, the case reflecting whether the NS5-branes wrap toric subspaces in the heterotic base.

We then verified various aspects of the duality for these models. We first treated the matches concerning the $\mathrm{E}_8$ fluxes: matching of anomaly conditions and bundle stability conditions, and multiplets in the two $\mathrm{E}_8$ sectors including massless $\mathrm{U}(1)$s. We then examined the aspects related to NS5-brane configuration / F-theory base geometry: we gave the vector and chiral multiplet counts in this sector, and proved for a broad class of models that these matches hold quite generally. We also discussed an interesting subtlety: when NS5-branes wrap base curves that don't intersect the discriminant locus, the heterotic chiral multiplet count is naively larger than the F-theory count. We argued that the resolution has two parts: the first is that some heterotic moduli are actually massive, and the second is that some F-theory moduli may become massless only in the stable degeneration limit. For the latter, one may expect that the corresponding heterotic chiral multiplets are lifted away from the limit by corrections to the supergravity descriptions that are valid only in the stable degeneration limit.

Finally, having constructed F-theory duals of heterotic models with arbitrary NS5-brane content, we used these to explore the F-theory duals of coincident and intersecting NS5-branes. We found a satisfying picture of how the toric description of the F-theory base reflects the heterotic NS5-brane configuration: the ordering of these NS5-branes as M5-branes in the Ho\v{r}ava-Witten interval is neatly reflected in the triangulation of the toric polytope, and the coincidence or intersection of NS5-branes is reflected in an obvious removal of cones in the triangulation. In order to have a description of the transitions catalysed by coincident NS5-branes in terms of effective field theory, we have confined ourselves to a discussion of the three-dimensional theory resulting from compactification of M-theory on this singular fourfold.

\bigskip

The constructions in this work allow for various possible extensions. First, we have not discussed intersections of horizontal and vertical NS5-branes. Whereas coincident horizontal NS5-branes correspond to singularities in the F-theory base, vertical branes correspond to D3-branes. Intersections between these sectors hence give a heterotic (or F-Theory) description of D3-branes at singularities in Type IIB. Investigation of this aspect may reveal new tools for model building in heterotic string theory. Second, F-theory models with surfaces of singularities in the base provide an alternative description of the four-dimensional theory arising from coincident heterotic NS5-branes, and it would be very interesting to use F-Theory to elucidate the physics of such configurations. Third, it would be very interesting to use these dual models, which in some respects are quite simple and `clean', as a basis for attempting to extend the duality to heterotic bundles that are not flat on the elliptic fiber. Note that this would allow the construction of F-theory duals of chiral heterotic line bundle models. Fourth, we would also like to extend the duality to cases where the heterotic elliptic threefold has multiple sections. In particular this would allow the construction of F-theory duals of the heterotic line bundle models constructed in Ref.~\cite{Braun:2017feb} that realise the Standard Model gauge group by exploiting the possibility of a quotient under exchange of sections to introduce Wilson lines.

\section*{Acknowledgements}
We thank Lara Anderson, Evgeny Buchbinder, Thomas Grimm, Hans Jockers, Dave Morrison, and Wati Taylor for useful discussions. A.~P.~B.~and A.~L.~would like to acknowledge support by the STFC grant~ST/L000474/1. A.~P.~B~is furthermore supported by the ERC Consolidator Grant 682608 ``Higgs bundles: Supersymmetric Gauge Theories and Geometry'' (HIGGSBNDL). C.~R.~B.~is supported by an STFC studentship. The work of A.~L.~and F.~R.~is supported by the EPSRC network grant EP/N007158/1.
\newpage

\newpage

\appendix

\section{Explicit blow-ups in 6d case}
\label{app:explicit_blowups_6d}

In this appendix we explicitly perform the blow-ups discussed in Section~\ref{sec:glob_models_in_6d}, for the example in the left of Figure~\ref{fig:fullyblownup_fan}, verifying that this toric procedure indeed removes any singularities worse than $\mathrm{E}_8$. We recall the setup. We are interested in global F-theory models dual to six-dimensional heterotic line bundle models. In this case there are $\mathrm{E}_8$ singularities at the two poles of the F-theory $\mbb{P}^1$, and hence the intersection of the remaining brane locus with these $\mathrm{E}_8$ singularities produces even more severe singularities, specifically with vanishing orders $(f,g,\Delta)\sim(4,6,12)$, which we call $\tilde{\mathrm{E}}_8$ singularities. The locus of these intersections has class $-12K_{\fbs}$, and these singularities require blow-ups in the F-theory base. We proposed that a possible F-theory threefold after these blow-ups is the fibration over the base given in the leftmost diagram of Figure~\ref{fig:fullyblownup_fan}. However it is not immediately obvious that this is so, since we may expect that torically we do not have enough freedom to blow up over all the intersection points. We see that if we want to use toric blow-ups to resolve the $\tilde{\mathrm{E}}_8$ singularities we must collect these singularities together in single points on each $\mathrm{E}_8$ stack, then perform repeated blow-ups over this point. We do this explicitly below, showing the result is a crepant resolution of these two singularities, leaving in the end only the two $\mathrm{E}_8$ singularities.

We write $\fpc_i$ for the homogeneous coordinates of the F-theory $\mbb{P}^1$, and $\bc_i$ for the homogeneous coordinates of the heterotic base, which forms the other $\mbb{P}^1$ of the F-theory base $\fbs$. After tuning $\mathrm{E}_8$ singularities at $\fpc_1=0$ and $\fpc_2=0$ we have for the Weierstrass model the expressions
\be
\wsf = \fpc_1^4\fpc_2^4\wsf_4 \,, \quad \wsg = \fpc_1^5\fpc_2^5\left( \fpc_1^2\wsg_7+\fpc_1\fpc_2\wsg_6+\fpc_2^2\wsg_5\right) \,, \quad \Delta = \fpc_1^{10}\fpc_2^{10}\Delta_r \,,
\ee
\be
\Delta_r = 4\fpc_1^2\fpc_2^2\wsf_4^3+27\left(\fpc_1^2\wsg_7+\fpc_1\fpc_2\wsg_6+\fpc_2^2\wsg_5\right)^2 \,.
\ee
Now we tune such that all the intersections of $\{\Delta_r=0\}$ with $\{\fpc_1=0\}$ are at $\bc_1=0$, and all the intersections with $\{\fpc_2=0\}$ are at $\bc_1=0$ too. These intersections are clearly governed by $\wsg_7$ and $\wsg_5$, so we are restricting to the case
\be
\wsf = \fpc_1^4\fpc_2^4\wsf_4 \,, \quad \wsg = \fpc_1^5\fpc_2^5\left( \alpha \fpc_1^2 \bc_1^{12}+\fpc_1\fpc_2\wsg_6+\beta \fpc_2^2\bc_1^{12} \right) \,,
\ee
where $\alpha$ and $\beta$ are arbitrary non-zero constants, and $\wsf_4$ and $\wsg_6$ remain arbitrary polynomials in $\{\bc_1,\bc_2\}$. We now blow up at $\{\fpc_1=0,\bc_1=0,\tacx=0,\tacy=0\}$, adding a new coordinate $\ep_1$, and taking the appropriate proper transform, giving a crepant resolution. We put hats on the resulting coordinates to indicate they are the coordinates after resolution. We then have the identifications
\be
\fpc_1 = \h{\fpc}_1\ep_1 \,, \quad \bc_1 = \h{\bc}_1 \ep_1 \,, \quad \tacx = \h{\tacx}\ep_1^2 \,, \quad \tacy = \h{\tacy}\ep_1^3 \,,
\ee
\be
\hat{\wsf} = \ep_1^{-4}\wsf = \h{\fpc}_1^4\h{\fpc}_2^4\wsf_4 \,, \quad
\hat{\wsg} = \ep_1^{-6}\wsg = \h{\fpc}_1^5\h{\fpc}_2^5\left( \alpha \h{\fpc}_1^2 \h{\bc}_1^{12}\cdot \ep_1^{13}+\h{\fpc}_1\h{\fpc}_2\wsg_6+\beta \h{\fpc}_2^2\h{\bc}_1^{12}\cdot \ep_1^{11} \right) \,.
\ee
We can now perform another similar blow-up, this time at $\{\fpc_1=0,\ep_1=0,\tacx=0,\tacy=0\}$. We will continue to write a single hat on a coordinate to indicate it is the coordinate after the blow-up. Then we have the following,
\be
\fpc_1 = \h{\fpc}_1\ep_1\ep_2^2 \,, \quad \fpc_1 = \h{\fpc}_1 \ep_1\ep_2 \,, \quad \tacx = \h{\tacx}\ep_1^2\ep_2^4 \,, \quad \tacy = \h{\tacy}\ep_1^3\ep_2^6 \,,
\ee
\be
\hat{\wsf} = \ep_2^{-4}\ep_1^{-4}\wsf = \h{\fpc}_1^4\h{\fpc}_2^4\wsf_4 \,, \quad
\hat{\wsg} = \ep_2^{-6}\ep_1^{-6}\wsg = \h{\fpc}_1^5\h{\fpc}_2^5\left( \alpha \h{\fpc}_1^2 \h{\bc}_1^{12}\cdot \ep_1^{13}\ep_2^{14}+\h{\fpc}_1\h{\fpc}_2\wsg_6+\beta \h{\fpc}_2^2\h{\bc}_1^{12}\cdot \ep_1^{11}\ep_2^{10} \right) \,.
\ee
We continue in this way until we have blown-up all the way to $\ep_{12}$. It is straightforward to see that we will then have the following expressions,
\be
\fpc_1 = \h{\fpc}_1 L_{\ep} \,, \quad \bc_1 = \h{\bc}_1 C_{\ep} \,, \quad \tacx = \h{\tacx}L_{\ep}^2 \,, \quad \tacy = \h{\tacy}L_{\ep}^3 \,,
\ee
\be
\hat{\wsf} = C_{\ep}^{-4}\wsf = \h{\fpc}_1^4\h{\fpc}_2^4\wsf_4 \,, \quad
\hat{\wsg} = C_{\ep}^{-6}\wsg = \h{\fpc}_1^5\h{\fpc}_2^5\left( \alpha \h{\fpc}_1^2 \h{\bc}_1^{12}\cdot L_{\ep}C_{\ep}^{12}+\h{\fpc}_1\h{\fpc}_2\wsg_6+\beta \h{\fpc}_2^2\h{\bc}_1^{12}\cdot L_{\ep}^{-1}C_{\ep}^{12} \right)\,.
\ee
where $L_{\ep} = \ep_1\ep_2^2\ep_3^3\ldots\ep_{12}^{12}$ and $C_{\ep} = \ep_1\ldots\ep_{12}$. We see that in the term with a $\beta$ coefficient, there are no powers of $\ep_{12}$. Hence we cannot perform any more of these blow-ups, as we would not be able to divide $\ep_{13}^6$ out of $g$. It is also clear that in order to be able to perform the 12 blow-ups, we did need all 12 of the points of intersection of $\fpc_1=0$ with $\Delta_r=0$ to sit on top of one another, since if $\wsg_5$ had a lower power of $\h{\bc}_1$ we would have run out of powers of $\ep$ factors earlier.

We can now perform the blow-ups on the other $\mathrm{E}_8$ brane. Analogously to above, we start with the point $\{\fpc_2=0,\bc_1=0,\tacx=0,\tacy=0\}$, and perform successive blow-ups, with new coordinates $\ep_{-1},\ep_{-2},\ldots$. Clearly after this process is complete we will have the following expressions,
\be
\fpc_1 = \h{\fpc}_1 L_{\ep_+} \,, \quad \fpc_2=\hat{\fpc}_2 L_{\ep_-} \,, \quad \bc_1 = \h{\bc}_1 C_{\ep_+}C_{\ep_-} \,, \quad \tacx = \h{\tacx}L_{\ep_+}^2L_{\ep_-}^2 \,, \quad \tacy = \h{\tacy}L_{\ep_+}^3L_{\ep_-}^3 \,,
\ee
\be
\hat{\wsf} = C_{\ep_+}^{-4}C_{\ep_-}^{-4}\wsf = \h{\fpc}_1^4\h{\fpc}_2^4\wsf_4 \,, \quad
\hat{\wsg} = C_{\ep_+}^{-6}C_{\ep_-}^{-6}\wsg = \h{\fpc}_1^5\h{\fpc}_2^5\left( \alpha \h{\fpc}_1^2 \h{\bc}_1^{12}\cdot A+\h{\fpc}_1\h{\fpc}_2\wsg_6+\beta \h{\fpc}_2^2\h{\bc}_1^{12}\cdot B \right) \,,
\ee
where
\be
A = L_{\ep_+}L_{\ep_-}^{-1}C_{\ep_+}^{12}C_{\ep_-}^{12} \,, \quad 
B = L_{\ep_+}^{-1}L_{\ep_-}C_{\ep_+}^{12}C_{\ep_-}^{12} \,,
\ee
in which
\be
L_{\ep_{\pm}} = \ep_{\pm1}\ep_{\pm2}^2\ldots\ep_{\pm12}^{12} \,, \quad C_{\ep_{\pm}} = \ep_{\pm1}\ldots\ep_{\pm12} \,.
\ee
These are the expressions in the final situation, after all blow-ups have been performed. The hypersurface is determined by the standard Weierstrass equation in the new coordinates,
\be
\h{\tacy}^2 = \h{\tacx}^3 + \h{f}\h{\tacx}\h{\tacz}^4 + \h{g}\h{\tacz}^6 \,.
\ee

It is clear that we still have two $\mathrm{E}_8$ singularities, which now sit at $\h{\fpc}_1=0$ and $\h{\fpc}_2=0$ (and it is straightforward to check from the ray diagram that $\h{\fpc}_1=0$ and $\h{\fpc}_2=0$ remain $\mbb{P}^1$s in the threefold base). It is also clear that the $\mathrm{E}_8$ singularities are geometrically non-Higgsable, by the following argument. The weight system has a row
\be
\begin{tabular}{c c c c c c c c c c c c c c}
$\hat{\tacy}$ & $\hat{\tacx}$ & $\hat{\tacz}$ & $\hat{\fpc}_1$ & $\hat{\fpc}_2$ & $\hat{\bc}_1$ & $\hat{\bc}_2$ & $\ep_1$ & \ldots & $\ep_{11}$ & $\ep_{12}$ & $\ep_{-1}$ & $\ldots $ & $\ep_{-12}$\\
\hline
42 & 28 & 0 & 0 & 12 & 0 & 1 & 0 & \ldots & 0 & 1 & 0 & \ldots & 0\\
\end{tabular} \,.
\ee
It is easy to check that each term in $g$ has degree $84$ under this scaling. We see we could not have a term with 8 or more powers of $\h{\fpc}_2$ in $g$, as it would overshoot the degree for the scaling above. Since we also know every term in $g$ contains a factor $\h{\fpc}_1^a\h{\fpc}_2^b$ with $a+b=12$, we see that we cannot have a term with fewer than 5 powers of $\h{\fpc}_1$. Analogously we cannot have a term with fewer than 5 powers of $\h{\fpc}_2$, and hence we have that $g \propto \h{\fpc}_1^5\h{\fpc}_2^5$. We can also make the analogous statement for $f$, so that $f \propto \h{\fpc}_1^4\h{\fpc}_2^4$. Hence we always have that $\Delta \propto\h{\fpc}_1^{10}\h{\fpc}_2^{10}$, so the $\mathrm{E}_8$ singularities are geometrically non-Higgsable. It is also clear that we have removed all intersections of the remaining brane locus with the $\mathrm{E}_8$ stacks, as follows. At $\h{\fpc}_1=0$ we have
\be
\left.\Delta_r\right|_{\h{\fpc}_1=0} = 27(\beta \h{\fpc}_2^2\h{\bc}_1^{12} B)^2 \,.
\ee
This expression involves only $\h{\fpc}_2$, $\h{\bc}_1$, and the $\ep_{-i},\ep_i$. From the fan it is clear that of these only $\ep_{12}$ is allowed to vanish when $\h{\fpc}_1=0$. However this is precisely the coordinate that does not appear in $B$. Hence we have that $\left.\Delta_r\right|_{\h{\fpc}_1=0} \neq 0$. Clearly we can make an analogous statement for the $\h{\fpc}_2=0$ surface. Hence the remaining brane locus does not intersect either of the $\mathrm{E}_8$ singularities. Finally we also note that for a generic $\wsf_4$ and $\wsg_6$ there should be no other singularities.

\section{Explicit blow-ups in 4d case}
\label{app:explicit_blowups_4d}

In this appendix, analogously to the six-dimensional case in Appendix \ref{app:explicit_blowups_6d}, we show in an example that toric blow-ups are sufficient to remove all singularities worse than $\mathrm{E}_8$ in the F-theory fourfold dual to a four-dimensional heterotic line bundle model. This example was discussed above in Section~\ref{sec:toric_global_models}, and the result upon performing all toric blow-ups will be the F-theory base fan shown in Figure~\ref{fig:fullyblownup_P1P1P1_fan}. 

In this example, the heterotic base is $\hb = \mbb{P}^1 \times \mbb{P}^1$, where the two $\mbb{P}^1$s have coordinates $\bc_i$ and $\bct_i$ respectively. Additionally we specialise to a particular tuning\footnote{While we are here treating an example case with a particular $\hb$ and a particular tuning of $\wsg_7$ and $\wsg_5$, it will be clear that the computations apply also to an arbitrary $\hb$ and to different choices of tuning allowing toric blow-ups.} of the functions in the Weierstrass equation,
\be
\wsg_7 = \alpha \bc_1^{12}\bct_1^{12} \,, \quad \wsg_5 = \beta \bc_1^{12}\bct_1^{12} \,.
\ee
We then proceed to blow up multiple times over $\bc_1=0$ and $\bct_1=0$ in each of the two $\fpc_{1,2}=0$ surfaces, each time blowing up on a previous exceptional divisor, and each time taking the proper transform to describe the crepant resolution. This is exactly analogous to the procedure in Appendix \ref{app:explicit_blowups_6d}, so we do not repeat it. Altogether we will blow up 12 times over each of the four severely-singular loci. After all these blow-ups and proper transforms, it is straightforward to see that we end up with the following expressions for the Weierstrass model,
\be
\hat{f} = \h{\fpc}_1^4\h{\fpc}_2^4\wsf_4 \,, \quad
\hat{g} = \h{\fpc}_1^5\h{\fpc}_2^5\left( 
\alpha \h{\fpc}_1^2 \h{\bc}_1^{12}\h{\bct}_1^{12}\cdot A + 
\h{\fpc}_1\h{\fpc}_2\wsg_6+
\beta \h{\fpc}_2^2\h{\bc}_1^{12}\h{\bct}_1^{12}\cdot B \right)\,.
\ee
where
\be
A = \prod_{i=1}^{12} \left(\ep_{-i}\rho_{-i}\right)^{12-i} \left(\ep_i\rho_i\right)^{12+i} \,, \quad 
B = \prod_{i=1}^{12} \left(\ep_{-i}\rho_{-i}\right)^{12+i} \left(\ep_i\rho_i\right)^{12-i} \,,
\ee
again analogously to the six-dimensional case.

We can also verify that each of $\h{\fpc}_1=0$ and $\h{\fpc}_2=0$, intersected with the section, are $\mbb{P}^1\times\mbb{P}^1$s, so they are diffeomorphic to the heterotic base. Let us focus on the $\h{\fpc}_1=0$ case. We know that none of the following coordinates can vanish, $\{\ep_{-i},\rho_{-i},\ep_1,\ldots,\ep_{11},\rho_1,\ldots,\rho_{11},\bc_1,\bct_1,\fpc_2\}$, so we can set them to 1 using scaling relations. Noting that
\be
\vec{\bct}_2+\vec{\rho}_{12}+12\vec{\h{\fpc}}_2=0 \quad \mathrm{and} \quad \vec{\bc}_2+\vec{\ep}_{12}+12\vec{\h{\fpc}}_2=0 \,,
\ee
we see that can rework the weight system to look as follows.
\be
\begin{tabular}{c c c c c c}
$\h{\bc}_1$ & $\h{\bc}_2$ & $\h{\bct}_1$ & $\h{\bct}_2$ & $\ep_{12}$ & $\rho_{12}$ \\
\hline
1 & 1 & 0 & 0 & 0 & 0 \\
0 & 0 & 1 & 1 & 0 & 0 \\
0 & 1 & 0 & 0 & 1 & 0 \\
0 & 0 & 0 & 1 & 0 & 1 \\
\end{tabular}
\ee
We can then use the first two rows to set $\h{\bc}_1=\h{\bct}_1=1$, giving the weight system for a $\mbb{P}^1\times\mbb{P}^1$,
\be
\begin{tabular}{c c c c c c}
$\h{\bc_2}$ & $\h{\bct}_2$ & $\ep_{12}$ & $\rho_{12}$ \\
\hline
1 & 0 & 1 & 0 \\
0 & 1 & 0 & 1 \\
\end{tabular}
\label{eq:gut_p1xp1}
\ee
Clearly we can make the same computation for $\h{\fpc}_2=0$. We can also ask whether the remaining brane locus $\Delta_r=0$ intersects either of the $\mathrm{E}_8$ stacks at $\h{z}_1=0$ and $\h{z}_2=0$. This computation is exactly analogous to the six-dimensional computation in Appendix \ref{app:explicit_blowups_6d}. At $\h{z}_1=0$ we have
\be
\left.\Delta_r\right|_{\fpc_1=0} = 27(\beta \h{\fpc}_2^2\h{\bc}_1^{12}\h{\bct}_1^{12} B)^2 \,.
\ee
This expression involves only $\h{\fpc}_2$, $\h{\bc}_1$, $\h{\bct}_1$ and the $\ep_{-i},\ep_i,\rho_{-i},\rho_i$. From the fan it is clear that of these only $\ep_{12}$ and $\rho_{12}$ are allowed to vanish when $\h{\fpc}_1=0$. However these are precisely the coordinates that do not appear in $B$, and hence we have that $\left.\Delta_r\right|_{\fpc_1=0} \neq 0$. Clearly we can make an analogous statement for the $\h{\fpc}_2=0$ surface. So we see that the remaining brane locus does not intersect either of the $\mathrm{E}_8$s. We can also ask whether the $\mathrm{E}_8$ singularities at $\h{\fpc}_1=0$ and $\h{\fpc}_2=0$ are now geometrically non-Higgsable. The result is that indeed they are, and one can verify this with an exactly analogous computation as we used in the case of compactification to six dimensions in Appendix \ref{app:explicit_blowups_6d}. Additionally, for a generic $\wsf_4$ and $\wsg_6$, there should be no other singularities.

\newpage

\bibliographystyle{JHEP}
\bibliography{bibliography}

\end{document}